\documentclass[reprint,showpacs,superscriptaddress, english, longbibliography]{revtex4-1}
\usepackage[utf8]{inputenc}
\usepackage{color}
\usepackage{array}
\usepackage{verbatim}
\usepackage{multirow}
\usepackage{amsmath}
\usepackage{amssymb}
\usepackage{dsfont}
\usepackage{graphicx}
\usepackage{esint}
\usepackage[unicode=true,
 bookmarks=true,bookmarksnumbered=true,bookmarksopen=false,
 breaklinks=true,pdfborder={0 0 0},backref=false,colorlinks=true]
 {hyperref}
\usepackage{cleveref}
\usepackage{enumitem}
\usepackage{ntheorem}
\usepackage{cleveref}
\usepackage{longtable}
\usepackage{bbm}

\hypersetup{
citecolor={blue},
urlcolor={magenta}
}

\def\ie{{\it i.e.},\ }
\def\eg{{\it e.g.},\ }
\def\etc{{\it etc.}\ }
\def\etal{{\it et al.}\ }

\newtheorem{mythm}{Theorem}
\newtheorem{mydef}[mythm]{Definition}

\crefname{appendix}{Appendix}{Appendices}
\crefname{equation}{Eq.}{Eqs.}
\crefname{figure}{Fig.}{Figs.}
\crefname{section}{Section}{Sections}
\crefname{table}{Table}{Tables}
\crefname{mythm}{Theorem}{Theorems}
\crefname{mydef}{Definition}{Definitions}
\crefname{mycnj}{Conjecture}{Conjectures}
\crefname{mylem}{Lemma}{Lemmas}

\newcommand{\mbf}{\mathbf}

\newcommand{\mbb}{\mathbb}

\newcommand{\mrm}{\mathrm}

\newcommand{\ovl}{\overline}

\def\beq#1\eeq{\begin{equation}#1\end{equation}}
\def\beqs#1\eeqs{\begin{align}#1\end{align}}

\def\pare#1{\left( #1 \right)}

\def\brace#1{\left\{#1\right\}}

\def\nono{\nonumber}

\def\pr{\prime}
\def\prpr{{\prime\prime}}
\def\kk{\mbf{k}}

\def\dim{\mrm{dim}}
\begin{document} 
\title{Fragile Phases As Affine Monoids: Classification and Material Examples} 
\author{Zhi-Da Song}
\thanks{These authors contributed equally to this work}
\affiliation{Department of Physics, Princeton University, Princeton, New Jersey 08544, USA}
\author{Luis Elcoro}
\thanks{These authors contributed equally to this work}
\affiliation{Department of Condensed Matter Physics, University of the Basque Country UPV/EHU, Apartado 644, 48080 Bilbao, Spain}
\author{Yuan-Feng Xu}
\affiliation{Max Planck Institute of Microstructure Physics, 06120 Halle, Germany}
\author{Nicolas Regnault}
\affiliation{Laboratoire  de  Physique  de  l'Ecole  normale  sup\'erieure PSL  University,  CNRS,  Sorbonne  Universit\'e,  Universit\'e  Paris  Diderot, Sorbonne  Paris  Cit\'e,  24  rue  Lhomond,  75005  Paris  France}
\affiliation{Department of Physics, Princeton University, Princeton, New Jersey 08544, USA}
\author{B. Andrei Bernevig}
\email{bernevig@princeton.edu}
\affiliation{Department of Physics, Princeton University, Princeton, New Jersey 08544, USA}
\affiliation{Max Planck Institute of Microstructure Physics, 06120 Halle, Germany}
\affiliation{Physics Department, Freie Universitat Berlin, Arnimallee 14, 14195 Berlin, Germany}

\date{\today}

\begin{abstract}
Topological phases in electronic structures contain a new type of topology, called fragile, which can arise, for example, when an Elementary Band Representation (Atomic Limit Band) splits into a particular set of bands. We obtain, for the first time, a complete classification of the fragile topological phases which can be diagnosed by symmetry eigenvalues, to find an incredibly rich structure which far surpasses that of stable/strong topological states. We find and enumerate all hundreds of thousands of different fragile topological phases diagnosed by symmetry eigenvalues (available at \href{https://www.cryst.ehu.es/html/doc/FragileRoots.pdf}{this http URL}), and link the mathematical structure of these phases to that of Affine Monoids in mathematics. Furthermore, we predict and calculate, for the first time, (hundred of realistic) materials where fragile topological bands appear, and show-case the very best ones. 
\end{abstract}

\maketitle

\section{Introduction}
Since the birth of topological insulators (TIs) \cite{Kane2005_QSH,Kane2005_Z2,Bernevig2006,Konig2007,Fu2007,Zhang2009,Chen2009,Xia2009,kitaev_periodic_2009}, researchers have found topological states of matter to be a theoretically and experimentally versatile field where new phenomena are uncovered every year \cite{Zhang2011RMP,Kane2010RMP,Moore2010}. 
From topological semimetals \cite{Murakami2007,Wan2011,Xu2011Weyl,Burkov2016,Weng2015Weyl,Xu2015Weyl,Yang2015Weyl,Wang2012,Young2012,Yang2014}, to topological crystalline insulators with symmorphic and non-symmorphic symmetries \cite{Teo2008,Fu2011,Hsieh2012,slager_space_2013,Ken2014TCI,Liu2014,Fang2015}, to higher order topological insulators \cite{Benalcazar2017,Benalcazar2017b,Schindler2018,Langbehn2017,Song2017,Fang2017,Ezawa2018}, the field of topological electronic phases of matter keeps evolving. 
As researchers steadily theoretically solve and experimentally find materials for several topological phases, new, further unknown types of  topological phases arise.

\begin{figure*}
\includegraphics[width=0.8\linewidth]{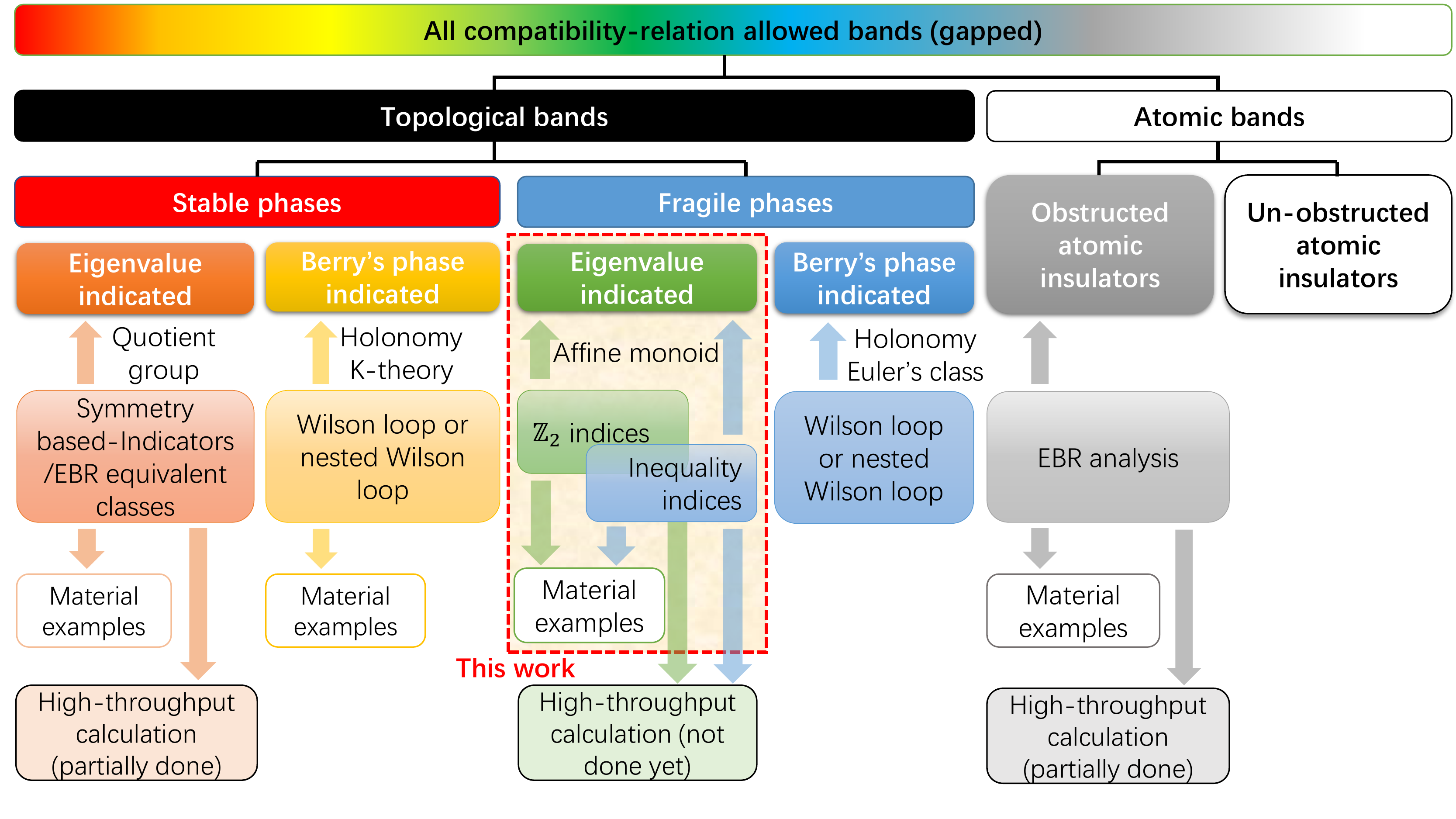}
\protect\caption{ The classification of topological bands, where the shaded area represent the contents of the present work.
All the band structures are classified into three categories: stable topological bands, fragile topological bands, and atomic (trivial) bands.
The stable or fragile topological bands are further classified into two sub-categories: those indicated by symmetry eigenvalues and those not indicated by symmetry eigenvalues.
The atomic bands are also classified into two subcategories depending on whether the Wannier functions locate at the positions of atoms. 
The not-eigenvalue-indicated topological states are usually identified by the Wilson loop method \cite{yu_equivalent_2011,alexandradinata_2014,Benalcazar2017,SlagerAndBouhon,Song2018c}; but the general framework to calculate their topological invariants is still unknown.
The eigenvalue-indicated stable topological states are classified by the TQC \cite{Bradlyn2017} and other theories \cite{Po2017,cano_building_2018}.
The present work finishes the classification of eigenvalue-indicated fragile topological states (EFPs).
\label{fig:workflow}}
\end{figure*}

Recent substantial progress in the field has led to the development of techniques  \cite{Bradlyn2017,Po2017,kruthoff_topological_2017,Elcoro2017,Vergniory2017,Cano2018,Khalaf2018,Song2018a,Song2018b} which can be used for a high-throughput discovery of topological materials, the beginning of which has been undertaken in  \cite{Vergniory2019,Zhang2019,Tang2019,Tang2019b}. 
Topological Quantum Chemistry  (TQC) \cite{Bradlyn2017,cano_building_2018,Vergniory2017,Elcoro2017} and the associated \href{http://www.cryst.ehu.es}{Bilbao Crystallographic Server (BCS)} \cite{Bradlyn2017,Elcoro2017,BCS,*Aroyo2006a,*Aroyo2006b}, have provided for a classification of \textit{all} the atomic limits - whose basis are the so-called Elementary Band Representations (EBRs) -  existent in the 230 non-magnetic space groups (SGs). TQC defines topological phases as the phases not adiabatically continuable to a sum of EBRs. 
This leads to different large series of topological states. The first series are the so-called eigenvalue stable (strong, weak and crystalline) topological states, whose characters at high symmetry points cannot be expressed as linear combination (sum or difference) of characters of EBRs. These have been fully classified and progress towards material high-throughput has been made \cite{Bradlyn2017,Po2017,Khalaf2018,Song2018a,Vergniory2019,Zhang2019,Tang2019}, with several partial catalogues of topological materials already existed. 
The second series are the so-called fragile states of matter - which here we call eigenvalue fragile phases (EFPs) - they cannot be written purely as a sum of characters (traces of representations) of EBRs but can be written as sums and differences of characters of EBRs. Last, there exist stable/fragile states also not characterizable by characters or irreps -  they are characterized by the flow of Berry phases. These fragile phases currently lack classifications - and lack any material examples.   A schematic of the classifications is shown in Fig. \ref{fig:workflow}.

%We refer to fragile topological states that can be diagnosed from symmetry eigenvalues of the corresponding band structure as eigenvalue fragile phases (EFPs). 
Fragile states show up in the examples of TQC \cite{Bradlyn2017,Bradlyn2019} although their potential has only been fully identified after \cite{Po2017, Po2018, Cano2018, SlagerAndBouhon, Ahn2018, Po2018b, Song2018c, de_paz2019, manes_fragile_2019, Haruki2019}.  
Refs.~\cite{Po2018,Bradlyn2019,Ahn2018,Po2018b,Song2018c} have discovered a small number of models of EFP by applying the methods of TQC but neither a general, complete,  (or even partial) classification, nor \emph{any} material examples for these phases are known.
This leaves us in the un-enviable situation of being far from a theoretical understanding of a so-far purely theoretical phase of matter. 
In this work, we perform ---for the first time--- three separate tasks. \textbf{(1)} We provide an elegant, mathematical framework to \textit{fully} classify and diagnose \emph{all} the EFPs. \textbf{(2)} We apply this formalism to the spin-orbit coupling (SOC) doubled groups with time-reversal symmetry (TRS) and classify all the 340,590 EFPs that can exist - a much richer structure than in stable/strong topological phases. \textbf{(3)} We provide examples of one hundred fragile bands in different real materials, some of them extremely well isolated in energy from other bands. 
Our framework is closely related to the mathematical theories of polyhedra and affine monoids, bringing highly esoteric mathematical concepts into real material structures. (See \cref{sec:math} for a mathematical definition.) Thus we call our method the polyhedron method. 
To underscore the importance of fragile topology, the low-lying states of twisted bilayer graphene (TBG), a wonder-material engineered of two twisted layers of graphene \cite{Bistritzer2011,Kim2017-TBG,cao_TBG1,cao_TBG2,Huang2018-TBG,Yankowitz2019-TBG}, is predicted to exhibit a fragile topology \cite{Po2018b,Ahn2018,Song2018c,Tarnopolsky2019,Liu2018_TBG}. 

\section{EFPs in viewpoint of TQC}
To obtain the mathematical structure of EFPs we first review their definition from the viewpoint of TQC.
A band structure is \textit{partially} indexed by its decomposition into irreducible representations (irreps) at the high symmetry momenta \cite{Bradlyn2017,kruthoff_topological_2017,Po2017}.
Such a decomposition is described by a ``symmetry data vector", where entries give the multiplicities of the irreps in the decomposition.
To be specific, we write the symmetry data vector as
\begin{equation}
B = (m_{1_{K_1}},\ m_{2_{K_1}},\ \cdots, m_{1_{K_2}},\ m_{2_{K_2}},\ \cdots)^T.\label{eq:band-structure}
\end{equation}
Here $K_1$, $K_2$, $\cdots$ are a known set of sufficient high symmetry momenta (maximal k-vectors in \cite{Elcoro2017}), and $m_{i_K}$ is the multiplicity of the $i$-th irrep of the little group at $K$.
(Different maximal k-vectors can have different irreps.)
For example, for a one-dimensional system with only inversion symmetry, the symmetry data vector is written as $B=(m_{+_0}, m_{-_0}, m_{+_\pi}, m_{-_\pi})$, where the four entries represent the multiplicities of the inversion even(+)/odd(-) irrep at $k=0,\pi$, respectively. The symmetry data vector of a gapped band structure necessarily satisfies a set of rules called ``compatibility relations" (available for all SGs on the BCS \cite{Bradlyn2017,Po2017,kruthoff_topological_2017,Elcoro2017,Vergniory2017,BCS,*Aroyo2006a,*Aroyo2006b}), which dictate if a given band structure can exist in the Brillouin Zone (BZ).  In the 1D example, the compatibility relation is trivial and enforces an equal band number at $k=0$, $\pi$: $m_{+_0}+m_{-_0} = m_{+_\pi} + m_{-\pi}$. We always assume that the symmetry data vector satisfies the compatibility relations.
For the symmetry data to be consistent with a \textit{trivial} insulator, it should be induced by local orbitals forming representations of the SG in real space. Such trivial insulators are labeled as band representations (BRs); their symmetry data vectors are defined as ``trivial". The generators of BRs are the EBRs \cite{Zak1982,Zak2001}. 

%The EBRs in all SGs are also available on the BCS \cite{Bradlyn2017,Elcoro2017,Vergniory2017,BCS}.
In the 1D  example, there are four EBRs, induced by the even(+)/odd(-) orbital at $x=0/\frac12$, respectively. They are $\mathrm{ebr}_1 = (1,0,1,0)^T$, $\mathrm{ebr}_2 = (0,1,0,1)^T$, $\mathrm{ebr}_3 = (1,0,0,1)^T$, and $\mathrm{ebr}_4 = (0,1,1,0)^T$, respectively.
In general, we can define the EBR matrix as $EBR = (\mathrm{ebr}_1, \mathrm{ebr}_2,\cdots)$, where the $i$-th column of the EBR matrix $\mathrm{ebr}_i$ is the $i$-th EBR of the corresponding SG.
A symmetry data vector $B$ is trivial if and only if there exit $p_1, p_2, \cdots \in \mathbb{N}$ ($\mathbb{N}$ stands for the non-negative integers) such that $B = p_1\mathrm{ebr}_1 + p_2\mathrm{ebr}_2 +\cdots $, or, equivalently,
\begin{equation}
\exists p = (p_1, p_2, \ldots )^T \in \mathbb{N}^{N_{EBR}}\quad s.t.\quad B = EBR\cdot  p.
\end{equation}
Here $N_{EBR}$ is the number of EBRs in the SG. Crucially, the EBRs may not be linearly independent: given $B$ the corresponding $p$ may not be unique.  

Because bands are only partially defined by symmetry data vectors, not all trivial symmetry data vectors imply trivial insulators (topological insulator in space group $P1$ have trivial symmetry data vectors). Hence nontrivial symmetry data is a sufficient but not necessary condition for a band to be topological nontrivial.

For any symmetry data vector $B$ satisfying compatibility relations, there always exists $p \in \mathbb{Q}^{N_{EBR}}$ (rational number) such that $B = EBR \cdot p$ (for a proof, see \cite{Po2017} and also in more detail \cite{PaperOnTheInductionMethor}). Due to this property, \textit{nontrivial} symmetry data vectors can be further classified into two cases, both included in the TQC formalism \cite{Bradlyn2017, Cano2018}: (i) $B$ cannot be written as an integer combination of EBRs but can only be written as fractional rational combination of EBRs, (ii) $B$ can be written as an integer combination of EBRs, and at least one of the coefficients is \emph{necessarily} negative. Case-(i) is characterized by topological indices: symmetry-based indicators \cite{Po2017, Song2018a} or EBR equivalence classes \cite{Bradlyn2017}, and implies robust topology \cite{Po2017,Khalaf2018,Song2018a}. Case-(ii), on the other hand, implies fragile topology \cite{Po2018,Bradlyn2019, Cano2018, SlagerAndBouhon, Song2018c}, and no classification, indices, or material examples are known for it. We provide a full classification and 100 material examples below. A symmetry data vector in case-(ii) can be generally written as $B = \sum_i p_i \mathrm{ebr}_i - \sum_j q_j \mathrm{ebr}_j$, with $p_i, q_j \in \mathbb{N}$ and $p_i q_i=0$ for all $i$. 
(Although not all vectors written in such form represent fragile phases. For example, if $\mathrm{ebr}_1 + \mathrm{ebr}_2 = \mathrm{ebr}_3$, then $\mathrm{ebr}_3 - \mathrm{ebr}_2 = \mathrm{ebr}_1$ is not fragile.)
The topologically nontrivial restriction is that $B$ does not decompose to a sum of EBRs. Once coupled to an atomic insulator BR $\sum_j q_j \mathrm{ebr}_j$, the total band structure, i.e., $\sum_i p_i \mathrm{ebr}_i$, represents a trivial symmetry data vector removing the topology imposed by symmetry eigenvalues - thus ``fragile". 

%\textbf{for BAB: I need to explain better the strategy and the flow of the paper, what is exceptional about it, right now it reads like a laundry list; need to do this in the intro better; mention we link this to a geometric problem, and relation to Hilbert basis. Emphasize the mathematical structure.}

We now introduce a convenient parametrization of the symmetry data. We can always write the Smith Decomposition of the EBR matrix as $EBR = L \Lambda R$, with $L$ (correspondingly $R$) an $N_B \times N_B$ (correspondingly $N_{EBR} \times N_{EBR}$) unimodular integer matrices, $N_B$ the length of symmetry data vector, $\Lambda$ a $N_B \times N_{EBR}$ matrix with diagonal integer entries $\Lambda_{ij}=\delta_{ij}\lambda_i$ for $i=1,2\cdots N_B$, $j=1,2\cdots N_{EBR}$, where $\lambda_i>0$ for $r=1\cdots r$ and $\lambda_i=0$ for $i>r$, with $r$ the EBR matrix rank (\cref{sec:inequality}).
For $B = EBR \cdot p$, for some $p \in \mathbb{Q}^{N_{EBR}}$, we can equivalently write the symmetry data vector as 
\begin{equation}
B_i= \sum_{j=1}^r L_{ij} \lambda_{j} y_j, \label{eq:parametrization}
\end{equation}
where $y$ is defined as $y_j = (Rp)_j$ ($j=1\cdots r$). 
While, in general, the map from $p$ to $y$ is many-to-one due to linear dependence of EBRs, the map from $y$ to $B$ is one-to-one: if $y$ and $y^\prime$ map to the same $B$, ($\sum_j L_{ij} \lambda_j y_j = \sum_{j} L_{ij} \lambda_j y^\prime_j$), multiplying $L^{-1}$ on both sides gives $y=y^\prime$.
Table S1 in \cite{SM} tabulates the parametrizations in all SGs.

As the symmetry data vector $B$ entries represent the multiplicities of the irreps they should be integer non-negative valued for any physical band structure, conditions which are not automatically guaranteed by the parametrization in \cref{eq:parametrization}. We ask: what conditions should the $y$-vector satisfy so that $B$ is: \textbf{(1)} non-negative (zero and strictly positive), \textbf{(2)} integer. For \textbf{(2)}, since the $L$ matrix is unimodular,  $B$ is an integer vector iff  $\lambda_i y_i$ ($i=1\cdots r$) are all integers, or  $y_i= c_i/\lambda_i$ where $c_i = (L^{-1} B)_i  \in  \mathbb{Z}$.
For the trivial and for the nontrivial fragile (case-(ii)) symmetry data vectors, both of which can be written as integer combinations of EBRs - $p \in \mathbb{Z}^{N_{EBR}}$- , the corresponding $y_i=(Rp)_i$ vector must be an integer. 
A fractional $y$, where $c_i\neq0\ \mathrm{mod}\ \lambda_i$ for some $i$,  corresponds to a symmetry data vector $B$ in the nontrivial case-(i). 
In fact, $c_i=(L^{-1} B)_i\ \mathrm{mod}\ \lambda_i$ are the symmetry-based indicators \cite{Po2017,Khalaf2018,Song2018a} or equivalently, the distinct EBR equivalence classes of \cite{Bradlyn2017}. 
In this article, we take the $y$ vector always integer to consider trivial and nontrivial fragile (case-(ii)) symmetry data vectors $B$. 

All the EFPs  (nontrivial case-(ii)) have trivial symmetry-based indicators. Instead, the EFPs are diagnosed by the \emph{fragile indices}. 
We prove that all band structures with time-reversal symmetry and SOC have only two kinds of fragile indices: a $\mathbb{Z}_2$-type (modulo 2) and an inequality-type (\cref{sec:index}).
We give examples of both these cases.

\section{Examples of fragile indices}
\subsection{Example of $\mathbb{Z}_2$-type fragile indices}

\begin{figure*}
\includegraphics[width=1.0\linewidth]{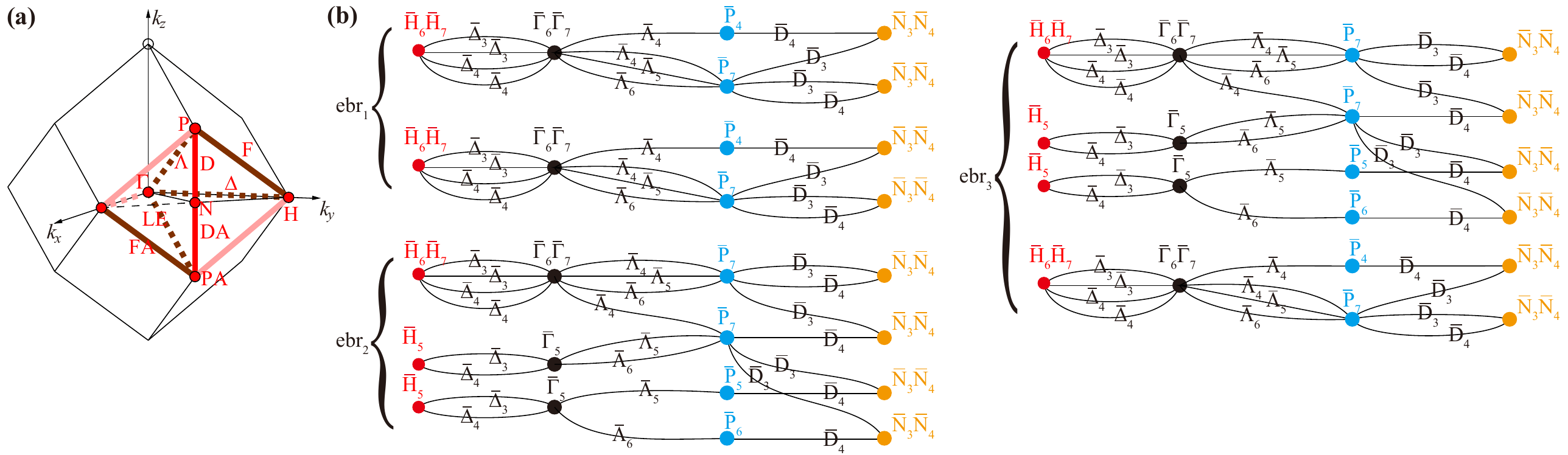}
\protect\caption{(a) BZ of SG 199 ($I2_13$). 
(b) The EBRs of SG 199. The dots and lines represent high symmetry points and the high symmetry lines connecting the high symmetry points, respectively. 
The symbols of irreps, \eg $\Gamma_6\Gamma_7$, are defined on the \href{https://www.cryst.ehu.es/cgi-bin/cryst/programs/representations.pl?tipogrupo=dbg}{REPRESENTATIONS DSG} tool of the BCS \cite{Bradlyn2017,Elcoro2017,BCS}.
\label{fig:SG199BZ}}
\end{figure*}

We consider SG 199 ($I2_13$). 
SG 199 has three EBRs, as shown in \cref{fig:SG199BZ}.
ebr$_1$ and ebr$_3$ split into disconnected branches. 
According to Ref.~\cite{Bradlyn2017}, in each of the splitted EBRs, at least one branch is topological.
For example, the upper branch in ebr$_1$ is not an EBR and hence must be topological.
Since it can be written as $\mrm{ebr}_3-\mrm{ebr}_2$, it at least has a fragile topology.

Now we apply a complete analysis on the EFPs in SG 199.
Since there are only three EBRs, the EBR matrix has three columns.
Arranging the irreps in order of $\overline{\Gamma}_5$, $\overline{\Gamma}_6\overline{\Gamma}_7$, $\overline{\mathrm{H}}_5$, $\overline{\mathrm{H}}_6\overline{\mathrm{H}}_7$, $\overline{\mathrm{P}}_4$, $\overline{\mathrm{P}}_5$, $\overline{\mathrm{P}}_6$, $\overline{\mathrm{P}}_7$, we can write the EBR matrix as
\begin{equation}
EBR = 
\left(\begin{array}{rrr}
0 & 2 & 2 \\
2 & 1 & 2 \\
0 & 2 & 2 \\
2 & 1 & 2 \\
2 & 0 & 1 \\
0 & 1 & 1 \\
0 & 1 & 1 \\
2 & 2 & 3
\end{array}\right).
\end{equation}
Here we have omitted the N point because N has only one type of irrep.
The Smith Decomposition of the $EBR= L \Lambda R$  matrix is
\begin{equation}
\left(\begin{array}{rrrrrrrrrrrrr}
2 & 0 & 1 & 0 & 0 & 0 & 0 & 0 \\
-1 & 1 & 0 & 0 & 0 & 0 & 0 & 0 \\
2 & 0 & 0 & 1 & 0 & 0 & 0 & 0 \\
-1 & 1 & 0 & 0 & 1 & 0 & 0 & 0 \\
-2 & 1 & 0 & 0 & 0 & 1 & 0 & 0 \\
1 & 0 & 0 & 0 & 0 & 0 & 1 & 0 \\
1 & 0 & 0 & 0 & 0 & 0 & 0 & 1 \\
0 & 1 & 0 & 0 & 0 & 0 & 0 & 0
\end{array}\right)
\left(\begin{array}{rrr}
1 & 0 & 0 \\
0 & 1 & 0 \\
0 & 0 & 0 \\
0 & 0 & 0 \\
0 & 0 & 0 \\
0 & 0 & 0 \\
0 & 0 & 0 \\
0 & 0 & 0
\end{array}\right)
\left(\begin{array}{rrr}
0 & 1 & 1 \\
2 & 2 & 3 \\
-1 & -1 & -1
\end{array}\right), \label{eq:EBR-199}
\end{equation}

\begin{figure}
\includegraphics[width=0.9\linewidth]{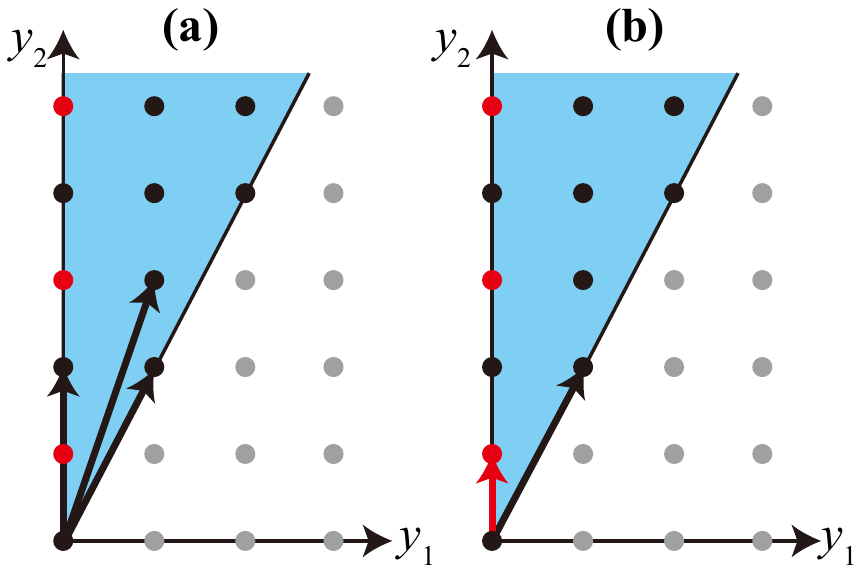}
\protect\caption{(a) EFPs in SG 199 ($I2_13$). The shaded area represents the cone $Y$, the black points represent the trivial points in $\overline{Y}$, the red points represent the EFPs, and the grey points correspond to un-physical symmetry data.
The three bold vectors $(0,2)^T$, $(1,2)^T$, $(1,3)^T$ are the generators of trivial points. 
(b) The Hilbert bases of $\overline{Y}$. The two bold vectors, i.e., $(0,1)^T$, $(1,2)^T$, generates all the points in $\overline{Y}$. $(0,1)^T$ is nontrivial and corresponds to a fragile root, and $(1,2)^T$ is trivial and corresponds to an EBR.
\label{fig:199}}
\end{figure}

The $\Lambda$ matrix has only two nonzero elements, meaning $r=2$, so the symmetry data is parameterized by a two-component integer vector $y=(y_1,y_2)^T$. From  \cref{eq:parametrization}, the symmetry data vector is then given by
\begin{equation}
B = (2y_1, -y_1+y_2, 2y_1, -y_1+y_2, -2y_1+y_2, y_1, y_1, y_2)^T.
\end{equation}
To ensure $B\ge0$, the $y$-vector should satisfy $y_2\ge 2y_1\ge 0$.
Therefore, the physical symmetry data vectors, i.e., $B$'s, belong to the set of integer points $\overline{Y}$ in the 2D cone  (open triangle) $Y = \{ y\in \mathbb{R}^2 | y_2\ge 2y_1 \ge 0\}$ defined as
\begin{equation}
\overline{Y} = \mathbb{Z}^2\cap Y = \{y\in \mathbb{Z}^2 | y_2\ge 2y_1\ge 0\}. \label{cone2D}
\end{equation}
and shown in Fig. \ref{fig:199}a The trivial symmetry data vectors can be written as sums of EBRs, i.e., $B = EBR \cdot p$ for $p \in \mathbb{N}^{N_{EBR}}$. 
They are represented by the $y$ vectors belonging to
\begin{equation}
\overline{X} =\{y\in \mathbb{Z}^r | y_i=(R p)_i\quad p \in \mathbb{N}^{N_{EBR}} \}. \label{eq:Xbar}
\end{equation}
In the case of \cref{eq:EBR-199}, we can write the trivial points as
\begin{equation}
\overline{X} = \{p_1 (0,2)^T + p_2 (1,2)^T + p_3(1,3)^T\;|\; p_{1,2,3}\in \mathbb{N}\},
\end{equation}
i.e., the black points in Fig. \ref{fig:199}a, generated by non-negative $p$ combinations of the three vectors $(0,2)^T, (1,2)^T, (1,3)^T$.
One can find that $(0,2)^T, (1,2)^T, (1,3)^T$ correspond to the ebr$_1$, ebr$_2$, and ebr$_3$ shown in \cref{fig:SG199BZ}, respectively.
We deduce that the nontrivial fragile symmetry data vectors, or the EFPs, are represented by the points in $\overline{Y}-\overline{X}$; these are the red points in Fig. \ref{fig:199}a.

We provide the explicit index for EFPs in SG 199. Consider the subset $y_1=0$ of $\overline{Y}$. Only one generator, i.e., $(0,2)^T$, satisfies this constraint. All the \textit{trivial} points in the $y_1=0$ subset of $\overline{Y}$ are generated by it.
The points  $(0,2p+1)^T$ ($p\in \mathbb{N}$) cannot be reached by non-negative combinations $p$ of EBRs and are nontrivial (fragile).
Thus one fragile criterion of SG 199 is given by
\begin{equation}
y_1=0,\quad \mathrm{and}\quad y_2=1\ \mathrm{mod}\ 2, \label{eq:criterion-199}
\end{equation}
where $y_2\ \mathrm{mod}\ 2$ is the $\mathbb{Z}_2$ index, and $y_1=0$ is the condition for the EFP to be diagnosable.
Are there any other fragile indices (for other points in $\overline{Y}$)? For $y_2$  even (remembering $y_2\ge 2y_1$), we can rewrite $y$ as $y = y_1 (1,2)^T + (\frac12 y_2-y_1)(0,2)^T$ and reach all points in this subspace of $\overline{Y}$;
for $y_2$ odd and $y_1\ge 1$, \cref{cone2D} implies $y_2\ge 2y_1+1$, and we can rewrite $y$ as $y = (1,3)^T + (y_1-1)(1,2)^T + (\frac12 y_2 -\frac12 - y_1)(0,2)^T $ and reach all points in this subspace of $\overline{Y}$. In both cases, we find the points are trivial. Hence only $y_2$ odd and $y_1=0$ are fragile and \cref{eq:criterion-199} is the only fragile index in SG 199. 
In Fig. \ref{fig:199}, we present a diagrammatic illustration of the points in $\overline{Y}$ and points in $\overline{X}$, from which one immediately concludes \cref{eq:criterion-199}. 
A (Hilbert) basis for all points in  $\overline{Y}$ - will be provided later.

\begin{table*}[!ht]
\small
%\begin{tabular}{|p{0.05cm}|p{0.2cm}|p{0.2cm}p{0.2cm}p{0.2cm}p{0.2cm}|p{0.05cm}|p{0.2cm}|p{0.2cm}p{0.2cm}p{0.2cm}p{0.2cm}|p{0.05cm}|l|llll|p{0.05cm}|l|llll|}
\begin{tabular}{|c|c|llll|c|c|llll|c|c|llll|c|l|llll|}
\hline 
\multirow{2}{*}{r} & \multirow{2}{*}{SG} & \multicolumn{2}{c|}{Basis} & \multicolumn{2}{c|}{Index} & 
\multirow{2}{*}{r} & \multirow{2}{*}{SG} & \multicolumn{2}{c|}{Basis} & \multicolumn{2}{c|}{Index} & 
\multirow{2}{*}{r} & \multirow{2}{*}{SG} & \multicolumn{2}{c|}{Basis} & \multicolumn{2}{c|}{Index} & 
\multirow{2}{*}{r} & \multirow{2}{*}{SG} & \multicolumn{2}{c|}{Basis} & \multicolumn{2}{c|}{Index}\tabularnewline
\cline{3-6} \cline{9-12} \cline{15-18} \cline{21-24} 
 &  & \multicolumn{1}{c|}{ebr} & \multicolumn{1}{c|}{root} & \multicolumn{1}{c|}{ieq} & $\mathbb{Z}_{2}$ &  &  & \multicolumn{1}{c|}{ebr} & \multicolumn{1}{c|}{root} & \multicolumn{1}{c|}{ieq} & $\mathbb{Z}_{2}$ &  &  & \multicolumn{1}{c|}{ebr} & \multicolumn{1}{c|}{root} & \multicolumn{1}{c|}{ieq} & $\mathbb{Z}_{2}$ &  &  & \multicolumn{1}{c|}{ebr} & \multicolumn{1}{c|}{root} & \multicolumn{1}{c|}{ieq} & $\mathbb{Z}_{2}$\tabularnewline
\hline 
\hline 
\multirow{4}{*}{2} & 199 & 1 & 1 & 0 & 1 & \multirow{3}{*}{4} & 218 & 2 & 6 & 6 & 0 & \multirow{16}{*}{6} & 69 & 10 & 52 & 24 & 0 & \multirow{7}{*}{8} & 148 & 8 & 140 & 24 & 0\tabularnewline
\cline{2-6} \cline{8-12} \cline{14-18} \cline{20-24} 
 & 208 & 1 & 1 & 0 & 1 &  & 219 & 2 & 6 & 6 & 0 &  & 71 & 10 & 132 & 28 & 0 &  & 166 & 8 & 140 & 24 & 0\tabularnewline
\cline{2-6} \cline{8-12} \cline{14-18} \cline{20-24} 
 & 210 & 1 & 1 & 0 & 1 &  & 220 & 5 & 5 & 4 & 4 &  & 85 & 10 & 16 & 24 & 0 &  & 193 & 9 & 975 & 30 & 24\tabularnewline
\cline{2-12} \cline{14-18} \cline{20-24} 
 & 214 & 1 & 1 & 0 & 1 & \multirow{26}{*}{5} & 11 & 9 & 8 & 8 & 0 &  & 125 & 10 & 16 & 24 & 0 &  & 200 & 8 & 64 & 24 & 0\tabularnewline
\cline{1-6} \cline{8-12} \cline{14-18} \cline{20-24} 
\multirow{8}{*}{3} & 70 & 5 & 10 & 2 & 0 &  & 13 & 9 & 8 & 8 & 0 &  & 129 & 10 & 16 & 24 & 0 &  & 224 & 11 & 90 & 40 & 0\tabularnewline
\cline{2-6} \cline{8-12} \cline{14-18} \cline{20-24} 
 & 150 & 2 & 2 & 1 & 3 &  & 14 & 8 & 8 & 8 & 0 &  & 132 & 10 & 92 & 24 & 0 &  & 226 & 11 & 334 & 28 & 0\tabularnewline
\cline{2-6} \cline{8-12} \cline{14-18} \cline{20-24} 
 & 157 & 2 & 2 & 1 & 3 &  & 15 & 9 & 60 & 12 & 0 &  & 163 & 8 & 68 & 12 & 4 &  & 227 & 13 & 464 & 26 & 0\tabularnewline
\cline{2-6} \cline{8-12} \cline{14-24} 
 & 159 & 4 & 2 & 1 & 1 &  & 49 & 9 & 8 & 8 & 0 &  & 165 & 6 & 40 & 12 & 12 & \multirow{11}{*}{9} & 2 & 16 & 1136 & 240 & 0\tabularnewline
\cline{2-6} \cline{8-12} \cline{14-18} \cline{20-24} 
 & 173 & 4 & 2 & 1 & 1 &  & 51 & 9 & 8 & 8 & 0 &  & 190 & 9 & 51 & 10 & 0 &  & 10 & 16 & 1136 & 240 & 0\tabularnewline
\cline{2-6} \cline{8-12} \cline{14-18} \cline{20-24} 
 & 182 & 4 & 2 & 1 & 1 &  & 53 & 8 & 8 & 8 & 0 &  & 201 & 10 & 8 & 12 & 0 &  & 47 & 16 & 1136 & 240 & 0\tabularnewline
\cline{2-6} \cline{8-12} \cline{14-18} \cline{20-24} 
 & 185 & 2 & 2 & 1 & 3 &  & 55 & 8 & 8 & 8 & 0 &  & 203 & 10 & 84 & 16 & 0 &  & 87 & 14 & 1188 & 56 & 0\tabularnewline
\cline{2-6} \cline{8-12} \cline{14-18} \cline{20-24} 
 & 186 & 4 & 2 & 1 & 1 &  & 58 & 8 & 8 & 8 & 0 &  & 205 & 8 & 6 & 2 & 0 &  & 139 & 14 & 1188 & 56 & 0\tabularnewline
\cline{1-6} \cline{8-12} \cline{14-18} \cline{20-24} 
\multirow{17}{*}{4} & 63 & 7 & 4 & 4 & 0 &  & 66 & 9 & 60 & 12 & 0 &  & 206 & 8 & 13 & 4 & 4 &  & 147 & 8 & 668 & 56 & 16\tabularnewline
\cline{2-6} \cline{8-12} \cline{14-18} \cline{20-24} 
 & 64 & 6 & 4 & 4 & 0 &  & 67 & 9 & 8 & 8 & 0 &  & 215 & 5 & 16 & 16 & 0 &  & 162 & 8 & 668 & 56 & 16\tabularnewline
\cline{2-6} \cline{8-12} \cline{14-18} \cline{20-24} 
 & 72 & 7 & 4 & 4 & 0 &  & 74 & 9 & 60 & 12 & 0 &  & 216 & 9 & 36 & 14 & 0 &  & 164 & 8 & 668 & 56 & 16\tabularnewline
\cline{2-6} \cline{8-12} \cline{14-18} \cline{20-24} 
 & 121 & 6 & 4 & 4 & 0 &  & 81 & 8 & 8 & 8 & 0 &  & 222 & 7 & 22 & 12 & 0 &  & 176 & 12 & 3070 & 54 & 0\tabularnewline
\cline{2-6} \cline{8-18} \cline{20-24} 
 & 126 & 6 & 8 & 4 & 0 &  & 82 & 8 & 8 & 8 & 0 & \multirow{11}{*}{7} & 12 & 12 & 224 & 56 & 0 &  & 192 & 11 & 723 & 30 & 24\tabularnewline
\cline{2-6} \cline{8-12} \cline{14-18} \cline{20-24} 
 & 130 & 7 & 8 & 4 & 0 &  & 86 & 8 & 16 & 8 & 0 &  & 65 & 12 & 224 & 56 & 0 &  & 194 & 12 & 3070 & 54 & 0\tabularnewline
\cline{2-6} \cline{8-12} \cline{14-24} 
 & 135 & 7 & 8 & 4 & 0 &  & 88 & 8 & 78 & 12 & 0 &  & 84 & 12 & 700 & 56 & 0 & \multirow{2}{*}{10} & 174 & 15 & 615 & 108 & 0\tabularnewline
\cline{2-6} \cline{8-12} \cline{14-18} \cline{20-24} 
 & 137 & 6 & 8 & 4 & 0 &  & 111 & 8 & 8 & 8 & 0 &  & 128 & 11 & 128 & 12 & 0 &  & 187 & 15 & 615 & 108 & 0\tabularnewline
\cline{2-6} \cline{8-12} \cline{14-24} 
 & 138 & 7 & 8 & 4 & 0 &  & 115 & 8 & 8 & 8 & 0 &  & 131 & 12 & 700 & 56 & 0 & \multirow{2}{*}{11} & 225 & 14 & 3208 & 34 & 0\tabularnewline
\cline{2-6} \cline{8-12} \cline{14-18} \cline{20-24} 
 & 143 & 6 & 6 & 3 & 5 &  & 119 & 8 & 8 & 8 & 0 &  & 140 & 12 & 220 & 24 & 0 &  & 229 & 14 & 868 & 88 & 0\tabularnewline
\cline{2-6} \cline{8-12} \cline{14-24} 
 & 149 & 6 & 6 & 3 & 5 &  & 134 & 8 & 16 & 8 & 0 &  & 188 & 12 & 102 & 18 & 28 & \multirow{2}{*}{13} & 83 & 20 & 58840 & 240 & 0\tabularnewline
\cline{2-6} \cline{8-12} \cline{14-18} \cline{20-24} 
 & 156 & 6 & 6 & 3 & 5 &  & 136 & 8 & 44 & 12 & 0 &  & 189 & 10 & 49 & 20 & 0 &  & 123 & 20 & 58840 & 240 & 0\tabularnewline
\cline{2-6} \cline{8-12} \cline{14-24} 
 & 158 & 6 & 6 & 3 & 5 &  & 141 & 8 & 78 & 12 & 0 &  & 202 & 8 & 48 & 12 & 0 & \multirow{3}{*}{14} & 175 & 17 & 72598 & 228 & 0\tabularnewline
\cline{2-6} \cline{8-12} \cline{14-18} \cline{20-24} 
 & 168 & 3 & 3 & 2 & 3 &  & 167 & 6 & 10 & 4 & 0 &  & 204 & 8 & 24 & 16 & 0 &  & 191 & 17 & 72598 & 228 & 0\tabularnewline
\cline{2-6} \cline{8-12} \cline{14-18} \cline{20-24} 
 & 177 & 3 & 3 & 2 & 3 &  & 217 & 3 & 8 & 8 & 0 &  & 223 & 4 & 57 & 28 & 16 &  & 221 & 20 & 51308 & 116 & 0\tabularnewline
\cline{2-6} \cline{8-24} 
 & 183 & 3 & 3 & 2 & 3 &  & 228 & 5 & 7 & 8 & 4 & \multirow{2}{*}{8} & 124 & 14 & 252 & 24 & 0 &  &  &  &  &  & \tabularnewline
\cline{2-6} \cline{8-12} \cline{14-24} 
 & 184 & 3 & 3 & 2 & 3 &  & 230 & 5 & 19 & 4 & 4 &  & 127 & 12 & 328 & 24 & 0 &  &  &  &  &  & \tabularnewline
\hline 
\end{tabular}
\caption{Sizes of Hilbert bases and numbers of fragile indices in SGs with time-reversal symmetry and significant SOC. SGs that do not have fragile indices are not tabulated. ``rank'' represents the rank of the EBR matrix. ``ebr'' and ``root'' represent the numbers of EBRs and fragile roots in the Hilbert bases of $\overline{Y}$, respectively. And, ``ieq'' and ``$\mathbb{Z}_2$'' represent the numbers of inequality-type indices and $\mathbb{Z}_2$-type indices, respectively. \label{tab:summary}} 
\end{table*}

\subsection{Example of inequality-type fragile indices} 
We consider SG 70 ($Fddd$). The Smith Decomposition of the $EBR= L\Lambda R$ matrix is
\begin{equation} \scriptsize
\left(\begin{array}{rrrrrrr}
-2 & -1 & -1 & 0 & 1\\
2 & 3 & 1 & 0 & 0 \\
0 & 1 & 0 & 0 & 1 \\
-1 & 2 & 0 & 0 & 0\\
1 & 0 & 0 & 1 & 0
\end{array}\right)
\left(\begin{array}{rrrrr}
1 & 0 & 0 & 0 & 0 \\
0 & 1 & 0 & 0 & 0 \\
0 & 0 & 4 & 0 & 0 \\
0 & 0 & 0 & 0 & 0 \\
0 & 0 & 0 & 0 & 0
\end{array}\right)
\left(\begin{array}{rrrrr}
1 & 1 & 3 & 3 & 1 \\
1 & 2 & 2 & 2 & 2 \\
-1 & -2 & -2 & -3 & -1 \\
0 & 0 & 0 & 0 & 1 \\
0 & 0 & 1 & 0 & 0
\end{array}\right), \label{eq:EBR-70}
\end{equation}
where the corresponding TRS and double-group irreps are $\overline{\Gamma}_5$, $\overline{\Gamma}_6$, $\overline{\mathrm{T}}_3\overline{\mathrm{T}}_4$, $\overline{\mathrm{L}}_2\overline{\mathrm{L}}_2$,   $\overline{\mathrm{L}}_3\overline{\mathrm{L}}_3$, respectively. Irreps at the maximal $k$-vectors $\mathrm{Y}$ and $\mathrm{Z}$ (not shown) are determined by the irreps at $\Gamma$ using TRS and compatibility relations (see \cite{Bradlyn2017,Po2017,kruthoff_topological_2017,Elcoro2017,Vergniory2017,BCS}). 

The $\Lambda$ matrix has only three nonzero elements ($r=3$), so the symmetry data is parameterized by a three-component integer vector $y=(y_1,y_2, y_3)^T$.
By requiring the symmetry data vector $B \ge 0$,  we obtain a set of inequalities for $y$  defining a 3D cone
\begin{equation}
Y = \{y \in \mathbb{R}^3| \ 2y_2\ge y_1\ge0, -2y_1-y_2\ge 4y_3 \ge-2y_1-3y_2 \}. \label{eq:Y-70}
\end{equation}
The physical symmetry data vectors $B$'s are represented by integer points in $Y$, i.e., $\overline{Y} = \mathbb{Z}^3 \cap Y$. 
Each inequality in \cref{eq:Y-70} specifies a plane in $\mathbb{R}^3$: the plane separates the points that do or do not satisfy the inequality. The cone $Y$ is cut out by four such planes specified by the four inequalities in \cref{eq:Y-70}. As shown in Eq. 3a, these planes cross each other at four rays contained in this cone. We obtain the directions of the four rays as $\mathbf{r}_1=(0,4,-1)^T$, $\mathbf{r}_2=(0,4,-3)^T$, $\mathbf{r}_3=(8,4,-7)^T$, $\mathbf{r}_4=(8,4,-5)^T$. For example, the planes $y_1=0$ and $- 2y_1- y_2= 4 y_3$ intersect each other on the line $t\cdot \mathbf{r}_1$, $t \in \mathbb{R}$; the planes $y_1=0$ and $2y_2= y_1$ intersect on the line $t\cdot (0,0,1)$, $t \in \mathbb{R}$, which (except for $t=0$, a point which is already included in the other ray $t\cdot \mathbf{r}_1$) does not satisfy the second inequality in \cref{eq:Y-70} and hence do not provide a separate ray.

\begin{figure}
\includegraphics[width=0.9\linewidth]{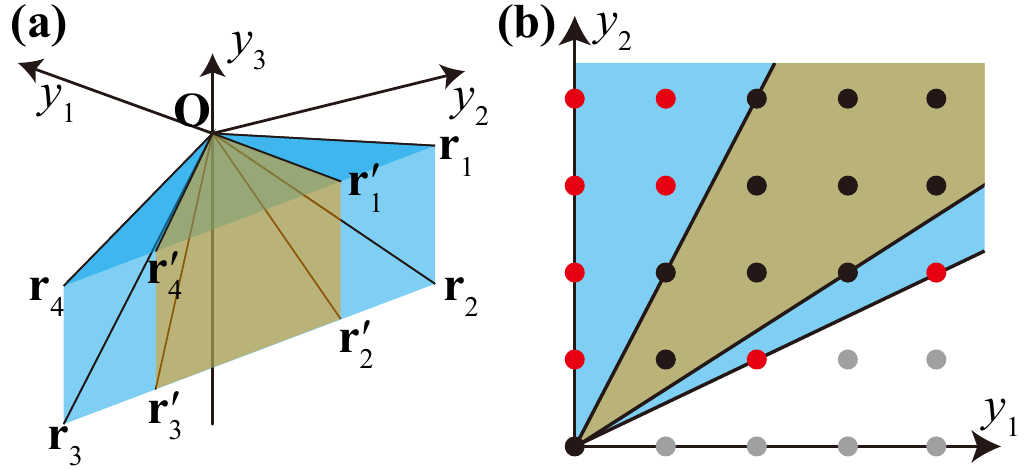}
\protect\caption{(a) The cones $Y$ and $X$ for SG 70 ($Fddd$).
The blue and yellow regions represent $Y$ and $X$, respectively. 
The ray vectors for $Y/X$ are $\mathbf{r}_{1,2,3,4}/\mathbf{r}_{1,2,3,4}^\prime$ respectively, given in the main text. (b) The projections of $Y$ and $X$ in the $y_1y_2$ plane.
The black points correspond to trivial symmetry data vectors $B$, the red points correspond to EFPs, and the grey points are un-physical.
\label{fig:70}}
\end{figure}

The trivial points in $Y$ (and $\overline{Y}$) are given by \cref{eq:Xbar}.
For simplicity, we first consider points in the cone 
\begin{equation}
X = \{y\in \mathbb{R}^r | y_i=(R p)_i\quad p \in \mathbb{R}^{N_{EBR}}_+ \}. \label{eq:X-def}
\end{equation}
In the general case $\mathbb{Z}^r \cap X$ is a superset of $\overline{X}$ (their difference being the non-integer $p>0$, such that $y \in \mathbb{Z}^r$). Due to the definition of $X$, and since $r=3$ in \cref{eq:X-def} for SG 70, it seems that (the first 3 rows of) each column of $R$ corresponds to a generator of $X$. However, in \cref{eq:EBR-70}, (the first 3 rows of) the first column of $R$, i.e., $(1,1,-1)^T$, can be spanned by the second and third columns as $\frac14 (1,2,-2)^T + \frac14 (3,2,-2)^T$, thus $X$ is generated by the last four columns of $R$. (There  exists a linear dependence between the 4 vectors defined by the first 3 rows of each of the last 4 columns of $R$, but it involves negative coefficients, and hence the vectors are linearly independent in $X$.) As shown in Fig. \ref{fig:70}a, each of the four generators corresponds to a ray of $X$:  $\mathbf{r}_1^\prime=(2,4,-2)^T$, $\mathbf{r}_2^\prime=(2,4,-4)^T$, $\mathbf{r}_3^\prime=(6,4,-6)^T$, $\mathbf{r}_4^\prime=(6,4,-4)^T$ (the rays are chosen as twice the generators for aesthetical purposes in Fig. \ref{fig:70}a ). 
Using elementary vector algebra, as explained in \cref{sec:polyhedron}, one finds the inequalities defining $X$,
\begin{equation}
X = \{y \in \mathbb{R}^3| \ 3y_2\ge 2y_1\ge y_2,-2y_1-y_2\ge 4y_3 \ge-2y_1-3y_2 \}.
\end{equation}
Illustrated in Fig. \ref{fig:70}a, $X$ is,  as it should be,  a subset of $Y$. 
The first/last two defining inequalities of $X$ are tighter than/identical to the first/last two of $Y$, respectively. For a point in $Y$ to be outside the trivial $X$, at least one of the first two inequalities of $X$ should be violated, i.e.,
\begin{equation}
2y_1-3y_2>0\quad \mathrm{or}\quad y_2-2y_1>0. \label{eq:criteria-70}
\end{equation}
\cref{eq:criteria-70} gives two inequality-type fragile indices for SG 70, which can also be obtained in a diagrammatic method.
From Fig. \ref{fig:70}a, we see that the boundaries separating $X$ and $Y$, i.e., $\mathbf{Or}_1^\prime\mathbf{r}_2^\prime$ and $\mathbf{Or}_3^\prime\mathbf{r}_4^\prime$, are parallel to the $y_3$-axis, where $\mathbf{O}$ stands for the origin. (Notice that all $\mathbf{r}_1 - \mathbf{r}_2, \mathbf{r}_3 - \mathbf{r}_4, \mathbf{r}_1^\prime - \mathbf{r}_2^\prime, \mathbf{r}_3^\prime - \mathbf{r}_4^\prime$ are parallel to $y_3$.) 
We project the cones to the $y_1y_2$ plane to obtain Fig. \ref{fig:70}b, from which we immediately conclude \cref{eq:criteria-70}.

%To prove that \cref{eq:criteria-70} is the sufficient condition, 
Since in general $\mathbb{Z}^r\cap X$ is a superset of $\overline{X}$, in principle there can be nontrivial points in $\mathbb{Z}^r\cap X - \overline{X}$. For example, for SG 199, $X$ (defined by \cref{eq:X-def}) is spanned by the rays $(0,1)^T$ and $(1,2)^T$ and is hence identical to $Y$ (Fig. \ref{fig:70}a).
Thus \cref{eq:criterion-199} identifies points in $\mathbb{Z}^2 \cap X - \overline{X}$.
However, for SG 70, there is no such point in $\mathbb{Z}^3 \cap X - \overline{X}$.
We verify this by explicitly listing the integer points in $X$ and by applying a general technique we introduce in \cref{sec:index} to calculate $\mathbb{Z}^r\cap X - \overline{X}$.

\section{The polyhedron method} \label{sec:main-polyhedron}
%After the two prior examples outlining our polyhedron method, 
We have outlined the polyhedron method through the two previous examples.
Now we summarize its general principle.

\subsection{Polyhedron description of symmetry data vectors} \label{sec:polyhedron-method}
In \cref{eq:parametrization} we parameterized the symmetry data as $B =\sum_j (L \Lambda)_{j} y_j$, where $(L \Lambda)_{j}$ represents the $j$th column of $(L \Lambda)$.
Now we define a polyhedral cone as
\beq
Y = \Big\{ y\in \mbb{R}^r\ \Big|\ \sum_j (L \Lambda)_{j} y_j \ge 0 \Big\}. \label{eq:Y-def}
\eeq 
The physical symmetry data vectors $B\in \mathbb{N}$ are faithfully represented by integer points $\overline{Y} = \mathbb{Z}^r \cap Y$.
%There is a one-to-one maping between points in $\ovl{Y}$ and symmetry data having trivial stable/strong phase indicators.
Due to \cref{thm:cone} in \cref{sec:math}, $Y$ can be represented by its rays and lines as
\begin{align}
Y =& \brace { Ray \cdot p + Line \cdot q| p \in \mbb{R}_+^m, q \in \mbb{R}^n}. \label{eq:Y-def-V}
\end{align}
where $Ray = (Ray_1,Ray_2,\cdots)$ is an $r\times m$ matrix, and $Line=(Line_1,Line_2,\cdots)$ is an $r\times n$ matrix. 
The difference between rays and lines is that rays have directions but lines do not. 
Correspondingly, the coefficients on rays ($p_i$'s) are nonnegative but the coefficients on lines ($q_j$'s) can be either nonnegative or negative.
For example, in \cref{fig:199}, $Y$ has two rays $(0,1)^T$ and $(1,2)^T$, but no lines.
A polyhedral cone is called \textit{pointed} if it does not contain lines.
Now we show that for any space group $Y$ is a pointed polyhedral cone.
Choosing $p=0$ and arbitrary $q$, due to \cref{eq:Y-def} we have $ L \Lambda_{:,1:r} Line\cdot q \ge 0$ as well as $ L \Lambda_{:,1:r} Line\cdot q \le 0$ since we can replace $q$ with $-q$, and thus $\forall q \in \mbb{R}^n,\; L \Lambda_{:,1:r} Line \cdot q = 0$.
As $L\Lambda_{:,1:r}$ is a rank-$r$ matrix, there must be $Line=0$.
In this paper, \cref{eq:Y-def} is called the H-representation of polyhedron and \cref{eq:Y-def-V} is called the V-representation.
The algorithm to determine the V-representation from the H-representation and vice versa is available in many mathematics packages such as the \href{http://www.sagemath.org/}{\it SageMath} package \cite{Sage}.

The trivial $B$-vectors, which decompose into positive sums of EBRs, are given as $\overline{X}$ (\cref{eq:Xbar}). 
%There is a one-to-one mapping between the sums of $B$-vectors of EBRs and points in $\ovl{X}$: on the one hand, for a band representation $ B =EBR\cdot p $, where $p\in\mbb{N}^{N_{EBR}}$, the corresponding point in $\ovl{X}$ is given by $ y_i = (L^{-1} B)_i/\Lambda_{i,i} = (R p)_i$; on the other hand, for a point in $\ovl{X}$, $y = R_{1:r,:} p$, the corresponding band representation is given by $B = L \Lambda_{:,1:r}y = EBR\cdot p$.
$\ovl{X}$ is a subset of $\ovl{Y}$, thus fragile symmetry data vectors are generated from points in $\ovl{Y}-\ovl{X}$. 
To classify them, we introduce an auxiliary polyhedral cone $X$ (\cref{eq:X-def}), which can be thought as an extension of $\overline{X}$ to allow non-negative real (not only integer) combination coefficients $p_i$.
The nontrivial points can then be divided into two parts: $\overline{Y} - \mathbb{Z}^r\cap X$ and $\mathbb{Z}^r\cap X - \overline{X}$. 
Points in $\overline{Y} - \mathbb{Z}^r\cap X = \mathbb{Z}^r\cap (Y-X)$ correspond to symmetry data vectors $B$ that cannot be written as non-negative combinations of the EBRs, even if the combination coefficients $p_i$ are allowed to be rational numbers.
Points in $\mathbb{Z}^r\cap X - \overline{X}$, on the other hand, correspond to symmetry data that can be written as non-negative rational combinations of EBRs but cannot be written as non-negative integer combinations of EBRs.

% It should be emphasized that in general $\ovl{X} \neq \mbb{Z}^r \cap X$. 
% In other words, there are such integer points in $X$ that do not belong to $\ovl{X}$.
% (As shown in \cref{sec:rootX,sec:Zn-type}, SG 150 is an example where $\ovl{X} \neq \mbb{Z}^r \cap X$.)
% Then we can classify fragile phases into two types: (type-I) $y\in \ovl{Y}$ and $y\not\in X$, and (type-II) $y\in X$ but $y\not\in \ovl{X}$.
% As will be discussed in \cref{sec:N-type,sec:Zn-type}, type-I and type-II fragile phases are diagnosed by the inequality-type fragile indices and the $\mbb{Z}_n$-type fragile indices, respectively.
% (In practice, the only $n\neq0$ case we find is $n=2$.)

Readers might refer to \cref{sec:example-Y-X} for more examples of $X$ and $Y$.

Points in $\mathbb{Z}^r \cap (Y-X)$ are outside $X$ and so violate the inequalities of $X$.
Thus, in general, points in $\mathbb{Z}^r\cap (Y-X)$ are diagnosed by the inequality-type indices, as in SG 70.
On the other hand, points in $\mathbb{Z}^r\cap X - \overline{X}$ are always near the boundary of $X$ and are diagnosed by the $\mbb{Z}_2$-type indices.
In the following we discuss these two types of indices in detail.

\subsection{Inequality-type fragile indices} \label{sec:N-type}
Now let us work out the inequality-type fragile criteria for $\mathbb{Z}^r \cap (Y-X)$.
First, we assume the H-representation of $X$ as $X = \{y\in \mbb{R}^r | Ay\ge 0,\ Bx=0 \}$, where $A \in \mbb{Q}^{n\times r}$ $B \in \mbb{Q}^{m\times r}$, $r$ is the rank of $EBR$, and $n,m$ some positive integers. 
(See \cref{thm:cone} for the general form of H-representation.)
Now we show that $B$ must be zero.
Since the first $r$ rows of $R$ has the rank $r$, $X$ is an $r$-dimensional object.
Presence of nonzero $B$ implies constraints on the points in $X$ and hence a lower-dimension of $X$.
Thus $B$ has to be zero and the H-representation of $X$ can always be written as
\begin{equation}
X = \{x\in \mbb{R}^r | Ax\ge 0 \}. \label{eq:X-Hrep}
\end{equation}
%where $A$ can be obtained algorithmically for all SGs through the method implemented in the \href{http://www.sagemath.org/}{\it SageMath} package \cite{Sage}.
For a point $y$ in $\mathbb{Z}^r \cap (Y-X)$, there should be some row in $A$, denoted as $a$, such that $ay<0$, so that $y\not\in X$.
Therefore, in principle, each row of $A$ gives an inequality-type fragile index $-ay$ and $-ay>0$ implies fragile phase.
One needs to check whether $ay<0$ is allowed in $Y$: if not allowed, there is no need to introduce the corresponding fragile index.
The method (with an example) to remove such unallowed criteria is given in \cref{sec:removing-inequality}.

\subsection{$\mbb{Z}_2$-type fragile indices}
As proved in \cref{sec:index}, points in $\mathbb{Z}^r\cap X - \overline{X}$ are always near the boundary of $X$ - the distances from these points to the boundary are always 0 or 1 - and so belong to some lower-dimensional sub-polyhedron of $X$.
Integer sums (per \cref{eq:Xbar}) of the generators of $\overline{X}$ belonging to this lower-dimensional subspace may not reach every integer point in this subspace.
For example, in SG 199, all the points in $\mathbb{Z}^r\cap X - \overline{X}$ also belong to the sub-cone $\{y\in \mathbb{Z}^2\;|\; y_1=0,\ y_2\ge 0\}$. The only $\overline{X}$ generator in this sub-cone is $(0,2)^T$. Thus $(0,y_2)$ cannot be generated if $y_2$ is odd, and  $ \mathbb{Z}^r\cap X - \overline{X}= \{y\in \mathbb{Z}^2\;|\; y_1=0,\ y_2\ge 0, \; y_2 \; \mathrm{\ mod\ } 2 = 1\}$.
As detailed in \ref{sec:index}, points in $\mathbb{Z}^r\cap X - \overline{X}$ can always be characterized by the decomposition coefficients of the $\overline{X}$ generators that are restricted in some lower-dimensional subspace.  If these coefficients are fractional, the corresponding symmetry data vectors $B$ are nontrivial.  Due to these fractional coefficients, in principle such diagnosis involves the modulo operation  (see \cref{eq:criterion-199}). We find that only the modulo 2 operation is involved (see \cref{eq:criterion-199}), and  we call these indices $\mathbb{Z}_2$-type.

\subsection{Fragile indices in all SGs}

In \cref{tab:summary} we summarize the numbers of inequality-type and $\mathbb{Z}_2$-type indices in all SGs; in Table S2 of \cite{SM} we explicitly tabulate all the fragile indices.

The mathematics of EFPs is much richer than that of robust topology. The latter usually forms a group.
For example in absence of TRS, according to the Chern number, the band insulators form an additive group $\mathbb{Z}$ \cite{Thouless1982,Haldane1988,Hatsugai1993}; in the presence of TRS (and without any other symmetries), according to the $\mathbb{Z}_2$ invariant, the band insulators form a $\mathbb{Z}_2$ group \cite{Kane2005_QSH,Kane2005_Z2,Bernevig2006}.
However, neither $\overline{Y}$ nor $\overline{X}$ form a group. 
Instead, $\overline{Y}$ and $\overline{X}$ are semigroups: A general element, except the identity, does not have an inverse.
To be specific, both $\overline{Y}$ and $\overline{X}$ can be written as $M = \{r_1p_1 + r_2p_2+\cdots r_n p_n | p_1\cdots p_n \in \mathbb{N}\}$, for some $n$, and $r_1p_1 + r_2p_2+\cdots r_n p_n = 0 \Rightarrow p=0$, where $r_i$'s are the generators which are no longer constrained to all be the columns of $R$. 
(See \cref{sec:polyhedron} for the proof that $\overline{Y}$ can be written in this form.)
For example in SG 199, $\overline{Y}$ can be written as $\{p_1(0,1)^T + p_2(1,2)^T| p_{1,2}\in \mathbb{N}\}$ (see Fig. \ref{fig:199}).  
$M$ is a \textit{positive affine monoid} in mathematics, and we make use of monoid properties to obtain the EFP roots.

\section{The EFP roots} \label{sec:main-root}
We find that the EFPs and the trivial states are always generated by a finite set of generators. ($\overline{Y}$ in SG 199 is generated by $(0,1)^T$ and $(1,2)^T$ (Fig. \ref{fig:199}b). We call the nontrivial states in the generators of $\overline{Y}$ the \textit{EFP roots}.
(Trivial states in the generators of $\overline{Y}$ are EBRs.)
By definition, an EFP can always be written as a sum of EFP roots plus some EBRs.
Thus the EFP roots can be thought as the \textit{non-redundant} representatives of the EFPs.
We worked out all the EFP roots in all SGs in presence of the TRS and the SOC, as tabulated in Table S3 of \cite{SM}.
The numbers of EFP roots in all SGs are summarized in Table \ref{tab:summary}.
As discussed in \cref{sec:polyhedron}, as a positive affine monoid, $\overline{Y}$ is generated by the \textit{Hilbert bases}: all of the elements $\overline{Y}$ can be written as a sum of Hilbert bases with positive coefficients, and none of the Hilbert bases can be written as a sum of other elements with positive coefficients.
The Hilbert bases form a \textit{unique} minimal set of generators of $\overline{Y}$. 
%\textbf{BAB: Zhida, we have to be a little less vague: what do you mean unique? If I have some linear dependence between EBRs, how do i have a unique decomposition}
%\textbf{SZD: Andrei, the Hilbert bases are unique. One cannot replace anyone of them by using the linear-dependence.}
For a given SG, we first calculate the Hilbert bases and then identify the topological nontrivial ones as the EFP roots.
There are two commonly used algorithms to get the Hilbert bases---the Normaliz algorithm \cite{Bruns2010} and the Hemmecke algorithm \cite{Hemmecke2002}.
In this work, we mainly use the Hemmecke algorithm.
In Table \ref{tab:summary} we tabulate the numbers of EBRs and EFP roots in the Hilbert bases in all SGs.

\begin{table}
\centering
\begin{tabular}{lrrr|lrrr|lr}
\hline 
& $\overline{\Gamma}_{7}$ & $\overline{\Gamma}_{8}$ & $\overline{\Gamma}_{9}$ &  & $\overline{\mathrm{K}}_4$ & $\overline{\mathrm{K}}_5$ &  $\overline{\mathrm{K}}_6$ & & $\overline{\mathrm{M}}_5$ \\
\hline 
$E$ & 2 & 2 & 2 & $E$ & 1 & 1 & 2 & $E$ & 2 \\
$2C_{6}$ & 0 & $-\sqrt3$ & $\sqrt3$ & $2C_3$ & -1 & -1 & 1 & $C_2$ & 0\\
$2C_{3}$ & -2 & 1 & 1 & $3\sigma_v$ & $-i$ & $i$ & 0 & $\sigma_d$ & 0\\
$C_{2}$ & 0 & 0 & 0 & & & & & $\sigma_v$ & 0\\ 
$3\sigma_d$ & 0 & 0 & 0 & & & & &\\
$3\sigma_v$ & 0 & 0 & 0 & & & & &\\
\hline
\end{tabular}
\protect\caption{\label{tab:irreps-P6mm} Character tables of irreps at high symmetry momenta in SG $P6mm$ (with TRS). The little co-groups at $\Gamma$, $\mathrm{K}$, and $\mathrm{M}$ are $C_{6v}$, $C_{3v}$, and $C_{2v}$ respectively. $C_6$, $C_3$, $C_2$ represent the $6$-, $3$-, $2$-fold rotations, and $\sigma_d$ and $\sigma_v$ represent the two classes of mirrors. }
\end{table}

\begin{figure*}
\begin{centering}
\includegraphics[width=1.0\linewidth]{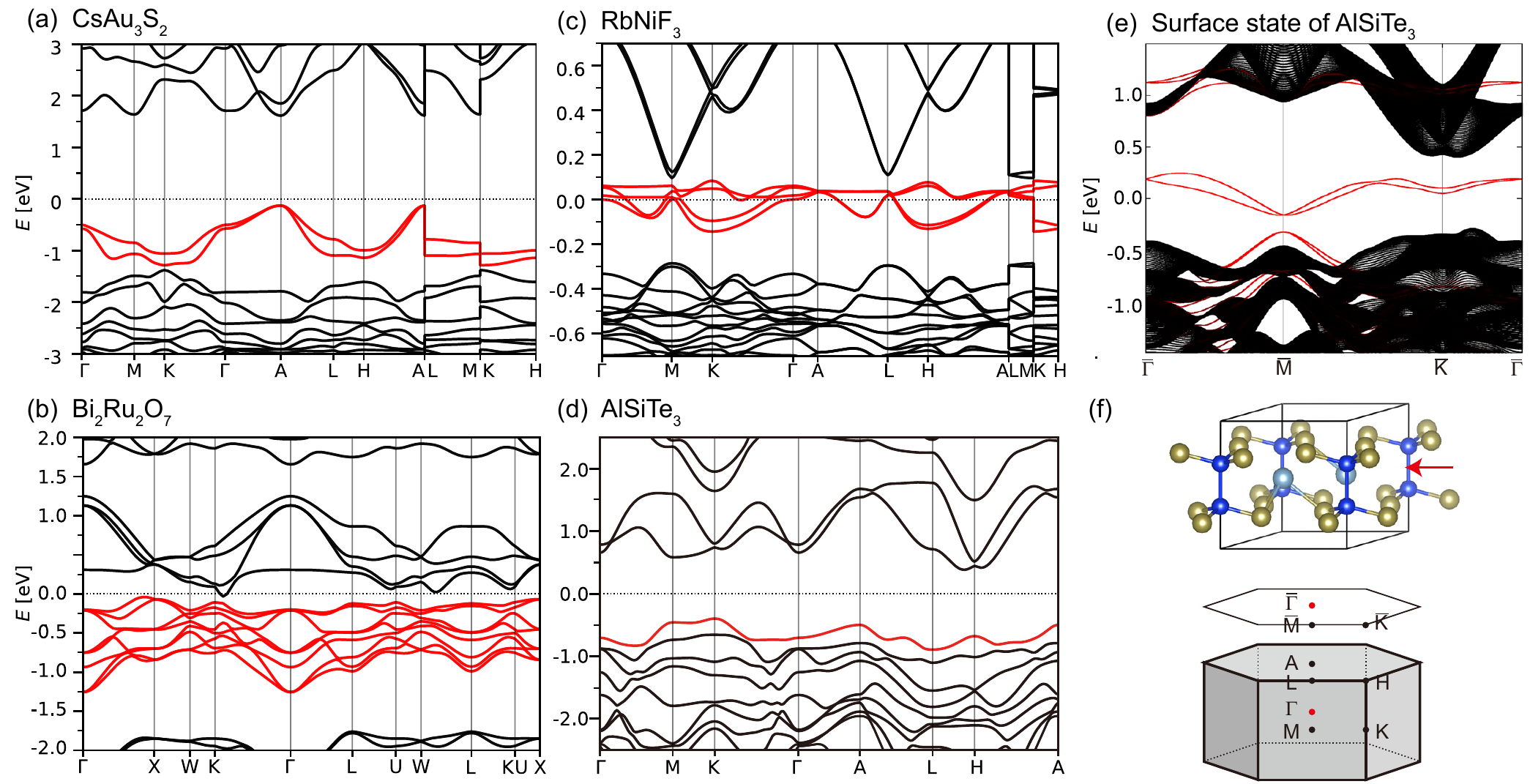}
\par\end{centering}
\protect\caption{ Fragile bands in materials.
(a) the band structure of CsAu$_3$S$_2$ (ICSD=82540) in SG 164 ($P\bar3m1$).
(b) Bi$_2$Ru$_2$O$_7$ (ICSD=166566) in SG 227 ($Fd\bar3m$).
(c) RbNiF$_3$ (ICSD=15090) in SG 194 ($P6_3/mmc$).
(d) AlSiTe$_2$ (ICSD=75001) in SG 147 ($P\bar3$).
(e) Top surface state of AlSiTe$_3$.
(f) The crystal structure of AlSiTe$_3$ and the bulk/surface BZ.
The red arrow shows the position of surface termination.
Here the fragile bands are colored as red, and the upper and lower bands are colored as black, and the Fermi-levels are represented by the horizontal dotted lines.
More information about the fragile bands, such as irreps and gaps from lower and upper bands as well as a hundred more band structures with fragile topology, can be found in \cite{SM}.
\label{fig:bands}}
\end{figure*}

We now present some examples of the roots for two known fragile phases. First we look at the SG 2 ($P\overline{1}$). $\Gamma$, $\mathrm{R}$, $\mathrm{T}$, $\mathrm{U}$, $\mathrm{V}$, $\mathrm{X}$, $\mathrm{Y}$, $\mathrm{Z}$ stand for the TRS momenta $(0,0,0)$, $(\pi,\pi,\pi)$, $(0,\pi,\pi)$, $(\pi,0,\pi)$, $(\pi,\pi,0)$, $(\pi,0,0)$, $(0,\pi,0)$, $(0,0,\pi)$, respectively. We find  1136 EFP roots in SG 2 (Table \ref{tab:summary}).
(We finally found that, upon coordinate rotation and gauge transformation, only 3 fragile roots are independent. See Table S8 of the SM of \cite{PaperOnTheInductionMethor}.)
Two of the roots are given by 
\begin{align}
& 2\overline{\Gamma}_2\overline{\Gamma}_2 + 2\overline{\mathrm{R}}_3\overline{\mathrm{R}}_3 + 2\overline{\mathrm{T}}_3\overline{\mathrm{T}}_3 + 2\overline{\mathrm{U}}_3\overline{\mathrm{U}}_3 \nonumber \\
+ & 2\overline{\mathrm{V}}_3\overline{\mathrm{V}}_3 + 2\overline{\mathrm{X}}_3\overline{\mathrm{X}}_3 + 2\overline{\mathrm{Y}}_3\overline{\mathrm{Y}}_3 + 2\overline{\mathrm{Z}}_2\overline{\mathrm{Z}}_2, \label{eq:root-2-1}
\end{align}
and
\begin{align}
& 4\overline{\Gamma}_2\overline{\Gamma}_2 + 4\overline{\mathrm{R}}_3\overline{\mathrm{R}}_3 + 4\overline{\mathrm{T}}_3\overline{\mathrm{T}}_3 + 4\overline{\mathrm{U}}_3\overline{\mathrm{U}}_3 \nonumber \\
+& 4\overline{\mathrm{V}}_3\overline{\mathrm{V}}_3 + 4\overline{\mathrm{X}}_3\overline{\mathrm{X}}_3 + 4\overline{\mathrm{Y}}_3\overline{\mathrm{Y}}_3 + 4\overline{\mathrm{Z}}_3\overline{\mathrm{Z}}_3, \label{eq:root-2-2}
\end{align}
The subscript $2$ ($3$) means that corresponding Kramer pair is odd (even) under inversion. Due to the Fu-Kane formula \cite{Fu2007b}, \cref{eq:root-2-1} is the double of a centrosymmetric weak TI with the index $(0;001)$.
In Ref. \cite{Taylor2011}, it has been shown that double of a 2D centro-symmetric TI remains nontrivial since its entanglement spectrum is gapless. \cref{eq:root-2-1} can be thought as a stacking of centrosymmetric double 2D TIs in the $001$-direction.
\cref{eq:root-2-2}, on the other hand, is four times of a 3D centrosymmetric TI. 
In Refs. \cite{Schindler2018,Fang2017,Khalaf2018,Song2018a} it was shown that double of a 3D centrosymmetric TI is an inversion-protected topological crystalline insulator HOTI. \cref{eq:root-2-2} shows that the double of a HOTI is a fragile phase.

Secondly we look at  SG 183 ($P6mm$) - the SG of graphene.
As discussed in Refs.~\cite{Ahn2018,Bradlyn2019,Song2018c,Po2018}, in presence of $C_2T$ symmetry ($(C_2T)^2=1$), a two-band system can host a fragile topology, and the topological invariant is given as the two-band Wilson loop winding. 
Refs.~\cite{Po2018,Bradlyn2019,Song2018c} have made use of TQC and similar methods to analyze the fragile phases in graphene. We relate the analysis of Refs.~\cite{Bradlyn2019,Song2018c} to our classification in which SG 183 has only three EFP roots $\overline{\Gamma}_7 + \overline{\mathrm{K}}_6 + \overline{\mathrm{M}}_5$, $\overline{\Gamma}_8 + \overline{\mathrm{K}}_4 + \overline{\mathrm{K}}_5 + \overline{\mathrm{M}}_5$, $\overline{\Gamma}_9 + \overline{\mathrm{K}}_4 + \overline{\mathrm{K}}_5 + \overline{\mathrm{M}}_5$.
Here $\Gamma$, $\mathrm{K}$, $\mathrm{M}$ stand for the high symmetry momenta $(0,0,0)$, $(\frac{2\pi}{3},\frac{2\pi}{3},0)$, $(\pi,0,0)$, respectively. Due to compatibility relations, the irreps in the $k_z=\pi$-plane are completely determined by the irreps in the $k_z=0$-plane. 
The irreps are defined in Table \ref{tab:irreps-P6mm}. 
For the first root, the $C_3$ representation matrices at $\Gamma$ and $\mathrm{K}$ can be written as $-\sigma_0$ and $e^{-i\frac{\pi}{3}\sigma_z}$, respectively.
And, for the second and third roots, the $C_3$ representation matrices at $\Gamma$ and $\mathrm{K}$ can be written as $e^{-i\frac{\pi}{3}\sigma_z}$ and $-\sigma_0$, respectively. 
Due to the correspondence between $C_3$ eigenvalues and Wilson loop winding in Refs. \cite{Song2018c, Bradlyn2019}, one can find that the Wilson loop winding in all the three cases are $3n\pm1$ for $n$ some integer, showing the topological nature of the state.

\section{Materials} 
Armed with our new complete classification, we set out to discover examples of topological bands in existing materials. This is a particularly challenging task, as one recent catalogue of high-throughput topological materials \cite{Vergniory2019} searched and found that \textit{no materials are topologically fragile at the Fermi level}, due to the fact that there are usually enough occupied EBR to turn any fragile set of bands into trivial. Hence, we have to settle for finding fragile sets of \textit{bands} hopefully close to the Fermi level. We have performed thousands of ab-initio calculations and have produced a list of 100 good materials which exhibit fragile topological bands close to the Fermi level (\cref{tab:material}). 
We showcase some of them in Fig. \ref{fig:bands}:  CsAu$_3$S$_2$ , Bi$_2$Ru$_2$O$_7$, RbNiF$_3$, and AlSiTe$_3$ have the fragile band right at or immediately below the Fermi level, well separated in the whole $k$-space sampled, from both the conduction and the valence bands. 
We computed our new fragile indices of these materials and confirmed them to be topological. 
This is the first time fragile topological bands have been predicted in crystalline systems.
The (relatively) flat fragile bands in RbNiF$_3$ may have interesting interacting physics since the bandwidths are smaller than the the on-site Hubbard interaction of Ni, which is usually 8-10eV.

Fragile topological bands have in-gap boundary states (the  ``filling anomaly'' \cite{wieder2018axion}) if the boundary cuts through empty Wyckoff positions which have nonzero coefficients in the EBR decomposition.
Here we take AlSiTe$_3$ as an example to show such in-gap states.
The SG 147 $P\bar3$ has four types of maximal Wyckoff positions: $1a\ (000)$, $1b\ (00\frac12)$, $3e\ (\frac1200)\ (0\frac120)\ (\frac12\frac120)$, $3f\ (\frac120\frac12)\ (0\frac12\frac12)\ (\frac12\frac12\frac12)$, and three types of non-maximal Wyckoff positions $2c\ (0,0,z)\ (0,0,-z)$, $2d\ (\frac13,\frac23,z)\ (\frac23,\frac13,-z)$, $6g\ (x,y,z)\ (-y,x-y,z)\ (-x+y,-x,z)\ (-x,-y,-z)\ (y,-x+y,-z)\ (x-y,x,-z)$.
The Al, Si, and Te atoms occupy the $2d$, $2c$, and $6g$ positions, respectively.
The BZ of SG 147 $P\bar3$, as shown in \cref{fig:bands}f, has six high symmetry momenta: $\Gamma$, $\mrm{K}$, $\mrm{M}$, $\mrm{A}$, $\mrm{H}$, $\mrm{L}$.
The irreps formed by the fragile band shown in \cref{fig:bands}d are
\begin{equation}
\ovl{\Gamma}_4\ovl{\Gamma}_4 + \ovl{\mrm{K}}_5\ovl{\mrm{K}}_6 
+ \ovl{\mrm{M}}_3\ovl{\mrm{M}}_3 +
\ovl{\mrm{A}}_7\ovl{\mrm{A}}_7 + \ovl{\mrm{H}}_5\ovl{\mrm{H}}_6 
+ \ovl{\mrm{L}}_2\ovl{\mrm{L}}_2.
\end{equation}
These irreps decompose into EBRs as
\begin{align}
& ({}^1\ovl{E}_g {}^2\ovl{E}_g)_b \uparrow SG 
\quad \oplus\quad ({}^1\ovl{E}_u {}^2\ovl{E}_u)_b \uparrow SG \nono\\
\quad \oplus\quad &  (\ovl{E}_u \ovl{E}_u)_b \uparrow SG 
\quad \ominus\quad ({}^1\ovl{E} {}^2\ovl{E})_{2d} \uparrow SG,
\end{align}
where $(\rho)_w\uparrow SG$ represent the EBR induced from the irrep $\rho$ of the site-symmetry-group of the Wyckoff position $w$.
Therefore, the fragile band is equivalent (in terms of irreps) to a combination of 3 Wannier functions at the $b$ position and ``$-1$'' Wannier functions at the $2d$ position.
We have checked that the trivial bands below the fragile band do not cancel the Wannier functions at $1b$.
Now we consider a surface terminating at the $b$ position (as shown in \cref{fig:bands}f).
Since the 3 Wannier states cannot be symmetrically divided into the two sides, there must be in-gap states on the surface.
We confirm the existence of such in-gap states by a first-principle calculation of a slab (\cref{fig:bands}e). 

We hope new experiments and predictions of responses in fragile states will follow our exciting discovery of fragile bands.

\renewcommand\arraystretch{0.65}
\newcommand{\PreserveBackslash}[1]{\let\temp=\\#1\let\\=\temp}
\newcolumntype{C}[1]{>{\PreserveBackslash\centering\tiny}p{#1}}
\newcolumntype{R}[1]{>{\PreserveBackslash\raggedleft\tiny}p{#1}}
\newcolumntype{L}[1]{>{\PreserveBackslash\raggedright\tiny}p{#1}}
\newcolumntype{D}[1]{>{\PreserveBackslash\centering}p{#1}}

\LTcapwidth=0.9\textwidth

\begin{longtable*}{|C{2cm}|C{0.5cm}|C{0.8cm}|C{0.8cm}|C{1cm}|C{7cm}|C{0.8cm}|C{0.8cm}|C{0.8cm}|C{0.8cm}|}
\caption{\label{tab:material}
Fragile bands in materials.
Fragile bands in materials. In the first three columns the chemical formulae, the space group numbers, and the
ICSD numbers of the materials are tabulated.
The fourth column gives the number of fragile branches in the band structure of the corresponding material.
In the fifth to tenth columns the information of the fragile branch closest to the Fermi level are tabulated.
``Bands'' gives the band indices of the fragile branch. 
Here we refer the highest occupied band as the $0$th band, and the lowest empty band as the 1st band, \etc
``Irreps'' gives the irreps formed by the fragile bands at high symmetry momenta.
$\Delta_l$ ($\Delta_u$) is the indirect gap between the fragile bands and the lower (upper) bands.
$\Delta_l^\pr$ ($\Delta_u^\pr$) is the direct gap between the fragile bands and the lower (upper) bands.}\\
\hline
Formula & SG & ICSD & NF & Bands & Irreps & $\Delta_l$(eV) & $\Delta_u$(eV) & $\Delta_l^\pr$(eV) & $\Delta_u^\pr$(eV)\\
\hline\hline
\endfirsthead

\multicolumn{10}{c}{\tablename\ \thetable{} (Continued.)}\\
\hline
Formula & SG & ICSD & NF & Bands & Irreps & $\Delta_l$(eV) & $\Delta_u$(eV) & $\Delta_l^\pr$(eV) & $\Delta_u^\pr$(eV) \\
\hline\hline
\endhead

\hline
\endfoot

\hline \hline \hline
\endlastfoot
    
$\mathrm{Cs(Au_{3}S_{2})}$ & 164 & 82540 & 2 & $-3\sim0$ & $\mathrm{\overline{A}_{4}\overline{A}_{5}}+\mathrm{\overline{A}_{8}}+\mathrm{\overline{\Gamma}_{4}\overline{\Gamma}_{5}}+\mathrm{\overline{\Gamma}_{8}}+\mathrm{\overline{H}_{4}\overline{H}_{5}}+\mathrm{\overline{H}_{6}}+\mathrm{\overline{K}_{4}\overline{K}_{5}}+\mathrm{\overline{K}_{6}}+2\mathrm{\overline{L}_{5}\overline{L}_{6}}+2\mathrm{\overline{M}_{5}\overline{M}_{6}}$ & 0.0946 & 1.7424 & 0.0946 & 1.7424 \\
\hline
$\mathrm{RbNiF_{3}}$ & 194 & 15090 & 3 & $-3\sim4$ & $\mathrm{\overline{A}_{4}\overline{A}_{5}}+\mathrm{\overline{A}_{6}}+2\mathrm{\overline{\Gamma}_{10}}+\mathrm{\overline{\Gamma}_{11}}+\mathrm{\overline{\Gamma}_{12}}+2\mathrm{\overline{H}_{8}}+2\mathrm{\overline{H}_{9}}+2\mathrm{\overline{K}_{8}}+2\mathrm{\overline{K}_{9}}+2\mathrm{\overline{L}_{3}\overline{L}_{4}}+4\mathrm{\overline{M}_{6}}$ & 0.1416 & 0.0121 & 0.2149 & 0.0336 \\
\hline
$\mathrm{Rb_{6}Ni_{6}F_{18}}$ & 194 & 410391 & 1 & $-3\sim4$ & $\mathrm{\overline{A}_{4}\overline{A}_{5}}+\mathrm{\overline{A}_{6}}+2\mathrm{\overline{\Gamma}_{10}}+\mathrm{\overline{\Gamma}_{11}}+\mathrm{\overline{\Gamma}_{12}}+2\mathrm{\overline{H}_{8}}+2\mathrm{\overline{H}_{9}}+2\mathrm{\overline{K}_{8}}+2\mathrm{\overline{K}_{9}}+2\mathrm{\overline{L}_{3}\overline{L}_{4}}+4\mathrm{\overline{M}_{6}}$ & 0.1561 & 0.012 & 0.2329 & 0.0121 \\
\hline
$\mathrm{Bi_{2}(Ru_{2}O_{7})}$ & 227 & 166566 & 2 & $-15\sim0$ & $\mathrm{\overline{\Gamma}_{6}}+\mathrm{\overline{\Gamma}_{7}}+3\mathrm{\overline{\Gamma}_{10}}+4\mathrm{\overline{X}_{5}}+3\mathrm{\overline{L}_{6}\overline{L}_{7}}+5\mathrm{\overline{L}_{9}}+2\mathrm{\overline{W}_{3}\overline{W}_{4}}+2\mathrm{\overline{W}_{5}\overline{W}_{6}}+4\mathrm{\overline{W}_{7}}$ & 0.5048 & 0.0065 & 0.771 & 0.0865 \\
\hline
$\mathrm{Bi_{2}O_{3}}$ & 164 & 186365 & 1 & $1\sim2$ & $\mathrm{\overline{A}_{8}}+\mathrm{\overline{\Gamma}_{8}}+\mathrm{\overline{H}_{4}\overline{H}_{5}}+\mathrm{\overline{K}_{4}\overline{K}_{5}}+\mathrm{\overline{L}_{3}\overline{L}_{4}}+\mathrm{\overline{M}_{3}\overline{M}_{4}}$ & 0.4934 & 0.0 & 1.1165 & 0.2286 \\
\hline
$\mathrm{Pb_{4}Se_{4}}$ & 225 & 238502 & 1 & $-3\sim0$ & $\mathrm{\overline{\Gamma}_{11}}+\mathrm{\overline{X}_{8}}+\mathrm{\overline{X}_{9}}+\mathrm{\overline{L}_{4}\overline{L}_{5}}+\mathrm{\overline{L}_{9}}+2\mathrm{\overline{W}_{7}}$ & 0.0 & 0.3062 & 0.2209 & 0.3085 \\
\hline
$\mathrm{PbSe}$ & 225 & 248492 & 1 & $-3\sim0$ & $\mathrm{\overline{\Gamma}_{11}}+\mathrm{\overline{X}_{8}}+\mathrm{\overline{X}_{9}}+\mathrm{\overline{L}_{4}\overline{L}_{5}}+\mathrm{\overline{L}_{9}}+2\mathrm{\overline{W}_{7}}$ & 0.0 & 0.224 & 0.2056 & 0.224 \\
\hline
$\mathrm{PbSe}$ & 225 & 62195 & 1 & $-3\sim0$ & $\mathrm{\overline{\Gamma}_{11}}+\mathrm{\overline{X}_{8}}+\mathrm{\overline{X}_{9}}+\mathrm{\overline{L}_{6}\overline{L}_{7}}+\mathrm{\overline{L}_{8}}+2\mathrm{\overline{W}_{6}}$ & 0.0 & 0.1305 & 0.212 & 0.1305 \\
\hline
$\mathrm{BiScO_{3}}$ & 221 & 181115 & 1 & $-15\sim0$ & $\mathrm{\overline{\Gamma}_{8}}+\mathrm{\overline{\Gamma}_{9}}+3\mathrm{\overline{\Gamma}_{11}}+2\mathrm{\overline{R}_{9}}+3\mathrm{\overline{R}_{11}}+3\mathrm{\overline{M}_{6}}+3\mathrm{\overline{M}_{7}}+2\mathrm{\overline{M}_{8}}+3\mathrm{\overline{X}_{6}}+3\mathrm{\overline{X}_{7}}+\mathrm{\overline{X}_{8}}+\mathrm{\overline{X}_{9}}$ & 0.0 & 0.6693 & 0.1211 & 0.8368 \\
\hline
$\mathrm{Ge}$ & 227 & 44841 & 1 & $-3\sim4$ & $\mathrm{\overline{\Gamma}_{8}}+\mathrm{\overline{\Gamma}_{9}}+\mathrm{\overline{\Gamma}_{10}}+2\mathrm{\overline{X}_{5}}+\mathrm{\overline{L}_{4}\overline{L}_{5}}+\mathrm{\overline{L}_{8}}+2\mathrm{\overline{L}_{9}}+\mathrm{\overline{W}_{3}\overline{W}_{4}}+\mathrm{\overline{W}_{5}\overline{W}_{6}}+2\mathrm{\overline{W}_{7}}$ & 0.0 & 0.0 & 0.2548 & 0.0874 \\
\hline
$\mathrm{MgCl_{2}}$ & 166 & 26157 & 1 & $-3\sim0$ & $\mathrm{\overline{\Gamma}_{6}\overline{\Gamma}_{7}}+\mathrm{\overline{\Gamma}_{9}}+\mathrm{\overline{T}_{6}\overline{T}_{7}}+\mathrm{\overline{T}_{9}}+2\mathrm{\overline{F}_{3}\overline{F}_{4}}+2\mathrm{\overline{L}_{3}\overline{L}_{4}}$ & 0.0 & 5.9915 & 0.085 & 6.1919 \\
\hline
$\mathrm{LiCdAs}$ & 216 & 609966 & 2 & $-1\sim2$ & $\mathrm{\overline{\Gamma}_{8}}+2\mathrm{\overline{X}_{7}}+\mathrm{\overline{L}_{4}\overline{L}_{5}}+\mathrm{\overline{L}_{6}}+2\mathrm{\overline{W}_{6}}+\mathrm{\overline{W}_{7}}+\mathrm{\overline{W}_{8}}$ & 0.0 & 0.0 & 0.1502 & 0.0849 \\
\hline
$\mathrm{Bi_{2}(Ru_{2}O_{7})}$ & 227 & 161102 & 2 & $-15\sim0$ & $\mathrm{\overline{\Gamma}_{6}}+\mathrm{\overline{\Gamma}_{7}}+3\mathrm{\overline{\Gamma}_{10}}+4\mathrm{\overline{X}_{5}}+3\mathrm{\overline{L}_{4}\overline{L}_{5}}+5\mathrm{\overline{L}_{8}}+2\mathrm{\overline{W}_{3}\overline{W}_{4}}+2\mathrm{\overline{W}_{5}\overline{W}_{6}}+4\mathrm{\overline{W}_{7}}$ & 0.5238 & 0.0 & 0.7923 & 0.0842 \\
\hline
$\mathrm{PbTe}$ & 225 & 648615 & 1 & $-3\sim0$ & $\mathrm{\overline{\Gamma}_{11}}+\mathrm{\overline{X}_{8}}+\mathrm{\overline{X}_{9}}+\mathrm{\overline{L}_{4}\overline{L}_{5}}+\mathrm{\overline{L}_{9}}+2\mathrm{\overline{W}_{7}}$ & 0.0 & 0.2613 & 0.0836 & 0.3544 \\
\hline
$\mathrm{Bi(ScO_{3})}$ & 221 & 158759 & 1 & $-15\sim0$ & $\mathrm{\overline{\Gamma}_{8}}+\mathrm{\overline{\Gamma}_{9}}+3\mathrm{\overline{\Gamma}_{11}}+2\mathrm{\overline{R}_{6}}+3\mathrm{\overline{R}_{10}}+3\mathrm{\overline{M}_{6}}+3\mathrm{\overline{M}_{7}}+2\mathrm{\overline{M}_{9}}+\mathrm{\overline{X}_{6}}+\mathrm{\overline{X}_{7}}+3\mathrm{\overline{X}_{8}}+3\mathrm{\overline{X}_{9}}$ & 0.0 & 0.422 & 0.0765 & 0.6957 \\
\hline
$\mathrm{AlSiTe_{3}}$ & 147 & 75001 & 4 & $-1\sim0$ & $\mathrm{\overline{A}_{7}\overline{A}_{7}}+\mathrm{\overline{\Gamma}_{4}\overline{\Gamma}_{4}}+\mathrm{\overline{H}_{5}\overline{H}_{6}}+\mathrm{\overline{K}_{5}\overline{K}_{6}}+\mathrm{\overline{L}_{2}\overline{L}_{2}}+\mathrm{\overline{M}_{3}\overline{M}_{3}}$ & 0.0 & 0.6883 & 0.0669 & 0.8923 \\
\hline
$\mathrm{CuIn}$ & 194 & 180112 & 1 & $1\sim8$ & $\mathrm{\overline{A}_{4}\overline{A}_{5}}+\mathrm{\overline{A}_{6}}+\mathrm{\overline{\Gamma}_{7}}+\mathrm{\overline{\Gamma}_{8}}+\mathrm{\overline{\Gamma}_{9}}+\mathrm{\overline{\Gamma}_{10}}+2\mathrm{\overline{H}_{6}\overline{H}_{7}}+2\mathrm{\overline{H}_{8}}+2\mathrm{\overline{K}_{7}}+\mathrm{\overline{K}_{8}}+\mathrm{\overline{K}_{9}}+2\mathrm{\overline{L}_{3}\overline{L}_{4}}+3\mathrm{\overline{M}_{5}}+\mathrm{\overline{M}_{6}}$ & 0.0 & 0.0 & 0.131 & 0.0669 \\
\hline
$\mathrm{PtO_{2}}$ & 164 & 24922 & 4 & $1\sim2$ & $\mathrm{\overline{A}_{4}\overline{A}_{5}}+\mathrm{\overline{\Gamma}_{4}\overline{\Gamma}_{5}}+\mathrm{\overline{H}_{6}}+\mathrm{\overline{K}_{6}}+\mathrm{\overline{L}_{3}\overline{L}_{4}}+\mathrm{\overline{M}_{3}\overline{M}_{4}}$ & 1.5328 & 0.0 & 1.757 & 0.0619 \\
\hline
$\mathrm{Hf_{3}Al_{3}C_{5}}$ & 194 & 161587 & 3 & $-1\sim6$ & $\mathrm{\overline{A}_{4}\overline{A}_{5}}+\mathrm{\overline{A}_{6}}+2\mathrm{\overline{\Gamma}_{7}}+\mathrm{\overline{\Gamma}_{8}}+\mathrm{\overline{\Gamma}_{9}}+\mathrm{\overline{H}_{4}\overline{H}_{5}}+\mathrm{\overline{H}_{6}\overline{H}_{7}}+\mathrm{\overline{H}_{8}}+\mathrm{\overline{H}_{9}}+2\mathrm{\overline{K}_{7}}+\mathrm{\overline{K}_{8}}+\mathrm{\overline{K}_{9}}+2\mathrm{\overline{L}_{3}\overline{L}_{4}}+4\mathrm{\overline{M}_{6}}$ & 0.0 & 0.0 & 0.0594 & 0.1486 \\
\hline
$\mathrm{SrAl_{2}Si_{2}}$ & 164 & 419886 & 3 & $-3\sim0$ & $\mathrm{\overline{A}_{4}\overline{A}_{5}}+\mathrm{\overline{A}_{8}}+\mathrm{\overline{\Gamma}_{6}\overline{\Gamma}_{7}}+\mathrm{\overline{\Gamma}_{9}}+\mathrm{\overline{H}_{4}\overline{H}_{5}}+\mathrm{\overline{H}_{6}}+\mathrm{\overline{K}_{4}\overline{K}_{5}}+\mathrm{\overline{K}_{6}}+2\mathrm{\overline{L}_{5}\overline{L}_{6}}+2\mathrm{\overline{M}_{3}\overline{M}_{4}}$ & 0.0 & 0.0 & 0.0559 & 0.3243 \\
\hline
$\mathrm{SnAs}$ & 225 & 611424 & 1 & $-2\sim1$ & $\mathrm{\overline{\Gamma}_{11}}+\mathrm{\overline{X}_{8}}+\mathrm{\overline{X}_{9}}+\mathrm{\overline{L}_{6}\overline{L}_{7}}+\mathrm{\overline{L}_{8}}+2\mathrm{\overline{W}_{6}}$ & 0.0 & 0.0 & 0.0518 & 0.1985 \\
\hline
$\mathrm{Bi_{2}(Os_{2}O_{7})}$ & 227 & 161105 & 1 & $-15\sim0$ & $\mathrm{\overline{\Gamma}_{6}}+\mathrm{\overline{\Gamma}_{7}}+3\mathrm{\overline{\Gamma}_{10}}+4\mathrm{\overline{X}_{5}}+3\mathrm{\overline{L}_{4}\overline{L}_{5}}+5\mathrm{\overline{L}_{8}}+2\mathrm{\overline{W}_{3}\overline{W}_{4}}+2\mathrm{\overline{W}_{5}\overline{W}_{6}}+4\mathrm{\overline{W}_{7}}$ & 0.8702 & 0.0 & 0.9833 & 0.0512 \\
\hline
$\mathrm{BaZr(PO_{4})_{2}}$ & 164 & 173842 & 9 & $1\sim2$ & $\mathrm{\overline{A}_{9}}+\mathrm{\overline{\Gamma}_{8}}+\mathrm{\overline{H}_{4}\overline{H}_{5}}+\mathrm{\overline{K}_{4}\overline{K}_{5}}+\mathrm{\overline{L}_{5}\overline{L}_{6}}+\mathrm{\overline{M}_{3}\overline{M}_{4}}$ & 3.905 & 0.0 & 3.9474 & 0.0502 \\
\hline
$\mathrm{SnS}$ & 225 & 52107 & 1 & $-3\sim0$ & $\mathrm{\overline{\Gamma}_{11}}+\mathrm{\overline{X}_{8}}+\mathrm{\overline{X}_{9}}+\mathrm{\overline{L}_{6}\overline{L}_{7}}+\mathrm{\overline{L}_{8}}+2\mathrm{\overline{W}_{6}}$ & 0.0 & 0.1121 & 0.0484 & 0.1328 \\
\hline
$\mathrm{NbSbSi}$ & 129 & 646436 & 1 & $-5\sim2$ & $2\mathrm{\overline{A}_{5}}+\mathrm{\overline{\Gamma}_{8}}+3\mathrm{\overline{\Gamma}_{9}}+2\mathrm{\overline{M}_{5}}+\mathrm{\overline{Z}_{8}}+3\mathrm{\overline{Z}_{9}}+2\mathrm{\overline{R}_{3}\overline{R}_{4}}+2\mathrm{\overline{X}_{3}\overline{X}_{4}}$ & 0.0 & 0.0 & 0.0465 & 0.0639 \\
\hline
$\mathrm{Na_{2}BaMgP_{2}O_{8}}$ & 147 & 262716 & 10 & $1\sim2$ & $\mathrm{\overline{A}_{8}\overline{A}_{9}}+\mathrm{\overline{\Gamma}_{5}\overline{\Gamma}_{6}}+\mathrm{\overline{H}_{4}\overline{H}_{4}}+\mathrm{\overline{K}_{4}\overline{K}_{4}}+\mathrm{\overline{L}_{2}\overline{L}_{2}}+\mathrm{\overline{M}_{3}\overline{M}_{3}}$ & 5.2768 & 0.0 & 5.312 & 0.0458 \\
\hline
$\mathrm{BiTeI}$ & 156 & 79364 & 5 & $-1\sim0$ & $\mathrm{\overline{A}_{4}\overline{A}_{5}}+\mathrm{\overline{\Gamma}_{4}\overline{\Gamma}_{5}}+\mathrm{\overline{H}_{5}}+\mathrm{\overline{H}_{6}}+\mathrm{\overline{K}_{5}}+\mathrm{\overline{K}_{6}}+\mathrm{\overline{L}_{3}\overline{L}_{4}}+\mathrm{\overline{M}_{3}\overline{M}_{4}}$ & 0.0 & 0.0447 & 0.0879 & 0.0447 \\
\hline
$\mathrm{Zr(MoO_{4})_{2}}$ & 164 & 59999 & 10 & $1\sim2$ & $\mathrm{\overline{A}_{4}\overline{A}_{5}}+\mathrm{\overline{\Gamma}_{4}\overline{\Gamma}_{5}}+\mathrm{\overline{H}_{6}}+\mathrm{\overline{K}_{6}}+\mathrm{\overline{L}_{3}\overline{L}_{4}}+\mathrm{\overline{M}_{3}\overline{M}_{4}}$ & 3.1215 & 0.0 & 3.135 & 0.0431 \\
\hline
$\mathrm{MoGe_{2}}$ & 139 & 76139 & 1 & $-3\sim2$ & $2\mathrm{\overline{\Gamma}_{6}}+\mathrm{\overline{\Gamma}_{9}}+\mathrm{\overline{M}_{6}}+\mathrm{\overline{M}_{7}}+\mathrm{\overline{M}_{8}}+2\mathrm{\overline{P}_{6}}+\mathrm{\overline{P}_{7}}+2\mathrm{\overline{X}_{5}}+\mathrm{\overline{X}_{6}}+3\mathrm{\overline{N}_{3}\overline{N}_{4}}$ & 0.0 & 0.0 & 0.0389 & 0.0832 \\
\hline
$\mathrm{PbMg_{2}}$ & 225 & 151361 & 1 & $-1\sim6$ & $\mathrm{\overline{\Gamma}_{10}}+\mathrm{\overline{\Gamma}_{11}}+\mathrm{\overline{X}_{6}}+\mathrm{\overline{X}_{7}}+\mathrm{\overline{X}_{8}}+\mathrm{\overline{X}_{9}}+\mathrm{\overline{L}_{4}\overline{L}_{5}}+\mathrm{\overline{L}_{6}\overline{L}_{7}}+2\mathrm{\overline{L}_{8}}+3\mathrm{\overline{W}_{6}}+\mathrm{\overline{W}_{7}}$ & 0.0 & 0.0 & 0.1356 & 0.0338 \\
\hline
$\mathrm{Ni(OH)_{2}}$ & 164 & 28101 & 4 & $-1\sim0$ & $\mathrm{\overline{A}_{4}\overline{A}_{5}}+\mathrm{\overline{\Gamma}_{4}\overline{\Gamma}_{5}}+\mathrm{\overline{H}_{6}}+\mathrm{\overline{K}_{6}}+\mathrm{\overline{L}_{3}\overline{L}_{4}}+\mathrm{\overline{M}_{3}\overline{M}_{4}}$ & 0.4417 & 0.0 & 0.494 & 0.0336 \\
\hline
$\mathrm{KSc(MoO_{4})_{2}}$ & 164 & 28019 & 7 & $1\sim2$ & $\mathrm{\overline{A}_{4}\overline{A}_{5}}+\mathrm{\overline{\Gamma}_{4}\overline{\Gamma}_{5}}+\mathrm{\overline{H}_{6}}+\mathrm{\overline{K}_{6}}+\mathrm{\overline{L}_{3}\overline{L}_{4}}+\mathrm{\overline{M}_{3}\overline{M}_{4}}$ & 2.9384 & 0.0 & 2.9384 & 0.0331 \\
\hline
$\mathrm{PbS}$ & 225 & 250762 & 1 & $-3\sim0$ & $\mathrm{\overline{\Gamma}_{11}}+\mathrm{\overline{X}_{8}}+\mathrm{\overline{X}_{9}}+\mathrm{\overline{L}_{4}\overline{L}_{5}}+\mathrm{\overline{L}_{9}}+2\mathrm{\overline{W}_{7}}$ & 0.0 & 0.1131 & 0.0328 & 0.1131 \\
\hline
$\mathrm{SrSi_{2}Al_{2}}$ & 164 & 609338 & 3 & $-3\sim0$ & $\mathrm{\overline{A}_{4}\overline{A}_{5}}+\mathrm{\overline{A}_{8}}+\mathrm{\overline{\Gamma}_{6}\overline{\Gamma}_{7}}+\mathrm{\overline{\Gamma}_{9}}+\mathrm{\overline{H}_{4}\overline{H}_{5}}+\mathrm{\overline{H}_{6}}+\mathrm{\overline{K}_{4}\overline{K}_{5}}+\mathrm{\overline{K}_{6}}+2\mathrm{\overline{L}_{5}\overline{L}_{6}}+2\mathrm{\overline{M}_{3}\overline{M}_{4}}$ & 0.0 & 0.0 & 0.0327 & 0.3753 \\
\hline
$\mathrm{PtB}$ & 194 & 615210 & 1 & $-1\sim10$ & $\mathrm{\overline{A}_{4}\overline{A}_{5}}+2\mathrm{\overline{A}_{6}}+\mathrm{\overline{\Gamma}_{7}}+\mathrm{\overline{\Gamma}_{8}}+\mathrm{\overline{\Gamma}_{10}}+\mathrm{\overline{\Gamma}_{11}}+2\mathrm{\overline{\Gamma}_{12}}+\mathrm{\overline{H}_{4}\overline{H}_{5}}+2\mathrm{\overline{H}_{6}\overline{H}_{7}}+2\mathrm{\overline{H}_{8}}+\mathrm{\overline{H}_{9}}+3\mathrm{\overline{K}_{7}}+\mathrm{\overline{K}_{8}}+2\mathrm{\overline{K}_{9}}+3\mathrm{\overline{L}_{3}\overline{L}_{4}}+6\mathrm{\overline{M}_{5}}$ & 0.0 & 0.0 & 0.0321 & 0.2012 \\
\hline
$\mathrm{CaGaGeH}$ & 156 & 173567 & 1 & $-1\sim0$ & $\mathrm{\overline{A}_{4}\overline{A}_{5}}+\mathrm{\overline{\Gamma}_{4}\overline{\Gamma}_{5}}+\mathrm{\overline{H}_{4}}+\mathrm{\overline{H}_{6}}+\mathrm{\overline{K}_{4}}+\mathrm{\overline{K}_{6}}+\mathrm{\overline{L}_{3}\overline{L}_{4}}+\mathrm{\overline{M}_{3}\overline{M}_{4}}$ & 0.0 & 0.4363 & 0.0314 & 0.7898 \\
\hline
$\mathrm{MgCl_{2}}$ & 164 & 17063 & 2 & $-3\sim0$ & $\mathrm{\overline{A}_{6}\overline{A}_{7}}+\mathrm{\overline{A}_{9}}+\mathrm{\overline{\Gamma}_{6}\overline{\Gamma}_{7}}+\mathrm{\overline{\Gamma}_{9}}+\mathrm{\overline{H}_{4}\overline{H}_{5}}+\mathrm{\overline{H}_{6}}+\mathrm{\overline{K}_{4}\overline{K}_{5}}+\mathrm{\overline{K}_{6}}+2\mathrm{\overline{L}_{3}\overline{L}_{4}}+2\mathrm{\overline{M}_{3}\overline{M}_{4}}$ & 0.0 & 5.5186 & 0.0299 & 5.5186 \\
\hline
$\mathrm{Mg_{2}Al_{2}Se_{5}}$ & 164 & 41928 & 7 & $-1\sim0$ & $\mathrm{\overline{A}_{4}\overline{A}_{5}}+\mathrm{\overline{\Gamma}_{6}\overline{\Gamma}_{7}}+\mathrm{\overline{H}_{6}}+\mathrm{\overline{K}_{6}}+\mathrm{\overline{L}_{3}\overline{L}_{4}}+\mathrm{\overline{M}_{5}\overline{M}_{6}}$ & 0.0 & 0.7281 & 0.0292 & 0.7281 \\
\hline
$\mathrm{K_{3}V(VO_{4})_{2}}$ & 164 & 100782 & 7 & $-1\sim0$ & $\mathrm{\overline{A}_{6}\overline{A}_{7}}+\mathrm{\overline{\Gamma}_{4}\overline{\Gamma}_{5}}+\mathrm{\overline{H}_{6}}+\mathrm{\overline{K}_{6}}+\mathrm{\overline{L}_{5}\overline{L}_{6}}+\mathrm{\overline{M}_{3}\overline{M}_{4}}$ & 2.1516 & 0.0 & 2.1663 & 0.0289 \\
\hline
$\mathrm{PtS_{2}}$ & 164 & 41375 & 4 & $1\sim2$ & $\mathrm{\overline{A}_{4}\overline{A}_{5}}+\mathrm{\overline{\Gamma}_{4}\overline{\Gamma}_{5}}+\mathrm{\overline{H}_{6}}+\mathrm{\overline{K}_{6}}+\mathrm{\overline{L}_{3}\overline{L}_{4}}+\mathrm{\overline{M}_{3}\overline{M}_{4}}$ & 0.4445 & 0.0 & 1.1335 & 0.0278 \\
\hline
$\mathrm{PbS}$ & 225 & 62190 & 1 & $-3\sim0$ & $\mathrm{\overline{\Gamma}_{11}}+\mathrm{\overline{X}_{8}}+\mathrm{\overline{X}_{9}}+\mathrm{\overline{L}_{6}\overline{L}_{7}}+\mathrm{\overline{L}_{8}}+2\mathrm{\overline{W}_{6}}$ & 0.0 & 0.0277 & 0.0338 & 0.0277 \\
\hline
$\mathrm{Cd(OH)_{2}}$ & 164 & 165225 & 4 & $-1\sim0$ & $\mathrm{\overline{A}_{8}}+\mathrm{\overline{\Gamma}_{8}}+\mathrm{\overline{H}_{4}\overline{H}_{5}}+\mathrm{\overline{K}_{4}\overline{K}_{5}}+\mathrm{\overline{L}_{3}\overline{L}_{4}}+\mathrm{\overline{M}_{3}\overline{M}_{4}}$ & 0.0 & 1.747 & 0.0262 & 1.9915 \\
\hline
$\mathrm{Mg_{2}Pb}$ & 225 & 642745 & 1 & $-1\sim6$ & $\mathrm{\overline{\Gamma}_{10}}+\mathrm{\overline{\Gamma}_{11}}+\mathrm{\overline{X}_{6}}+\mathrm{\overline{X}_{7}}+\mathrm{\overline{X}_{8}}+\mathrm{\overline{X}_{9}}+\mathrm{\overline{L}_{4}\overline{L}_{5}}+\mathrm{\overline{L}_{6}\overline{L}_{7}}+2\mathrm{\overline{L}_{8}}+3\mathrm{\overline{W}_{6}}+\mathrm{\overline{W}_{7}}$ & 0.0 & 0.0 & 0.2433 & 0.0259 \\
\hline
$\mathrm{K(Ag(CN)_{2})}$ & 163 & 30275 & 2 & $-3\sim0$ & $\mathrm{\overline{A}_{4}\overline{A}_{4}}+\mathrm{\overline{\Gamma}_{4}\overline{\Gamma}_{5}}+\mathrm{\overline{\Gamma}_{6}\overline{\Gamma}_{7}}+\mathrm{\overline{H}_{4}\overline{H}_{5}}+\mathrm{\overline{H}_{6}}+\mathrm{\overline{K}_{4}\overline{K}_{5}}+\mathrm{\overline{K}_{6}}+\mathrm{\overline{L}_{2}\overline{L}_{2}}+\mathrm{\overline{M}_{3}\overline{M}_{4}}+\mathrm{\overline{M}_{5}\overline{M}_{6}}$ & 0.0 & 3.1161 & 0.0251 & 3.1161 \\
\hline
$\mathrm{SnP}$ & 225 & 77786 & 1 & $-2\sim1$ & $\mathrm{\overline{\Gamma}_{11}}+\mathrm{\overline{X}_{8}}+\mathrm{\overline{X}_{9}}+\mathrm{\overline{L}_{6}\overline{L}_{7}}+\mathrm{\overline{L}_{8}}+2\mathrm{\overline{W}_{6}}$ & 0.0 & 0.0 & 0.0251 & 0.2046 \\
\hline
$\mathrm{Mg_{2}O(OH)_{2}}$ & 164 & 95472 & 1 & $-1\sim0$ & $\mathrm{\overline{A}_{6}\overline{A}_{7}}+\mathrm{\overline{\Gamma}_{6}\overline{\Gamma}_{7}}+\mathrm{\overline{H}_{6}}+\mathrm{\overline{K}_{6}}+\mathrm{\overline{L}_{5}\overline{L}_{6}}+\mathrm{\overline{M}_{5}\overline{M}_{6}}$ & 0.0 & 3.6627 & 0.0242 & 3.6627 \\
\hline
$\mathrm{Cu_{4}O_{3}}$ & 141 & 100566 & 1 & $-7\sim0$ & $2\mathrm{\overline{\Gamma}_{6}}+2\mathrm{\overline{\Gamma}_{7}}+2\mathrm{\overline{M}_{5}}+2\mathrm{\overline{P}_{3}\overline{P}_{6}}+2\mathrm{\overline{P}_{7}}+2\mathrm{\overline{X}_{3}\overline{X}_{4}}+\mathrm{\overline{N}_{3}\overline{N}_{4}}+3\mathrm{\overline{N}_{5}\overline{N}_{6}}$ & 0.0 & 0.0 & 0.0466 & 0.0242 \\
\hline
$\mathrm{BaSr_{2}Mg(SiO_{4})_{2}}$ & 164 & 247861 & 4 & $1\sim2$ & $\mathrm{\overline{A}_{9}}+\mathrm{\overline{\Gamma}_{8}}+\mathrm{\overline{H}_{4}\overline{H}_{5}}+\mathrm{\overline{K}_{4}\overline{K}_{5}}+\mathrm{\overline{L}_{5}\overline{L}_{6}}+\mathrm{\overline{M}_{3}\overline{M}_{4}}$ & 5.0528 & 0.0 & 5.1228 & 0.0236 \\
\hline
$\mathrm{CuI}$ & 156 & 84217 & 5 & $-1\sim0$ & $\mathrm{\overline{A}_{4}\overline{A}_{5}}+\mathrm{\overline{\Gamma}_{4}\overline{\Gamma}_{5}}+\mathrm{\overline{H}_{4}}+\mathrm{\overline{H}_{5}}+\mathrm{\overline{K}_{4}}+\mathrm{\overline{K}_{5}}+\mathrm{\overline{L}_{3}\overline{L}_{4}}+\mathrm{\overline{M}_{3}\overline{M}_{4}}$ & 0.0 & 1.367 & 0.0234 & 1.367 \\
\hline
$\mathrm{Ag_{2}O}$ & 164 & 20368 & 5 & $-1\sim0$ & $\mathrm{\overline{A}_{6}\overline{A}_{7}}+\mathrm{\overline{\Gamma}_{6}\overline{\Gamma}_{7}}+\mathrm{\overline{H}_{6}}+\mathrm{\overline{K}_{6}}+\mathrm{\overline{L}_{5}\overline{L}_{6}}+\mathrm{\overline{M}_{5}\overline{M}_{6}}$ & 0.0 & 0.0 & 0.0542 & 0.0233 \\
\hline
$\mathrm{Nb_{3}Au_{2}}$ & 139 & 54403 & 1 & $0\sim3$ & $\mathrm{\overline{\Gamma}_{6}}+\mathrm{\overline{\Gamma}_{9}}+\mathrm{\overline{M}_{7}}+\mathrm{\overline{M}_{8}}+2\mathrm{\overline{P}_{7}}+\mathrm{\overline{X}_{5}}+\mathrm{\overline{X}_{6}}+2\mathrm{\overline{N}_{5}\overline{N}_{6}}$ & 0.0 & 0.0 & 0.0212 & 0.0369 \\
\hline
$\mathrm{W_{2}Zr}$ & 227 & 653435 & 1 & $-15\sim4$ & $\mathrm{\overline{\Gamma}_{7}}+\mathrm{\overline{\Gamma}_{8}}+2\mathrm{\overline{\Gamma}_{10}}+2\mathrm{\overline{\Gamma}_{11}}+5\mathrm{\overline{X}_{5}}+\mathrm{\overline{L}_{4}\overline{L}_{5}}+3\mathrm{\overline{L}_{6}\overline{L}_{7}}+2\mathrm{\overline{L}_{8}}+4\mathrm{\overline{L}_{9}}+3\mathrm{\overline{W}_{3}\overline{W}_{4}}+2\mathrm{\overline{W}_{5}\overline{W}_{6}}+5\mathrm{\overline{W}_{7}}$ & 0.0 & 0.0 & 0.021 & 0.0702 \\
\hline
$\mathrm{ZrW_{2}}$ & 227 & 151401 & 1 & $-15\sim4$ & $\mathrm{\overline{\Gamma}_{7}}+\mathrm{\overline{\Gamma}_{8}}+2\mathrm{\overline{\Gamma}_{10}}+2\mathrm{\overline{\Gamma}_{11}}+5\mathrm{\overline{X}_{5}}+\mathrm{\overline{L}_{4}\overline{L}_{5}}+3\mathrm{\overline{L}_{6}\overline{L}_{7}}+2\mathrm{\overline{L}_{8}}+4\mathrm{\overline{L}_{9}}+3\mathrm{\overline{W}_{3}\overline{W}_{4}}+2\mathrm{\overline{W}_{5}\overline{W}_{6}}+5\mathrm{\overline{W}_{7}}$ & 0.0 & 0.0 & 0.021 & 0.0702 \\
\hline
$\mathrm{MgSiN_{2}}$ & 166 & 186509 & 1 & $-3\sim0$ & $\mathrm{\overline{\Gamma}_{6}\overline{\Gamma}_{7}}+\mathrm{\overline{\Gamma}_{9}}+\mathrm{\overline{T}_{4}\overline{T}_{5}}+\mathrm{\overline{T}_{8}}+2\mathrm{\overline{F}_{3}\overline{F}_{4}}+2\mathrm{\overline{L}_{5}\overline{L}_{6}}$ & 0.0 & 4.2025 & 0.0206 & 5.0508 \\
\hline
$\mathrm{Ba(Ag_{2}S_{2})}$ & 164 & 50183 & 5 & $1\sim2$ & $\mathrm{\overline{A}_{8}}+\mathrm{\overline{\Gamma}_{8}}+\mathrm{\overline{H}_{4}\overline{H}_{5}}+\mathrm{\overline{K}_{4}\overline{K}_{5}}+\mathrm{\overline{L}_{3}\overline{L}_{4}}+\mathrm{\overline{M}_{3}\overline{M}_{4}}$ & 0.9298 & 0.0 & 0.9298 & 0.0201 \\
\hline
$\mathrm{In_{2}Se_{3}}$ & 164 & 602266 & 5 & $1\sim2$ & $\mathrm{\overline{A}_{8}}+\mathrm{\overline{\Gamma}_{8}}+\mathrm{\overline{H}_{4}\overline{H}_{5}}+\mathrm{\overline{K}_{4}\overline{K}_{5}}+\mathrm{\overline{L}_{3}\overline{L}_{4}}+\mathrm{\overline{M}_{3}\overline{M}_{4}}$ & 0.0 & 0.0 & 0.4934 & 0.0199 \\
\hline
$\mathrm{PbTe}$ & 225 & 153711 & 1 & $-3\sim0$ & $\mathrm{\overline{\Gamma}_{11}}+\mathrm{\overline{X}_{8}}+\mathrm{\overline{X}_{9}}+\mathrm{\overline{L}_{6}\overline{L}_{7}}+\mathrm{\overline{L}_{8}}+2\mathrm{\overline{W}_{6}}$ & 0.0 & 0.0199 & 0.0597 & 0.0199 \\
\hline
$\mathrm{RbYTe_{2}}$ & 194 & 419996 & 3 & $-3\sim0$ & $\mathrm{\overline{A}_{4}\overline{A}_{5}}+2\mathrm{\overline{\Gamma}_{10}}+\mathrm{\overline{H}_{8}}+\mathrm{\overline{H}_{9}}+\mathrm{\overline{K}_{8}}+\mathrm{\overline{K}_{9}}+\mathrm{\overline{L}_{3}\overline{L}_{4}}+2\mathrm{\overline{M}_{6}}$ & 0.0 & 0.9207 & 0.0196 & 1.604 \\
\hline
$\mathrm{W_{2}Zr}$ & 227 & 106218 & 1 & $-15\sim4$ & $\mathrm{\overline{\Gamma}_{7}}+\mathrm{\overline{\Gamma}_{8}}+2\mathrm{\overline{\Gamma}_{10}}+2\mathrm{\overline{\Gamma}_{11}}+5\mathrm{\overline{X}_{5}}+3\mathrm{\overline{L}_{4}\overline{L}_{5}}+\mathrm{\overline{L}_{6}\overline{L}_{7}}+4\mathrm{\overline{L}_{8}}+2\mathrm{\overline{L}_{9}}+2\mathrm{\overline{W}_{3}\overline{W}_{4}}+3\mathrm{\overline{W}_{5}\overline{W}_{6}}+5\mathrm{\overline{W}_{7}}$ & 0.0 & 0.0 & 0.0191 & 0.0676 \\
\hline
$\mathrm{Li_{2}CuSn_{2}}$ & 141 & 426084 & 1 & $-1\sim6$ & $2\mathrm{\overline{\Gamma}_{6}}+2\mathrm{\overline{\Gamma}_{7}}+2\mathrm{\overline{M}_{5}}+2\mathrm{\overline{P}_{3}\overline{P}_{6}}+2\mathrm{\overline{P}_{7}}+2\mathrm{\overline{X}_{3}\overline{X}_{4}}+\mathrm{\overline{N}_{3}\overline{N}_{4}}+3\mathrm{\overline{N}_{5}\overline{N}_{6}}$ & 0.0 & 0.0 & 0.0182 & 0.0336 \\
\hline
$\mathrm{BaGaGeH}$ & 156 & 246820 & 1 & $-1\sim0$ & $\mathrm{\overline{A}_{4}\overline{A}_{5}}+\mathrm{\overline{\Gamma}_{4}\overline{\Gamma}_{5}}+\mathrm{\overline{H}_{4}}+\mathrm{\overline{H}_{5}}+\mathrm{\overline{K}_{4}}+\mathrm{\overline{K}_{5}}+\mathrm{\overline{L}_{3}\overline{L}_{4}}+\mathrm{\overline{M}_{3}\overline{M}_{4}}$ & 0.0 & 0.0179 & 0.0335 & 0.0179 \\
\hline
$\mathrm{Ni_{3}Sn_{2}S_{2}}$ & 166 & 646379 & 1 & $1\sim4$ & $\mathrm{\overline{\Gamma}_{4}\overline{\Gamma}_{5}}+\mathrm{\overline{\Gamma}_{8}}+\mathrm{\overline{T}_{4}\overline{T}_{5}}+\mathrm{\overline{T}_{8}}+2\mathrm{\overline{F}_{5}\overline{F}_{6}}+2\mathrm{\overline{L}_{5}\overline{L}_{6}}$ & 0.0 & 0.0 & 0.0175 & 0.0501 \\
\hline
$\mathrm{CaBe_{2}P_{2}}$ & 164 & 616191 & 1 & $-3\sim0$ & $\mathrm{\overline{A}_{4}\overline{A}_{5}}+\mathrm{\overline{A}_{8}}+\mathrm{\overline{\Gamma}_{6}\overline{\Gamma}_{7}}+\mathrm{\overline{\Gamma}_{9}}+\mathrm{\overline{H}_{4}\overline{H}_{5}}+\mathrm{\overline{H}_{6}}+\mathrm{\overline{K}_{4}\overline{K}_{5}}+\mathrm{\overline{K}_{6}}+2\mathrm{\overline{L}_{5}\overline{L}_{6}}+2\mathrm{\overline{M}_{3}\overline{M}_{4}}$ & 0.0 & 0.8824 & 0.0171 & 1.5635 \\
\hline
$\mathrm{Ni_{3}Sn_{2}S_{2}}$ & 166 & 402458 & 1 & $1\sim4$ & $\mathrm{\overline{\Gamma}_{4}\overline{\Gamma}_{5}}+\mathrm{\overline{\Gamma}_{8}}+\mathrm{\overline{T}_{6}\overline{T}_{7}}+\mathrm{\overline{T}_{9}}+2\mathrm{\overline{F}_{5}\overline{F}_{6}}+2\mathrm{\overline{L}_{3}\overline{L}_{4}}$ & 0.0 & 0.0 & 0.017 & 0.0496 \\
\hline
$\mathrm{Li_{2}NiO_{2}}$ & 164 & 71421 & 3 & $-1\sim0$ & $\mathrm{\overline{A}_{4}\overline{A}_{5}}+\mathrm{\overline{\Gamma}_{4}\overline{\Gamma}_{5}}+\mathrm{\overline{H}_{6}}+\mathrm{\overline{K}_{6}}+\mathrm{\overline{L}_{3}\overline{L}_{4}}+\mathrm{\overline{M}_{3}\overline{M}_{4}}$ & 0.5262 & 0.0 & 0.5298 & 0.0169 \\
\hline
$\mathrm{BaGaGeH}$ & 156 & 173573 & 1 & $-1\sim0$ & $\mathrm{\overline{A}_{4}\overline{A}_{5}}+\mathrm{\overline{\Gamma}_{4}\overline{\Gamma}_{5}}+\mathrm{\overline{H}_{4}}+\mathrm{\overline{H}_{6}}+\mathrm{\overline{K}_{4}}+\mathrm{\overline{K}_{6}}+\mathrm{\overline{L}_{3}\overline{L}_{4}}+\mathrm{\overline{M}_{3}\overline{M}_{4}}$ & 0.0 & 0.0186 & 0.0167 & 0.0186 \\
\hline
$\mathrm{Au_{2}Nb_{3}}$ & 139 & 58559 & 1 & $0\sim3$ & $\mathrm{\overline{\Gamma}_{6}}+\mathrm{\overline{\Gamma}_{9}}+\mathrm{\overline{M}_{7}}+\mathrm{\overline{M}_{8}}+2\mathrm{\overline{P}_{7}}+\mathrm{\overline{X}_{5}}+\mathrm{\overline{X}_{6}}+2\mathrm{\overline{N}_{5}\overline{N}_{6}}$ & 0.0 & 0.0 & 0.0152 & 0.0275 \\
\hline
$\mathrm{BaSr(Fe_{4}O_{8})}$ & 162 & 1838 & 10 & $1\sim2$ & $\mathrm{\overline{A}_{6}\overline{A}_{7}}+\mathrm{\overline{\Gamma}_{6}\overline{\Gamma}_{7}}+\mathrm{\overline{H}_{6}}+\mathrm{\overline{K}_{6}}+\mathrm{\overline{L}_{5}\overline{L}_{6}}+\mathrm{\overline{M}_{5}\overline{M}_{6}}$ & 0.0 & 0.0 & 0.0152 & 0.0534 \\
\hline
$\mathrm{CaAl_{2}Si_{2}}$ & 164 & 20278 & 2 & $-3\sim0$ & $\mathrm{\overline{A}_{4}\overline{A}_{5}}+\mathrm{\overline{A}_{8}}+\mathrm{\overline{\Gamma}_{6}\overline{\Gamma}_{7}}+\mathrm{\overline{\Gamma}_{9}}+\mathrm{\overline{H}_{4}\overline{H}_{5}}+\mathrm{\overline{H}_{6}}+\mathrm{\overline{K}_{4}\overline{K}_{5}}+\mathrm{\overline{K}_{6}}+2\mathrm{\overline{L}_{5}\overline{L}_{6}}+2\mathrm{\overline{M}_{3}\overline{M}_{4}}$ & 0.0 & 0.0 & 0.0126 & 0.5233 \\
\hline
$\mathrm{Cr_{2}Ta}$ & 227 & 626854 & 1 & $-9\sim2$ & $\mathrm{\overline{\Gamma}_{7}}+\mathrm{\overline{\Gamma}_{8}}+\mathrm{\overline{\Gamma}_{10}}+\mathrm{\overline{\Gamma}_{11}}+3\mathrm{\overline{X}_{5}}+2\mathrm{\overline{L}_{4}\overline{L}_{5}}+3\mathrm{\overline{L}_{8}}+\mathrm{\overline{L}_{9}}+\mathrm{\overline{W}_{3}\overline{W}_{4}}+2\mathrm{\overline{W}_{5}\overline{W}_{6}}+3\mathrm{\overline{W}_{7}}$ & 0.0 & 0.0 & 0.0125 & 0.0857 \\
\hline
$\mathrm{CdTlTe_{2}}$ & 164 & 620548 & 3 & $-4\sim1$ & $\mathrm{\overline{A}_{4}\overline{A}_{5}}+\mathrm{\overline{A}_{6}\overline{A}_{7}}+\mathrm{\overline{A}_{9}}+\mathrm{\overline{\Gamma}_{4}\overline{\Gamma}_{5}}+\mathrm{\overline{\Gamma}_{6}\overline{\Gamma}_{7}}+\mathrm{\overline{\Gamma}_{8}}+\mathrm{\overline{H}_{4}\overline{H}_{5}}+2\mathrm{\overline{H}_{6}}+\mathrm{\overline{K}_{4}\overline{K}_{5}}+2\mathrm{\overline{K}_{6}}+\mathrm{\overline{L}_{3}\overline{L}_{4}}+2\mathrm{\overline{L}_{5}\overline{L}_{6}}+2\mathrm{\overline{M}_{3}\overline{M}_{4}}+\mathrm{\overline{M}_{5}\overline{M}_{6}}$ & 0.0 & 0.0 & 0.0123 & 0.9159 \\
\hline
$\mathrm{Ba(Sb_{2}O_{6})}$ & 162 & 74541 & 5 & $-1\sim0$ & $\mathrm{\overline{A}_{6}\overline{A}_{7}}+\mathrm{\overline{\Gamma}_{6}\overline{\Gamma}_{7}}+\mathrm{\overline{H}_{6}}+\mathrm{\overline{K}_{6}}+\mathrm{\overline{L}_{5}\overline{L}_{6}}+\mathrm{\overline{M}_{5}\overline{M}_{6}}$ & 0.0 & 3.3497 & 0.0119 & 3.4314 \\
\hline
$\mathrm{Ca(Al_{12}Si_{4}O_{27})}$ & 147 & 91233 & 6 & $1\sim2$ & $\mathrm{\overline{A}_{5}\overline{A}_{6}}+\mathrm{\overline{\Gamma}_{5}\overline{\Gamma}_{6}}+\mathrm{\overline{H}_{4}\overline{H}_{4}}+\mathrm{\overline{K}_{4}\overline{K}_{4}}+\mathrm{\overline{L}_{3}\overline{L}_{3}}+\mathrm{\overline{M}_{3}\overline{M}_{3}}$ & 5.4299 & 0.0 & 5.5036 & 0.0119 \\
\hline
$\mathrm{Au_{3}In_{2}}$ & 164 & 612019 & 2 & $0\sim1$ & $\mathrm{\overline{A}_{9}}+\mathrm{\overline{\Gamma}_{8}}+\mathrm{\overline{H}_{4}\overline{H}_{5}}+\mathrm{\overline{K}_{4}\overline{K}_{5}}+\mathrm{\overline{L}_{5}\overline{L}_{6}}+\mathrm{\overline{M}_{3}\overline{M}_{4}}$ & 0.0 & 0.0 & 0.0116 & 0.1515 \\
\hline
$\mathrm{Tl(Mo_{6}O_{17})}$ & 164 & 62699 & 9 & $0\sim1$ & $\mathrm{\overline{A}_{8}}+\mathrm{\overline{\Gamma}_{8}}+\mathrm{\overline{H}_{4}\overline{H}_{5}}+\mathrm{\overline{K}_{4}\overline{K}_{5}}+\mathrm{\overline{L}_{3}\overline{L}_{4}}+\mathrm{\overline{M}_{3}\overline{M}_{4}}$ & 0.0 & 0.0 & 0.0188 & 0.0115 \\
\hline
$\mathrm{NbSe_{2}}$ & 164 & 76576 & 6 & $0\sim1$ & $\mathrm{\overline{A}_{4}\overline{A}_{5}}+\mathrm{\overline{\Gamma}_{4}\overline{\Gamma}_{5}}+\mathrm{\overline{H}_{6}}+\mathrm{\overline{K}_{6}}+\mathrm{\overline{L}_{3}\overline{L}_{4}}+\mathrm{\overline{M}_{3}\overline{M}_{4}}$ & 0.0 & 0.0 & 0.011 & 0.0667 \\
\hline
$\mathrm{CdLi_{2}Ge}$ & 225 & 52803 & 2 & $-1\sim2$ & $\mathrm{\overline{\Gamma}_{11}}+\mathrm{\overline{X}_{8}}+\mathrm{\overline{X}_{9}}+\mathrm{\overline{L}_{6}\overline{L}_{7}}+\mathrm{\overline{L}_{8}}+2\mathrm{\overline{W}_{6}}$ & 0.0 & 0.0 & 0.0109 & 0.1314 \\
\hline
$\mathrm{Sr(As_{2}O_{6})}$ & 162 & 420296 & 3 & $-1\sim0$ & $\mathrm{\overline{A}_{6}\overline{A}_{7}}+\mathrm{\overline{\Gamma}_{4}\overline{\Gamma}_{5}}+\mathrm{\overline{H}_{6}}+\mathrm{\overline{K}_{6}}+\mathrm{\overline{L}_{5}\overline{L}_{6}}+\mathrm{\overline{M}_{3}\overline{M}_{4}}$ & 0.0 & 3.9384 & 0.0106 & 4.2275 \\
\hline
$\mathrm{Mg(Cr_{2}O_{4})}$ & 227 & 167459 & 1 & $-7\sim0$ & $\mathrm{\overline{\Gamma}_{6}}+\mathrm{\overline{\Gamma}_{7}}+\mathrm{\overline{\Gamma}_{10}}+2\mathrm{\overline{X}_{5}}+\mathrm{\overline{L}_{6}\overline{L}_{7}}+\mathrm{\overline{L}_{8}}+2\mathrm{\overline{L}_{9}}+2\mathrm{\overline{W}_{5}\overline{W}_{6}}+2\mathrm{\overline{W}_{7}}$ & 0.0 & 0.0 & 0.0132 & 0.0101 \\
\hline
$\mathrm{YRe_{2}}$ & 194 & 150517 & 1 & $-15\sim0$ & $2\mathrm{\overline{A}_{4}\overline{A}_{5}}+2\mathrm{\overline{A}_{6}}+2\mathrm{\overline{\Gamma}_{7}}+\mathrm{\overline{\Gamma}_{8}}+2\mathrm{\overline{\Gamma}_{10}}+\mathrm{\overline{\Gamma}_{11}}+2\mathrm{\overline{\Gamma}_{12}}+2\mathrm{\overline{H}_{6}\overline{H}_{7}}+4\mathrm{\overline{H}_{8}}+2\mathrm{\overline{H}_{9}}+2\mathrm{\overline{K}_{7}}+4\mathrm{\overline{K}_{8}}+2\mathrm{\overline{K}_{9}}+4\mathrm{\overline{L}_{3}\overline{L}_{4}}+3\mathrm{\overline{M}_{5}}+5\mathrm{\overline{M}_{6}}$ & 0.0 & 0.0 & 0.0099 & 0.0238 \\
\hline
$\mathrm{HfMo_{2}}$ & 227 & 638607 & 1 & $-11\sim8$ & $\mathrm{\overline{\Gamma}_{7}}+\mathrm{\overline{\Gamma}_{8}}+2\mathrm{\overline{\Gamma}_{10}}+2\mathrm{\overline{\Gamma}_{11}}+5\mathrm{\overline{X}_{5}}+3\mathrm{\overline{L}_{4}\overline{L}_{5}}+\mathrm{\overline{L}_{6}\overline{L}_{7}}+4\mathrm{\overline{L}_{8}}+2\mathrm{\overline{L}_{9}}+2\mathrm{\overline{W}_{3}\overline{W}_{4}}+3\mathrm{\overline{W}_{5}\overline{W}_{6}}+5\mathrm{\overline{W}_{7}}$ & 0.0 & 0.0 & 0.0168 & 0.0097 \\
\hline
$\mathrm{Ba_{3}Si_{6}O_{12}N_{2}}$ & 147 & 421322 & 3 & $-1\sim0$ & $\mathrm{\overline{A}_{8}\overline{A}_{9}}+\mathrm{\overline{\Gamma}_{8}\overline{\Gamma}_{9}}+\mathrm{\overline{H}_{4}\overline{H}_{4}}+\mathrm{\overline{K}_{4}\overline{K}_{4}}+\mathrm{\overline{L}_{2}\overline{L}_{2}}+\mathrm{\overline{M}_{2}\overline{M}_{2}}$ & 0.0 & 4.766 & 0.0095 & 5.0373 \\
\hline
$\mathrm{MgCr_{2}O_{4}}$ & 227 & 290599 & 1 & $-7\sim0$ & $\mathrm{\overline{\Gamma}_{6}}+\mathrm{\overline{\Gamma}_{7}}+\mathrm{\overline{\Gamma}_{10}}+2\mathrm{\overline{X}_{5}}+\mathrm{\overline{L}_{6}\overline{L}_{7}}+\mathrm{\overline{L}_{8}}+2\mathrm{\overline{L}_{9}}+2\mathrm{\overline{W}_{5}\overline{W}_{6}}+2\mathrm{\overline{W}_{7}}$ & 0.0 & 0.0 & 0.013 & 0.0095 \\
\hline
$\mathrm{Al_{3}Pd_{2}}$ & 164 & 58117 & 1 & $0\sim1$ & $\mathrm{\overline{A}_{6}\overline{A}_{7}}+\mathrm{\overline{\Gamma}_{4}\overline{\Gamma}_{5}}+\mathrm{\overline{H}_{6}}+\mathrm{\overline{K}_{6}}+\mathrm{\overline{L}_{5}\overline{L}_{6}}+\mathrm{\overline{M}_{3}\overline{M}_{4}}$ & 0.0 & 0.0 & 0.0093 & 0.1289 \\
\hline
$\mathrm{Zr_{3}Al_{3}C_{5}}$ & 194 & 159412 & 2 & $-1\sim6$ & $\mathrm{\overline{A}_{4}\overline{A}_{5}}+\mathrm{\overline{A}_{6}}+2\mathrm{\overline{\Gamma}_{7}}+\mathrm{\overline{\Gamma}_{8}}+\mathrm{\overline{\Gamma}_{9}}+\mathrm{\overline{H}_{4}\overline{H}_{5}}+\mathrm{\overline{H}_{6}\overline{H}_{7}}+\mathrm{\overline{H}_{8}}+\mathrm{\overline{H}_{9}}+2\mathrm{\overline{K}_{7}}+\mathrm{\overline{K}_{8}}+\mathrm{\overline{K}_{9}}+2\mathrm{\overline{L}_{3}\overline{L}_{4}}+4\mathrm{\overline{M}_{6}}$ & 0.0 & 0.0 & 0.017 & 0.0093 \\
\hline
$\mathrm{Ca(As_{2}O_{6})}$ & 162 & 81064 & 1 & $-1\sim0$ & $\mathrm{\overline{A}_{6}\overline{A}_{7}}+\mathrm{\overline{\Gamma}_{4}\overline{\Gamma}_{5}}+\mathrm{\overline{H}_{6}}+\mathrm{\overline{K}_{6}}+\mathrm{\overline{L}_{5}\overline{L}_{6}}+\mathrm{\overline{M}_{3}\overline{M}_{4}}$ & 0.0 & 3.9469 & 0.0093 & 4.2667 \\
\hline
$\mathrm{CaSi_{2}}$ & 166 & 248517 & 1 & $1\sim4$ & $\mathrm{\overline{\Gamma}_{6}\overline{\Gamma}_{7}}+\mathrm{\overline{\Gamma}_{9}}+\mathrm{\overline{T}_{6}\overline{T}_{7}}+\mathrm{\overline{T}_{9}}+2\mathrm{\overline{F}_{3}\overline{F}_{4}}+2\mathrm{\overline{L}_{3}\overline{L}_{4}}$ & 0.0 & 0.0 & 0.0153 & 0.0092 \\
\hline
$\mathrm{CuV_{2}S_{4}}$ & 227 & 628953 & 1 & $-5\sim6$ & $\mathrm{\overline{\Gamma}_{7}}+\mathrm{\overline{\Gamma}_{8}}+\mathrm{\overline{\Gamma}_{10}}+\mathrm{\overline{\Gamma}_{11}}+3\mathrm{\overline{X}_{5}}+2\mathrm{\overline{L}_{6}\overline{L}_{7}}+\mathrm{\overline{L}_{8}}+3\mathrm{\overline{L}_{9}}+2\mathrm{\overline{W}_{3}\overline{W}_{4}}+\mathrm{\overline{W}_{5}\overline{W}_{6}}+3\mathrm{\overline{W}_{7}}$ & 0.0 & 0.0 & 0.0105 & 0.009 \\
\hline
$\mathrm{SiC}$ & 156 & 43827 & 10 & $-1\sim0$ & $\mathrm{\overline{A}_{4}\overline{A}_{5}}+\mathrm{\overline{\Gamma}_{4}\overline{\Gamma}_{5}}+\mathrm{\overline{H}_{4}}+\mathrm{\overline{H}_{6}}+\mathrm{\overline{K}_{4}}+\mathrm{\overline{K}_{6}}+\mathrm{\overline{L}_{3}\overline{L}_{4}}+\mathrm{\overline{M}_{3}\overline{M}_{4}}$ & 0.0 & 1.6239 & 0.0087 & 2.733 \\
\hline
$\mathrm{CsSnI_{3}}$ & 127 & 69995 & 1 & $1\sim12$ & $3\mathrm{\overline{A}_{6}\overline{A}_{7}}+\mathrm{\overline{\Gamma}_{6}}+3\mathrm{\overline{\Gamma}_{7}}+\mathrm{\overline{\Gamma}_{8}}+\mathrm{\overline{\Gamma}_{9}}+3\mathrm{\overline{M}_{8}\overline{M}_{9}}+2\mathrm{\overline{Z}_{6}}+4\mathrm{\overline{Z}_{7}}+3\mathrm{\overline{R}_{3}\overline{R}_{4}}+3\mathrm{\overline{X}_{3}\overline{X}_{4}}$ & 0.1877 & 0.0 & 0.1877 & 0.0086 \\
\hline
$\mathrm{ZnCr_{2}S_{4}}$ & 227 & 42019 & 2 & $-11\sim0$ & $\mathrm{\overline{\Gamma}_{7}}+\mathrm{\overline{\Gamma}_{8}}+\mathrm{\overline{\Gamma}_{10}}+\mathrm{\overline{\Gamma}_{11}}+3\mathrm{\overline{X}_{5}}+2\mathrm{\overline{L}_{6}\overline{L}_{7}}+\mathrm{\overline{L}_{8}}+3\mathrm{\overline{L}_{9}}+2\mathrm{\overline{W}_{3}\overline{W}_{4}}+\mathrm{\overline{W}_{5}\overline{W}_{6}}+3\mathrm{\overline{W}_{7}}$ & 0.0 & 0.0 & 0.0083 & 0.0099 \\
\hline
$\mathrm{LiSr(AlF_{6})}$ & 163 & 68905 & 1 & $1\sim4$ & $\mathrm{\overline{A}_{5}\overline{A}_{6}}+2\mathrm{\overline{\Gamma}_{8}}+\mathrm{\overline{H}_{4}\overline{H}_{5}}+\mathrm{\overline{H}_{6}}+\mathrm{\overline{K}_{4}\overline{K}_{5}}+\mathrm{\overline{K}_{6}}+\mathrm{\overline{L}_{2}\overline{L}_{2}}+2\mathrm{\overline{M}_{3}\overline{M}_{4}}$ & 7.6262 & 0.0 & 7.6405 & 0.0083 \\
\hline
$\mathrm{Ba_{2}NiOsO_{6}}$ & 164 & 16406 & 11 & $-1\sim0$ & $\mathrm{\overline{A}_{6}\overline{A}_{7}}+\mathrm{\overline{\Gamma}_{4}\overline{\Gamma}_{5}}+\mathrm{\overline{H}_{6}}+\mathrm{\overline{K}_{6}}+\mathrm{\overline{L}_{5}\overline{L}_{6}}+\mathrm{\overline{M}_{3}\overline{M}_{4}}$ & 0.0 & 0.0 & 0.0079 & 0.0096 \\
\hline
$\mathrm{BaSrFe_{4}O_{8}}$ & 162 & 37011 & 11 & $-1\sim0$ & $\mathrm{\overline{A}_{9}}+\mathrm{\overline{\Gamma}_{9}}+\mathrm{\overline{H}_{4}\overline{H}_{5}}+\mathrm{\overline{K}_{4}\overline{K}_{5}}+\mathrm{\overline{L}_{5}\overline{L}_{6}}+\mathrm{\overline{M}_{5}\overline{M}_{6}}$ & 0.0 & 0.0 & 0.0299 & 0.0079 \\
\hline
$\mathrm{SiC}$ & 156 & 107204 & 2 & $-1\sim0$ & $\mathrm{\overline{A}_{4}\overline{A}_{5}}+\mathrm{\overline{\Gamma}_{4}\overline{\Gamma}_{5}}+\mathrm{\overline{H}_{5}}+\mathrm{\overline{H}_{6}}+\mathrm{\overline{K}_{5}}+\mathrm{\overline{K}_{6}}+\mathrm{\overline{L}_{3}\overline{L}_{4}}+\mathrm{\overline{M}_{3}\overline{M}_{4}}$ & 0.0 & 1.7191 & 0.0079 & 2.8422 \\
\hline
$\mathrm{BC_{7}}$ & 156 & 181953 & 3 & $0\sim1$ & $\mathrm{\overline{A}_{4}\overline{A}_{5}}+\mathrm{\overline{\Gamma}_{4}\overline{\Gamma}_{5}}+\mathrm{\overline{H}_{4}}+\mathrm{\overline{H}_{6}}+\mathrm{\overline{K}_{4}}+\mathrm{\overline{K}_{6}}+\mathrm{\overline{L}_{3}\overline{L}_{4}}+\mathrm{\overline{M}_{3}\overline{M}_{4}}$ & 0.0 & 2.7989 & 0.0077 & 3.7572 \\
\hline
$\mathrm{Sn_{2}(Ta_{2}O_{7})}$ & 227 & 27119 & 4 & $-11\sim0$ & $\mathrm{\overline{\Gamma}_{7}}+\mathrm{\overline{\Gamma}_{8}}+\mathrm{\overline{\Gamma}_{10}}+\mathrm{\overline{\Gamma}_{11}}+3\mathrm{\overline{X}_{5}}+2\mathrm{\overline{L}_{6}\overline{L}_{7}}+\mathrm{\overline{L}_{8}}+3\mathrm{\overline{L}_{9}}+2\mathrm{\overline{W}_{3}\overline{W}_{4}}+\mathrm{\overline{W}_{5}\overline{W}_{6}}+3\mathrm{\overline{W}_{7}}$ & 0.0 & 0.8308 & 0.0077 & 0.8798 \\
\hline
$\mathrm{Fe(Cr_{2}O_{4})}$ & 227 & 183963 & 2 & $1\sim8$ & $\mathrm{\overline{\Gamma}_{6}}+\mathrm{\overline{\Gamma}_{7}}+\mathrm{\overline{\Gamma}_{10}}+2\mathrm{\overline{X}_{5}}+\mathrm{\overline{L}_{4}\overline{L}_{5}}+3\mathrm{\overline{L}_{9}}+\mathrm{\overline{W}_{3}\overline{W}_{4}}+\mathrm{\overline{W}_{5}\overline{W}_{6}}+2\mathrm{\overline{W}_{7}}$ & 0.0 & 0.0 & 0.0085 & 0.0075 \\
\hline
$\mathrm{Zn(Cr_{2}S_{4})}$ & 227 & 166481 & 1 & $-11\sim0$ & $\mathrm{\overline{\Gamma}_{7}}+\mathrm{\overline{\Gamma}_{8}}+\mathrm{\overline{\Gamma}_{10}}+\mathrm{\overline{\Gamma}_{11}}+3\mathrm{\overline{X}_{5}}+2\mathrm{\overline{L}_{6}\overline{L}_{7}}+\mathrm{\overline{L}_{8}}+3\mathrm{\overline{L}_{9}}+2\mathrm{\overline{W}_{3}\overline{W}_{4}}+\mathrm{\overline{W}_{5}\overline{W}_{6}}+3\mathrm{\overline{W}_{7}}$ & 0.0 & 0.0 & 0.0075 & 0.0141 \\
\hline
$\mathrm{BC_{5}}$ & 156 & 180770 & 3 & $0\sim1$ & $\mathrm{\overline{A}_{4}\overline{A}_{5}}+\mathrm{\overline{\Gamma}_{4}\overline{\Gamma}_{5}}+\mathrm{\overline{H}_{5}}+\mathrm{\overline{H}_{6}}+\mathrm{\overline{K}_{5}}+\mathrm{\overline{K}_{6}}+\mathrm{\overline{L}_{3}\overline{L}_{4}}+\mathrm{\overline{M}_{3}\overline{M}_{4}}$ & 0.0 & 2.9665 & 0.0074 & 4.2867 \\
\hline
\end{longtable*}
% 157

\section{Discussion}
In this section we discuss two examples related to our classification. 
The first is the TBG \cite{Po2018b,Ahn2018,Song2018c,Tarnopolsky2019,Liu2018_TBG}, and the second is Fu's topological crystalline insulator \cite{Fu2011}.
TBG can be successfully diagnosed through our framework; while Fu's model is beyond the symmetry eigenvalue classification.
Nevertheless, we develop a generalized symmetry eigenvalue criterion for Fu's state.

\subsection{Twisted bilayer graphene}

\begin{table}
\begin{tabular}{lrrr|lrr|lrr}
\hline 
& $\Gamma_{1}$ & $\Gamma_{2}$ & $\Gamma_{3}$ &  & $\rm M_1$ & $\rm M_2$ &  & $\rm K_1$ & $\rm K_2K_3$ \tabularnewline
\hline 
$E$ & 1 & 1 & 2 &        $E$ & 1 & 1 &        $E$ & 1 & 2\tabularnewline
$2C_{3}$ & 1 & 1 & -1 &  $C_{2}^\prime$ & 1 & -1 & $C_{3}$ & 1 & -1 \tabularnewline
$3C_{2}^\prime$ & 1 & -1 & 0 &     &   &   &       $C_{3}^{-1}$ & 1 & -1 \tabularnewline
\hline
\end{tabular}
\protect\caption{\label{tab:irreps-TBG} Character table of irreps at high symmetry momenta in magnetic space group $P6^\prime2^\prime2$ (\#177.151 in BNS settings) \cite{BCS}. For the little group of $\Gamma$, $E$, $C_3$, and $C_2^\prime$ represent the conjugation classes generated from identity, $C_{3z}$, and $C_{2x}$, respectively. The number before each conjugate class represents the number of operations in this class. Conjugate class symbols at M and K are defined in similar ways. }
\end{table}

TBG has an approximate valley-U(1) symmetry \cite{Bistritzer2011} and the single-valley Hamiltonian has the magnetic SG $P6^\pr 2^\pr 2$ (\#177.151 in the BNS notation).
The irreps of $P6^\pr 2^\pr 2$ are tabulated in \cref{tab:irreps-TBG}.
The nearly flat bands around the Fermi level form the irreps \cite{Song2018c}
\begin{equation}\rm 
\Gamma_1 + \Gamma_2+  M_1 + M_2+ K_2K_3. \label{eq:root1-TBG}
\end{equation}
Ref. \cite{Song2018c} found that these irreps cannot be obtained as a difference of EBRs and proved that bands having these irreps have $C_2T$-protected Wilson loop winding, with the winding number $3n\pm1$ ($n\in\mbb{Z}$).
Another EFP having Wilson loop winding given in the supplementary material of Ref. \cite{Song2018c} is
\begin{equation}\rm 
\Gamma_3+ M_1 + M_2+ 2K_1.\label{eq:root2-TBG}
\end{equation}

We apply the polyhedron method (\cref{sec:main-polyhedron}) to the magnetic SG $P6^\pr 2^\pr 2$ and obtain the complete eigenvalue criteria for the EFPs.
The details of calculations are given in \cref{app:TBG}.
Here we briefly describe the results.
We obtain a single inequality-type criterion
\begin{equation}
2m({\rm K_2K_3}) < m(\Gamma_3),\label{eq:criterion1-TBG}
\end{equation}
and a single $\mbb{Z}_2$-type criterion
\begin{align}
& m(\Gamma_1)+m(\Gamma_2)+2m(\Gamma_3)-2m({\rm K_2K_3})=0, \nono\\
& m(\Gamma_2)=1\mod2. \label{eq:criterion2-TBG}
\end{align}
The EFP (\ref{eq:root1-TBG}) is diagnosed by the $\mbb{Z}_2$-type criterion, and the other EFP (\ref{eq:root2-TBG}) is diagnosed by the inequality-type criterion.
We emphasize that \cref{eq:criterion1-TBG,eq:criterion2-TBG} go beyond the two-band eigenvalue criteria derived in Ref.~\cite{Song2018c} because they apply to many-band systems.
Applying the method introduced in \cref{sec:main-root}, we find that $P6^\pr 2^\pr 2$ has only two EFP roots and the two roots are just \cref{eq:root1-TBG,eq:root2-TBG}.

% \cref{eq:criterion1-TBG} implies that there are EFPs with band numbers larger than 2.
% For example, $\rm \Gamma_1+\Gamma_3+ 2M_1 + M_2 + 3K_1$ is a EFP with band number 3 because it satisfies \cref{eq:criterion1-TBG}.
% Thus such 

\subsection{Fu's topological crystalline insulator and a generalized symmetry eigenvalue criterion}
Fu's state is spinless and is protected by $C_4$ rotation and TRS.
%Its generalization to point groups $C_{3v}$, $C_{4v}$, $C_{6v}$ was proposed latter \cite{Alexandradinata2014}.
The topological invariant is well defined only if the Hilbert space is restricted to $p_{x,y}$ orbitals.
Correspondingly, the topological surface state is stable only if the model consists of $p_{x,y}$ orbitals.
%The surface state can be symmetrically gapped by being coupled to orbitals of other type (such as $s$ orbitals).
Therefore, this state has the defining character of fragile topology.
Recently, Alexandradinata \etal proved that the topology of Fu's model is indeed fragile \cite{Alexandradinata2014}.

This model has an accidental inversion symmetry.
In \cref{app:Fu-irrep} we show that the fragile topology cannot be diagnosed through the $C_4$ and inversion eigenvalues.
Nevertheless, we develop a ``generalized symmetry eigenvalue criterion'' for this state (\cref{app:Fu-criterion}). 
Usually, the diagnosis of topology involves additional symmetries.
For example, diagnosis of strong topological insulator, which is protected by TRS, involves the inversion symmetry.
We find that the additional symmetries diagnosing Fu's state are inversion and $C_{2x}$ rotation.
With these additional symmetries, a $\mbb{Z}_2$ invariant can be defined in terms of the symmetry eigenvalues.
However, the $\mbb{Z}_2$ nontrivial phase is indeed a topological nodal ring semimetal, where the nodal rings are stabilized by inversion (and/or $M_z$).
After the inversion is broken, the nodal rings are gapped.
If the inversion symmetry is broken in such a way that no gap closing happens at the high symmetry points, the obtained insulating phase has the topology of Fu's model.
Since the additional symmetries for diagnosis enforces the topological state to be semimetal, we call this eigenvalue criterion as ``generalized''.

\section{Summary} 
In this paper, we have obtained three major progresses in the field of topological phases. We have, for the first time, entirely mathematically classified the fragile topological states indicated by symmetry eigenvalues - EFPs. We found an extremely rich structure of these phases, linked to the mathematical classification of affine monoids, which surpasses the richness of stable topological phases. We have then, for the first time, provided examples of fragile bands in more than a hundred of realistic materials, showcasing some of the best, well-separated sets of bands. Our work finishes one other important sub-field of topological states of matter. It would be remarkably interesting to find a clear experimental consequence of the well-separated fragile sets of bands we have discovered. 
One such fragile band is the wonder-material of twisted bilayer graphene.

\begin{acknowledgements}
We acknowledge Maia G. Vergniory, Zhijun Wang, and Claudia Felser for collaboration on previous related works.
Z.S. and B.B. are supported by the Department of Energy Grant No. desc-0016239, the National Science Foundation EAGER Grant No. DMR 1643312, Simons Investigator Grants No. 404513, No. ONR N00014-14-1-0330, No. NSF-MRSECDMR DMR 1420541, the Packard Foundation No. 2016-65128, the Schmidt Fund for Development of Majorama Fermions funded by the Eric and Wendy Schmidt Transformative Technology Fund.
L.E. is supported by the Government of the Basque Country (project IT779-13), the Spanish Ministry of Economy and Competitiveness (MINECO), and the European Fund for Economic and Regional Development (FEDER; project MAT2015-66441-P). 
We are mostly grateful to the MPG computing center for allowing us access to the Cobra and Draco Supercomputers. 
This research also used resources of the National Energy Research Scientific Computing Center (NERSC), a U.S. Department of Energy Office of Science User Facility operated under Contract No. DE-AC02-05CH11231.
\end{acknowledgements}

\bibliography{ref}

\appendix
\onecolumngrid

\section{Diagnosis for fragile phases: the inequality method}\label{sec:inequality}
Fragile topological states \cite{Po2018,Cano2018,Bradlyn2017}, also referred to as fragile phases in this paper, are defined to be non-Wannierizable insulating states, where the Wannier obstruction can be removed by coupling the state to a particular set of trivial (Wannierizable) bands.
(A band structure is Wannierizable if a set of \textit{symmetric} Wannier functions can be constructed from the bands.)
In other words, if the number of Wannier functions is fixed to be the number of bands, the fragile phase is not Wannierizable; however, if more Wannier functions are allowed, the fragile bands can be realized as a subset of the bands constructed from all the Wannier functions; the bands outside of this subset are completely trivial (Wannierizable).
Physically, the bands outside the subset correspond to the trivial bands that are added to remove the Wannier obstruction.
In this paper, we restrict ourself to the fragile phases that can be diagnosed from symmetry eigenvalues.
These fragile phases cannot be diagnosed from the indicators introduced in Ref. \cite{Po2017,Khalaf2018,Song2018a,Song2018b}, but they can be written in terms of EBRs \cite{Bradlyn2017,Po2018,Po2018b,Ahn2018,SlagerAndBouhon,Song2018c}.
%as the indicator remains unchanged upon stacking trivial bands, whereas fragile topology can be removed by stacking trivial bands.

Thus we need a new framework to understand the symmetry data vector ($B$) of fragile phases.
Generally speaking, if the symmetry eigenvalues have the following property
\beq
\exists p \in \mbb{Z}^{N_{EBR}},\; s.t.\; B = EBR \cdot p,\qquad
\text{and}\quad \forall p \in \mbb{N}^{N_{EBR}}, B\neq EBR \cdot  p, \label{eq:EFP-def}
\eeq
then we say that the corresponding band structure has at least a fragile topology diagnosable by the symmetry eigenvalues or EBRs.
(It could also have robust/strong topology undiagnosable by symmetry eigenvalues.)
In other words, the symmetry data vector $B$ of a fragile phase cannot be written as a sum of EBRs, but only as a difference of two sums of EBRs, \ie $B = \sum_{i} p_i EBR_i - \sum_{i} q_j EBR_j$, where $p_i,q_j\ge0$ and $p_i q_i=0$ for all $i$.
Here and after we use $A_i$ to represent the $i$-th column of the matrix $A$.
Then adding the BR written as $\sum_{i} q_i EBR_i$ to the fragile phase makes the total symmetry data vector completely trivial.

\subsection{An example: SG 150}\label{sec:P321}
To familiarize ourselves with the symmetry data of fragile phases, here we take an example SG 150 ($P321$) in presence of SOC and TRS.
The EBR matrix of is given by
\beq
EBR = \begin{pmatrix}
0 & 1 & 2 & 0 \\
1 & 0 & 0 & 2 \\
0 & 1 & 2 & 0 \\
1 & 0 & 0 & 2 \\
0 & 1 & 0 & 1 \\
0 & 1 & 0 & 1 \\
1 & 0 & 2 & 1 \\
0 & 1 & 0 & 1 \\
0 & 1 & 0 & 1 \\
1 & 0 & 2 & 1 \\
0 & 1 & 0 & 1 \\
0 & 1 & 0 & 1 \\
1 & 0 & 2 & 1 \\
0 & 1 & 0 & 1 \\
0 & 1 & 0 & 1 \\
1 & 0 & 2 & 1 \\
1 & 1 & 2 & 2 \\
1 & 1 & 2 & 2
\end{pmatrix}
= \left(\begin{array}{rrrrrrrrrrrrrrrrrr}
1 & 0 & -1 & 0 & 0 & 0 & 1 & 0 & 0 & 0 & 0 & 0 & 0 & 0 & 0 & 0 & 0 & 0 \\
0 & 0 & 1 & 0 & 0 & 0 & 0 & 1 & 0 & 0 & 0 & 0 & 0 & 0 & 0 & 0 & 0 & 0 \\
1 & 0 & -1 & 0 & 0 & 0 & 0 & 0 & 0 & 0 & 0 & 0 & 0 & 0 & 0 & 0 & 0 & 0 \\
0 & 0 & 1 & 0 & 0 & 1 & 0 & 0 & 0 & 0 & 0 & 0 & 0 & 0 & 0 & 0 & 0 & 0 \\
1 & -1 & 0 & 0 & 0 & 0 & 0 & 0 & 1 & 0 & 0 & 0 & 0 & 0 & 0 & 0 & 0 & 0 \\
1 & -1 & 0 & 0 & 0 & 0 & 0 & 0 & 0 & 1 & 0 & 0 & 0 & 0 & 0 & 0 & 0 & 0 \\
0 & 1 & 0 & 0 & 0 & 0 & 0 & 0 & 0 & 0 & 1 & 0 & 0 & 0 & 0 & 0 & 0 & 0 \\
1 & -1 & 0 & 0 & 0 & 0 & 0 & 0 & 0 & 0 & 0 & 1 & 0 & 0 & 0 & 0 & 0 & 0 \\
1 & -1 & 0 & 0 & 0 & 0 & 0 & 0 & 0 & 0 & 0 & 0 & 1 & 0 & 0 & 0 & 0 & 0 \\
0 & 1 & 0 & 0 & 0 & 0 & 0 & 0 & 0 & 0 & 0 & 0 & 0 & 1 & 0 & 0 & 0 & 0 \\
1 & -1 & 0 & 0 & 0 & 0 & 0 & 0 & 0 & 0 & 0 & 0 & 0 & 0 & 1 & 0 & 0 & 0 \\
1 & -1 & 0 & 0 & 0 & 0 & 0 & 0 & 0 & 0 & 0 & 0 & 0 & 0 & 0 & 1 & 0 & 0 \\
0 & 1 & 0 & 0 & 0 & 0 & 0 & 0 & 0 & 0 & 0 & 0 & 0 & 0 & 0 & 0 & 1 & 0 \\
1 & -1 & 0 & 0 & 0 & 0 & 0 & 0 & 0 & 0 & 0 & 0 & 0 & 0 & 0 & 0 & 0 & 1 \\
1 & -1 & 0 & 0 & 0 & 0 & 0 & 0 & 0 & 0 & 0 & 0 & 0 & 0 & 0 & 0 & 0 & 0 \\
0 & 1 & 0 & 0 & 1 & 0 & 0 & 0 & 0 & 0 & 0 & 0 & 0 & 0 & 0 & 0 & 0 & 0 \\
1 & 0 & 0 & 0 & 0 & 0 & 0 & 0 & 0 & 0 & 0 & 0 & 0 & 0 & 0 & 0 & 0 & 0 \\
1 & 0 & 0 & 1 & 0 & 0 & 0 & 0 & 0 & 0 & 0 & 0 & 0 & 0 & 0 & 0 & 0 & 0
\end{array}\right)
\begin{pmatrix}
1 & 0 & 0 & 0\\
0 & 1 & 0 & 0\\
0 & 0 & 1 & 0\\
0 & 0 & 0 & 0\\
0 & 0 & 0 & 0\\
0 & 0 & 0 & 0\\
0 & 0 & 0 & 0\\
0 & 0 & 0 & 0\\
0 & 0 & 0 & 0\\
0 & 0 & 0 & 0\\
0 & 0 & 0 & 0\\
0 & 0 & 0 & 0\\
0 & 0 & 0 & 0\\
0 & 0 & 0 & 0\\
0 & 0 & 0 & 0\\
0 & 0 & 0 & 0\\
0 & 0 & 0 & 0\\
0 & 0 & 0 & 0
\end{pmatrix}
\left(\begin{array}{rrrr}
1 & 1 & 2 & 2 \\
1 & 0 & 2 & 1 \\
1 & 0 & 0 & 2 \\
0 & 0 & -1 & 0
\end{array}\right), \label{eq:EBR-150}
\eeq
where the Smith Decomposition $L\Lambda R$ is given after the second equal-sign.
Here each column of the matrix $EBR$ represents an EBR, and the irreps represented by the rows of the EBR matrix are $\ovl{\mrm{A}}_4 \ovl{\mrm{A}}_5$, $\ovl{\mrm{A}}_6$, $\ovl{\Gamma}_4\ovl{\Gamma}_5$, $\ovl{\Gamma}_6$, $\ovl{\mrm{H}}_4$, $\ovl{\mrm{H}}_5$,  $\ovl{\mrm{H}}_6$,  $\ovl{\mrm{HA}}_4$, $\ovl{\mrm{HA}}_5$, $\ovl{\mrm{HA}}_6$, $\ovl{\mrm{K}}_4$, $\ovl{\mrm{K}}_5$, $\ovl{\mrm{K}}_6$, $\ovl{\mrm{KA}}_4$, $\ovl{\mrm{KA}}_5$, $\ovl{\mrm{KA}}_6$, $\ovl{\mrm{L}}_3\ovl{\mrm{L}}_4$, $\ovl{\mrm{M}}_3\ovl{\mrm{M}}_4$, in the notation of the \href{http://www.cryst.ehu.es/}{\it Bilbao Crystallographic Server} (BCS) \cite{Elcoro2017,BCS,Aroyo2006a,Aroyo2006b}. 
(One can find the definitions of these irreps at the \href{http://www.cryst.ehu.es/cgi-bin/cryst/programs/representations.pl?tipogrupo=dbg}{\it Irreducible representations of the Double Space Groups} section of the BCS \cite{Elcoro2017}.)
Since the diagonal elements of $\Lambda$ are either 1 or 0, there is no indicator in this SG.
As described in Ref. \cite{Bradlyn2017,Elcoro2017,Po2017,PaperOnTheInductionMethor}, the space of compatibility-relation-allowed symmetry data can be generated from the first $r$ columns of the $L$ matrix, with $r$ the rank of $\Lambda$ (here $r=3$).
In other words, we can always write the symmetry data vector as $B = EBR\cdot p = L \Lambda R \cdot p$.
Thus, we can introduce the parameter's $y_i = ( R p)_i$ ($i=1,2\cdots r$) and write the symmetry data as
\beqs
B = \sum_{i=1}^3 (L \Lambda)_i y_i = (& y_1-y_3,\ y_3,\ y_1-y_3,\ y_3,\ y_1-y_2,\ y_1-y_2,\ y_2,\ y_1-y_2,\ \nono\\
  & y_1-y_2,\ y_2,\  y_1-y_2,\ y_1-y_2,\ y_2,\ y_1-y_2,\ y_1-y_2,\ y_2,\ y_1,\ y_1)^T. \label{eq:BS-150}
\eeqs
For the number of irreps to be nonnegative, \ie $B\ge 0$, the following inequalities should be satisfied
\beq
y_1\ge y_3\ge 0, \qquad y_1\ge y_2\ge 0. \label{eq:Y-150}
\eeq
Therefore, only the $y$'s that satisfy \cref{eq:Y-150} correspond to physical band structures.
In the following we will use $y$ to represent the band structures.
 
Now we decompose the symmetry data vector in \cref{eq:BS-150} as a combination of EBRs, \ie $B = \sum_i p_i EBR_i$ (\cref{eq:EFP-def}).
On one hand, as $y_i = (R\cdot p)_i$ ($i=1,2,3$) and $\Lambda = \mrm{diag}(1\ 1\ 1\ 0)$ (\cref{eq:EBR-150}), we can always write $p$ as $ p = y_1 R^{-1}_1 + y_2 R^{-1}_2 + y_3 R^{-1}_3$.
($R_i^{-1}$ is the $i$-th column of the matrix $R^{-1}$.)
On the other hand, if $p$ is a solution of $y_i = (Rp)_i $ ($i=1,2,3$), $p + k R^{-1}_4$ is also a solution, where $k$ is a free parameter, because $(R R^{-1}_4)_{1,2,3}=0$.
Therefore, the general solution of the equation $B = EBR \cdot p$ or $y_i =(\Lambda Rp)_i$ ($i=1,2,3$) takes the form of
\begin{equation}
p = y_1 R^{-1}_1 + y_2 R^{-1}_2 + y_3 R^{-1}_3 + k R^{-1}_4. \label{eq:p-solution-150}
\end{equation}
Substituting the $R$ matrix into \cref{eq:p-solution-150}, we obtain
\beq
p = (2y_2-y_3-4k,\ y_1-y_3-2k,\ k,\ -y_2+y_3+2k)^T.
\eeq

For $p$ to be an integer, vector $k$ need to be integer because $p_3=k$.
For a given $y$ vector, if there exits some integer $k$ such that each element of $p$ is nonnegative, then the corresponding symmetry data vector can be written as a sum positive EBRs, and so can be a trivial phase; otherwise, the corresponding band structure necessarily has a fragile topology.
Therefore we conclude that the equivalent condition for symmetry data associated with $y$ to be trivial is
\beq
\exists k \in \mbb{Z},\quad s.t.\quad
2y_2-y_3-4k\ge 0, \quad y_1-y_3-2k\ge 0,\quad k\ge 0,\quad -y_2+y_3+2k\ge 0. \label{eq:k-inequality-150}
\eeq
Solving the inequalities in \cref{eq:k-inequality-150}, we rewrite the trivial condition as
\beq
\exists k \in \mbb{Z},\quad s.t.\quad
\max\pare{0,\frac12 y_2-\frac12 y_3}\le k \le \min\pare{\frac12 y_2-\frac14 y_3, \frac12 y_1-\frac12 y_3 }.  \label{eq:k-inequality2-150}
\eeq
There are two possible cases where \cref{eq:k-inequality2-150} has no solution: (type-I) \cref{eq:k-inequality2-150} has no solution even where $k$ is allowed to be a rational number; (type-II) \cref{eq:k-inequality2-150} has rational solutions but no integer solution.
The two cases correspond to the inequality-type index and the $\mathbb{Z}_2$-type index defined in the main text, respectively.
Here we first consider type-I.
We directly see that type-I happens when any of the following four inequalities is satisfied (A) $0>\frac12 y_2 -\frac14 y_3$, (B) $\frac12 y_2 -\frac12 y_3> \frac12 y_2 -\frac14 y_3$, (C) $0>\frac12 y_1-\frac12 y_3$, (D) $\frac12 y_2 -\frac12 y_3 >\frac12 y_1-\frac12 y_3$.
For example $0>\frac12 y_2 -\frac14 y_3$ (A) and \cref{eq:k-inequality2-150} implies $0\le k < 0$, which has no solution.
Actually, (B), (C) and (D) cannot be satisfied by real band structures because they conflict with $B\ge 0$ (\cref{eq:Y-150}).
Therefore the only left possibility is (A), for which we get the fragile criterion as 
\beq
y_3-2y_2>0. \label{eq:150-inequality}
\eeq
In this paper we refer to $y_3-2y_2$ as an \textit{inequality-type fragile index}. 
A key difference between the inequality-type index for fragile phase and the symmetry-based indicator for stable/strong phase is that the latter can be become trivial upon stacking, whereas the former cannot.
For example, double of the generator state of a $\mbb{Z}_2$ indicator becomes a trivial state, whereas stacking of any positive number of the inequality-type fragile phase, for example, the $y_3-2y_2=1$ state, is still a fragile phase, because the inequality is still satisfied.

Now we consider type-II, where \cref{eq:k-inequality2-150} has solutions only if $k$ is allowed to be rational number.
Type-II happens if the interval set by \cref{eq:k-inequality2-150} is nonzero but does not contain any integer.
We notice that the solution of \cref{eq:k-inequality2-150} can be written as the intersection of the following three intervals
\begin{equation}
0\le k \le \min\pare{\frac12 y_2-\frac14 y_3,\frac12 y_1-\frac12 y_3},  \label{eq:k-inequality2-150-tmp1}
\end{equation}
\begin{equation}
\frac12y_2-\frac12y_3\le k \le \frac12 y_2-\frac14 y_3,  \label{eq:k-inequality2-150-tmp2}
\end{equation}
\begin{equation}
\frac12y_2-\frac12y_3\le k \le \frac12 y_1-\frac12 y_3. \label{eq:k-inequality2-150-tmp3}
\end{equation}
If \cref{eq:k-inequality2-150-tmp1} has solutions, the solutions must include the lower bound $0$, which violates our request that the interval does not contain integers.
Thus \cref{eq:k-inequality2-150-tmp1} does not bring new indices.
Therefore we only need to consider the case where \cref{eq:k-inequality2-150-tmp2} or (\ref{eq:k-inequality2-150-tmp3}) has only fractional solutions but no integer solution.
For \cref{eq:k-inequality2-150-tmp2} to have no integer solution, we set $y_2-y_3$ to be an odd integer such that the lower bound $\frac12y_2-\frac12y_3$ is fractional, and, at the same time, set the interval to be smaller than $\frac12$, \ie $0\le \frac12y_2 -\frac14 y_3 -(\frac12y_2-\frac12y_3)<\frac12$.
Considering $y$'s are integers, this condition can be realized when (A) $y_2-y_3=1 \mod 2$ and $\frac12 y_2-\frac14 y_3 - (\frac12y_2-\frac12y_3) = 0$, or (B) $y_2-y_3=1 \mod 2$ and $\frac12 y_2-\frac14 y_3 - (\frac12y_2-\frac12y_3) = \frac14$.
For \cref{eq:k-inequality2-150-tmp3} to have no integer solution, we set $y_2-y_3$ to be an odd integer such that the lower bound $\frac12y_2-\frac12y_3$ is fractional, and, at the same time, set the interval to be smaller than $\frac12$, \ie $0\le \frac12y_1 -\frac12 y_3 -(\frac12y_2-\frac12y_3)<\frac12$.
Considering $y$'s are integers, this condition can be realized when (C) $y_2-y_3=1 \mod 2$ and $\frac12y_1 -\frac12 y_3 - (\frac12y_2-\frac12y_3) = 0$.
Since all the three cases (A,B,C) are not inconsistent with \cref{eq:Y-150}, all of them can be realized by some physical band structures.
Therefore, we obtain fragile criteria for the three cases as
\beq
y_3=0\quad \text{and}\quad \delta_1 = y_2-y_3 =1 \mod 2, \label{eq:150-Z2-1}
\eeq
\beq
y_3=1\quad \text{and}\quad \delta_2 = y_2-y_3 =1 \mod 2, \label{eq:150-Z2-2}
\eeq
\beq
y_1-y_2=0\quad \text{and}\quad \delta_3 = y_2-y_3 =1 \mod 2. \label{eq:150-Z2-3}
\eeq
In this paper we refer to $\delta_{1,2,3}$ as \textit{$\mbb{Z}_2$-type fragile indices}, which are similar with the symmetry-based indicators in the sense that they also will become trivial upon stacking. 
For example, the double of the state $y=(1,1,0)$, where $y_3=0$ and $\delta_1=1$, is a trivial state, because it reads $2y=(2,2,0)$ and has trivial indices.

\subsection{The inequality method to get the fragile criteria}\label{sec:inequality-method}
The above method can be generalized to any SG. 
In this paper we refer to this method as the inequality method.
Here we present a summary of the inequality method.
First, making use of the Smith Decomposition of the EBR matrix ($EBR = L \Lambda R$), we parameterize the symmetry data as $B = \sum_{i=1}^r (L \Lambda)_i y_i$, with $r$ the rank of $\Lambda$ and $y=(y_1\cdots y_r)^T$ an integer vector, such that $B$ has vanishing indicators.
For the numbers of irreps to be nonnegative, there should be
\beq
B = \sum_{i=1}^r (L \Lambda)_i y_i \ge 0. \label{eq:B>=0}
\eeq
Second, we decompose the symmetry data vector associated with $y$ as a combination of EBRs, \ie $B = EBR \cdot p$ or $y_i = (Rp)_i$ ($i=1\cdots r$), where $p=(p_1\cdots p_{N_{EBR}})^T$ is the combination coefficient.
Clearly, the decomposition
\begin{equation}
B = \sum_{i=1}^{r} (L\Lambda)_i y_i + \sum_{i=1}^{N_{EBR}-r} (L\Lambda)_{r+i} k_i, 
\end{equation} 
where $k_i$'s are free (integer) parameters, gives the same symmetry data vector than \cref{eq:B>=0} because $\Lambda_{i} = 0$ for $i>r$. 
Thus the general solution of $B = EBR \cdot p$  can be written as $p = R^{-1}\begin{pmatrix} y \\ k\end{pmatrix}$.
As both $R$ and $R^{-1}$ are integer matrices, $p$ is integer vector iff $k$ is integer vector.
Therefore, the condition for the symmetry data to be trivial is equivalent to the existence of integer solutions of $k$ for the inequalities $p\ge 0$ subject to the constraints $B\ge 0$ (\cref{eq:B>=0}): If there exists such $k$ such that $p\ge 0$, then the symmetry data can be written as a sum of EBRs.
Now we describe the solution of $p\ge 0$.
At the first step, we consider $k_1$ as a variable and $k_2\cdots k_{N_{EBR}-1}$ and $y$ as fixed parameters. 
Then the solution of $p\ge0$ takes the form
\beq
\max\pare{f_1^{(-1)}(k_2\cdots y_r),f_1^{(-2)}(k_2\cdots y_r),\cdots}\le k_1 \le \min\pare{f_1^{(1)}(k_2\cdots y_r),f_1^{(2)}(k_2\cdots y_r),\cdots} 
 \label{eq:general-solution-k1}
\eeq
Here $f$'s are linear functions of $k$ and $y$ with rational coefficients.
At the second step, by requiring $f_1^{(i)}\le f_1^{(j)}$, where $i<0$ and $j>0$, such that $k_1$ has a nontrivial solution, we obtain a set of constraints about $k_2\cdots k_{N_{EBR}-r}$ and $y$. 
Solving these constraints by regarding $k_2$ as the variable and $k_3\cdots k_{N_{EBR}-1}$ and $y$ as fixed parameters, we can obtain the solution as
\beq
\max\pare{f_2^{(-1)}(k_3\cdots y_r),f_2^{(-2)}(k_3\cdots y_r),\cdots} \le k_2 \le  \min\pare{f_2^{(1)}(k_3 \cdots y_r),f_2^{(2)}(k_3\cdots y_r),\cdots} 
\label{eq:general-solution-k2}
\eeq
\beq \cdots \nono
\eeq
At the $(N_{EBR}-r)$-th step, solving the constraints that guarantee $k_{N_{EBR}-r-1}$ to have a nontrivial solution by considering $k_{N_{EBR}-r}$ as the variable and $y$ as fixed parameter, we obtain the solution
\beq
\max\pare{h_{0}^{(-1)}(y_1\cdots y_r),h_{0}^{(-2)}(y_1\cdots y_r),\cdots}
\le k_{N_{EBR}-r} \le \min\pare{h_{0}^{(1)}(y_1\cdots y_r),h_{0}^{(2)}(y_1\cdots y_r),\cdots} \label{eq:general-solution-kN},
\eeq
Here $h$'s are linear functions of $y$ with rational coefficients.
At the next step, we regard $y_1$ as the variable and $y_2\cdots y_r$ as fixed parameters.
By requiring $k_{N_{EBR}-r}$ to have a nontrivial solution, we obtain the constraints satisfied by $y_1$ as
\beq
\max\pare{h_1^{(-1)}(y_2\cdots y_r),h_1^{(-2)}(y_2\cdots y_r),\cdots} \le y_1 \le  \min\pare{h_1^{(1)}(y_2\cdots y_r),h_1^{(2)}(y_2\cdots y_r),\cdots}, \label{eq:general-solution-y1}
\eeq
Following the procedure, we can successfully obtain the constraints satisfied by $y_2\cdots y_r$ as
\beq
\max\pare{h_2^{(-1)}(y_3\cdots y_r),h_2^{(-2)}(y_3\cdots y_r),\cdots} \le y_2 \le  \min\pare{h_2^{(1)}(y_3\cdots y_r),h_2^{(2)}(y_3\cdots y_r),\cdots}, \label{eq:general-solution-y2}
\eeq
\beq \cdots \nono
\eeq
\beq
\max\pare{ h_{r-1}^{(-1)}(y_r), h_{r-1}^{(-2)}(y_r), \cdots} \le y_{r-1} \le \min\pare{ h_{r-1}^{(1)}(y_r), h_{r-1}^{(2)}(y_r), \cdots}. \label{eq:general-solution-yr-1}
\eeq

\cref{eq:general-solution-k1,eq:general-solution-k2,eq:general-solution-kN,eq:general-solution-y1,eq:general-solution-y2,eq:general-solution-yr-1} can be thought as an algorithm, where the $(n+1)$-th step is obtained by requiring that the $n$-th step has a nontrivial (rational) solution.
Therefore, to ensure $k_1\cdots k_{N_{EBR}}$ to have a nontrivial (rational) solution, we need and only need the constaints in \cref{eq:general-solution-y1} to (\ref{eq:general-solution-yr-1}) to be satisfied.
In other words, if $h^{(i)}_l>h^{(j)}_l$ for any $i<0$, $j>0$ ($l=0,1\cdots r$), the constraints \cref{eq:general-solution-y1} to (\ref{eq:general-solution-yr-1}) are violated, this would imply the non-existence of $k$ satisfying \cref{eq:general-solution-k1,eq:general-solution-k2,eq:general-solution-kN}.
Hence we can define the inequality-type indices as $h^{(i)}_l - h^{(j)}_l$ ($i<0$, $j>0$, $l=0,1\cdots r$), the positive values of which imply \cref{eq:general-solution-k1,eq:general-solution-k2,eq:general-solution-kN} not having a solution and hence $k_1\cdots k_{N_{EBR}-r}$ not having a solution and hence imply to a fragile topology.
In practice, we need to check whether $h^{(i)}_l - h^{(j)}_l>0$ is consistent with  the positivity of $B$ (\cref{eq:B>=0}).
If not, then there is no need to introduce such an index, as it cannot be realized by real band structure.
As discussed in the paragraph below in \cref{eq:k-inequality2-150} in the example of SG 150, $h^{(i)}_l$ and $h^{(j)}_l$ pairs set four possible inequality-type indices, but only one of them is consistent with \cref{eq:B>=0}.
By this method we can obtain all the inequality-type fragile indices in principle.
However, we emphasize that the computational time of solving $p\ge 0$ in form of \cref{eq:general-solution-k1,eq:general-solution-k2,eq:general-solution-kN,eq:general-solution-y1,eq:general-solution-yr-1} increases exponentially with the number of variables.
Therefore, it is very hard to solve SGs where $r$ is very large by the inequality method; several these groups are solved in Ref. \cite{PaperOnTheInductionMethor}.

Finding $\mbb{Z}_{n=2,3\cdots}$-type fragile indices is more complicated: one needs to check whether the solution of $k$ contains integer points. 
For simplicity, let us first check whether the $k_{N_{EBR}-r}$ component has integer solutions.
For a given $y=(y_1\cdots y_r)^T$, if there exist fractional $h^{(i)}_0$ and $h^{(j)}_0$ ($i<0, j>0$), then the conditions for $k_{N_{EBR}-r}$ to have no integer solutions are (i) $h^{(i)}_0 \in \mbb{Q}-\mbb{Z}$ (non-integer rational) and (ii) $ h^{(j)}_0 < \lceil h^{(i)}_0  \rceil $ such that $ h^{(i)}_0\le k_{N_{EBR} -r} \le  h^{(j)}_0 < \lceil h^{(i)}_0  \rceil$ has no integer solution sitting between a non-integer rational and the smallest integer larger or equal to this non-integer rational.
Here $\lceil x \rceil$ represents the smallest integer larger or equal to $x$.
Now let us write the condition (i) and (ii) more explicitly to get the $\mbb{Z}_n$-type indices.
As $h_0^{(i)}$ is a rational linear function of $y$, there exit a minimal integer $\kappa$ such that $\kappa h_0^{(i)} \in \mbb{Z}$ for arbitrary $y \in \mbb{Z}^r$. 
Therefore, for given $y$, condition-(i) is equivalent to
\beq
\delta = \kappa h^{(i)}_0(y) \neq 0 \mod \kappa. \label{eq:hi-mod}
\eeq 
When \cref{eq:hi-mod} is satisfied, $\lceil h^{(i)}_0  \rceil$ can be written as $h^{(i)}_0 + 1 - \frac1{\kappa}(\kappa h^{(i)}_0\ \mrm{mod}\ \kappa) $, and thus condition-(ii) is equivalent to 
\beq
h_0^{(j)} - h_0^{(i)} < 1- \frac{1}{\kappa}(\kappa h^{(i)}_0\ \mrm{mod}\ \kappa). \label{eq:hi-ieq}
\eeq
In the example of SG 150, picking $h_0^{(i)} $ as $\frac12 y_2 - \frac12 y_3$, $h_0^{(j)} $ as $ \frac12 y_2 -\frac14 y_3$, and $\kappa=2$, \cref{eq:hi-mod} and \cref{eq:hi-ieq} are given as $\delta = y_2 -y_3 =1 \mod 2$ and $ \frac14 y_3 < \frac12 $, respectively, which implies \cref{eq:150-Z2-1,eq:150-Z2-2}. 

So far we have derived the $\mbb{Z}_n$-type fragile indices given by the $k_{N_{EBR}-r}$ component.
Now we consider the $\mbb{Z}_n$-type fragile indices given by the $k_{N_{EBR}-r-1}$ component.
We can re-derive \cref{eq:general-solution-k1,eq:general-solution-k2,eq:general-solution-kN} in a different order of $k$'s components, where the last-solved component is $k_{N_{EBR}-r-1}$.
Then following \cref{eq:hi-mod,eq:hi-ieq} we can derive the $\mbb{Z}_n$-type fragile criteria given by $k_{N_{EBR}-r-1}$.
In \cref{sec:P3} we present an example of interchanging the order of $k_{N_{EBR}-r}$ and $k_{N_{EBR}-r-1}$.
By setting each $k_i$ as the last-solved component, we can get all the $\mbb{Z}_n$-type fragile criteria given by \textit{individual} $k$ components.
We present a case by case study of this method in Ref. \cite{PaperOnTheInductionMethor}.

%However, as shown in the end of \cref{sec:P3}, there are cases where each $k_i$ can be integer but $k_i$'s cannot be all integers at the same time due to the constraints between $k_i$'s. 
%Thus the above discussion is not complete.
%For now we do not know how to algorithmically derive the complete $\mbb{Z}_n$-type fragile indices by the inequality method in general case.
%(In a case by case study, this can be done \cite{PaperOnTheInductionMethor}).
%To solve this problem, and also to speed up calculating the inequality-type indices, we will introduce the polyhedron method in \cref{sec:polyhedron}.

\subsection{Another example: SG 143}\label{sec:P3}
Here we present the calculation of fragile indices in SG 143 ($P3$) as a nontrivial example of the the $\mbb{Z}_n$-type indices.
The Smith Decomposition of the EBR matrix is given by
\beq
EBR =
\left(\begin{array}{rrrrrrrrrrrrrrrrrr}
1 & 0 & 0 & -1 & 0 & 0 & 0 & 0 & 1 & 0 & 0 & 0 & 0 & 0 & 0 & 0 & 0 & 0 \\
0 & 0 & 0 & 1 & 0 & 0 & 0 & 0 & 0 & 1 & 0 & 0 & 0 & 0 & 0 & 0 & 0 & 0 \\
1 & 0 & 0 & -1 & 0 & 0 & 0 & 0 & 0 & 0 & 0 & 0 & 0 & 0 & 0 & 0 & 0 & 0 \\
0 & 0 & 0 & 1 & 0 & 0 & 0 & 1 & 0 & 0 & 0 & 0 & 0 & 0 & 0 & 0 & 0 & 0 \\
2 & -1 & -1 & 0 & 0 & 0 & 0 & 0 & 0 & 0 & 1 & 0 & 0 & 0 & 0 & 0 & 0 & 0 \\
0 & 0 & 1 & 0 & 0 & 0 & 0 & 0 & 0 & 0 & 0 & 1 & 0 & 0 & 0 & 0 & 0 & 0 \\
0 & 1 & 0 & 0 & 0 & 0 & 0 & 0 & 0 & 0 & 0 & 0 & 1 & 0 & 0 & 0 & 0 & 0 \\
2 & -1 & -1 & 0 & 0 & 0 & 0 & 0 & 0 & 0 & 0 & 0 & 0 & 1 & 0 & 0 & 0 & 0 \\
0 & 0 & 1 & 0 & 0 & 0 & 0 & 0 & 0 & 0 & 0 & 0 & 0 & 0 & 1 & 0 & 0 & 0 \\
0 & 1 & 0 & 0 & 0 & 0 & 0 & 0 & 0 & 0 & 0 & 0 & 0 & 0 & 0 & 1 & 0 & 0 \\
2 & -1 & -1 & 0 & 0 & 0 & 0 & 0 & 0 & 0 & 0 & 0 & 0 & 0 & 0 & 0 & 1 & 0 \\
0 & 0 & 1 & 0 & 0 & 0 & 0 & 0 & 0 & 0 & 0 & 0 & 0 & 0 & 0 & 0 & 0 & 1 \\
0 & 1 & 0 & 0 & 0 & 0 & 0 & 0 & 0 & 0 & 0 & 0 & 0 & 0 & 0 & 0 & 0 & 0 \\
2 & -1 & -1 & 0 & 0 & 0 & 0 & 0 & 0 & 0 & 0 & 0 & 0 & 0 & 0 & 0 & 0 & 0 \\
0 & 0 & 1 & 0 & 0 & 0 & 1 & 0 & 0 & 0 & 0 & 0 & 0 & 0 & 0 & 0 & 0 & 0 \\
0 & 1 & 0 & 0 & 0 & 1 & 0 & 0 & 0 & 0 & 0 & 0 & 0 & 0 & 0 & 0 & 0 & 0 \\
1 & 0 & 0 & 0 & 0 & 0 & 0 & 0 & 0 & 0 & 0 & 0 & 0 & 0 & 0 & 0 & 0 & 0 \\
1 & 0 & 0 & 0 & 1 & 0 & 0 & 0 & 0 & 0 & 0 & 0 & 0 & 0 & 0 & 0 & 0 & 0
\end{array}\right)
\left(\begin{array}{rrrrrr}
1 & 0 & 0 & 0 & 0 & 0 \\
0 & 1 & 0 & 0 & 0 & 0 \\
0 & 0 & 1 & 0 & 0 & 0 \\
0 & 0 & 0 & 1 & 0 & 0 \\
0 & 0 & 0 & 0 & 0 & 0 \\
0 & 0 & 0 & 0 & 0 & 0 \\
0 & 0 & 0 & 0 & 0 & 0 \\
0 & 0 & 0 & 0 & 0 & 0 \\
0 & 0 & 0 & 0 & 0 & 0 \\
0 & 0 & 0 & 0 & 0 & 0 \\
0 & 0 & 0 & 0 & 0 & 0 \\
0 & 0 & 0 & 0 & 0 & 0 \\
0 & 0 & 0 & 0 & 0 & 0 \\
0 & 0 & 0 & 0 & 0 & 0 \\
0 & 0 & 0 & 0 & 0 & 0 \\
0 & 0 & 0 & 0 & 0 & 0 \\
0 & 0 & 0 & 0 & 0 & 0 \\
0 & 0 & 0 & 0 & 0 & 0
\end{array}\right) 
\left(\begin{array}{rrrrrr}
1 & 1 & 1 & 1 & 1 & 1 \\
2 & 0 & 0 & 0 & 1 & 1 \\
0 & 2 & 0 & 1 & 0 & 1 \\
0 & 0 & 0 & 1 & 1 & 1 \\
-1 & -1 & 0 & 0 & 0 & 0 \\
1 & 0 & 0 & 0 & 0 & 0
\end{array}\right), \label{eq:EBR-143}
\eeq
where the order of irreps is $\ovl{\mrm{A}}_4\ovl{\mrm{A}}_4$, $\ovl{\mrm{A}}_5\ovl{\mrm{A}}_6$, $\ovl{\Gamma}_4\ovl{\Gamma}_4$, $\ovl{\Gamma}_5\ovl{\Gamma}_6$, $\ovl{\mrm{H}}_4$, $\ovl{\mrm{H}}_5$, $\ovl{\mrm{H}}_6$, $\ovl{\mrm{HA}}_4$, $\ovl{\mrm{HA}}_5$, $\ovl{\mrm{HA}}_6$, $\ovl{\mrm{K}}_4$, $\ovl{\mrm{K}}_5$, $\ovl{\mrm{K}}_6$, $\ovl{\mrm{KA}}_4$, $\ovl{\mrm{KA}}_5$, $\ovl{\mrm{KA}}_6$, $\ovl{\mrm{L}}_2\ovl{\mrm{L}}_2$, $\ovl{\mrm{M}}_2\ovl{\mrm{M}}_2$, in the BCS notation \cite{Elcoro2017,BCS,Aroyo2006a,Aroyo2006b}.
The rank of the EBR matrix is $r=4$, and the number of EBRs is $N_{EBR}=6$, thus $y$ has four components and $k$ has two components.
From \cref{eq:EBR-143}, it is direct to see that the solution of $ B = \sum_{i=1}^r (L\Lambda)_i y_i \ge 0$ is 
\beq
y_1\ge y_4\ge0,\qquad y_2\ge0,\qquad y_3\ge0,\qquad 2y_1-y_2-y_3\ge 0. \label{eq:143-ineq1}
\eeq
Relying on the discussion in \cref{sec:inequality-method}, we can write the $p$-vector as 
\beq
p = R^{-1} \begin{pmatrix}
y\\ k
\end{pmatrix} = \begin{pmatrix}
k_2 \\
-k_1-k_2 \\
y_1-y_4+k_1 \\
-y_2 + y_4 + 2k_2 \\
-y_3 + y_4 - 2k_1 - 2k_2 \\
y_2 + y_3 -y_4 + 2k_1
\end{pmatrix}.
\eeq
Now we solve the inequality by the method described in \cref{eq:general-solution-k1,eq:general-solution-k2,eq:general-solution-kN,eq:general-solution-y1,eq:general-solution-y2,eq:general-solution-yr-1}.
At the first step, we take $k_1$ as the variable, then $p\ge 0$ gives 
\begin{equation}
-k_1 -k_2\ge 0,\qquad y_1-y_4 + k_1\ge0,\qquad  -y_3 + y_4 -2k_1 -2k_2\ge0,\qquad  y_2 + y_3 - y_4 + 2k_1\ge 0. \label{eq:143-k1-tmp}
\end{equation}
(For now we temporarily omit the first and fourth components of $p$, \ie $k_2$ and $-y_2+y_4+2k_2$, where $k_1$ is not involved.)
The four constraints in \cref{eq:143-k1-tmp} should be satisfied at the same time, so we obtain
\beq
\max\pare{-\frac12 y_2 - \frac12 y_3 + \frac12 y_4, -y_1 + y_4} \le k_1 \le
\min\pare{-k_2, -k_2 - \frac12 y_3 + \frac12 y_4}, \label{eq:143-k1}
\eeq
At the second step, we regard $k_2$ as the variable and find the constraints satisfied by $k_2$ that guarantee (i) $p_1\ge 0$, $p_4\ge 0$, and (ii) $k_1$ has a nontrivial solution.
On one hand, for $p_1$ and $p_4$ to be nonnegative, we have
\begin{equation}
k_2\ge 0,\qquad  -y_2+y_4 + 2k_2 \ge 0. \label{eq:143-k2-tmp2}
\end{equation}
On the other hand, for $k_1$ to have nontrivial solutions, we should satisfy the inequalities
\begin{equation}
\begin{aligned}
-\frac12 y_2 -\frac12 y_3+ \frac12 y_4 &\le -k_2\\
-\frac12 y_2 -\frac12 y_3+ \frac12 y_4 &\le -k_2 - \frac12 y_3 + \frac12 y_4 \\
-y_1 + y_4 &\le -k_2 \\
-y_1 + y_4 &\le -k_2 - \frac12 y_3 + \frac12 y_4
\end{aligned}. \label{eq:143-k2-tmp1}
\end{equation}
Regarding $k_2$ as the variable, the constraints in \cref{eq:143-k2-tmp1,eq:143-k2-tmp2} can be equivalently written as
\beq
\max\pare{0, \frac12 y_2 - \frac12 y_4} \le k_2 \le \min\pare{\frac12 y_2, \frac12 y_2 + \frac12 y_3 - \frac12 y_4, y_1 - \frac12 y_3 - \frac12 y_4, y_1 - y_4}. \label{eq:143-k2}
\eeq
\cref{eq:143-k2} guarantees that (i) $p_{1,4}$ are nonnegative, and (ii) \cref{eq:143-k1} has a nontrivial solution, which guarantees that $p_{2,3,5,6}$ are nonnegative.
Thus, the sufficient and necessary condition for $p$ to be nonnegative is that \cref{eq:143-k2} has nontrivial solutions.
And, \cref{eq:143-k2} has nontrivial solutions if and only if the two lower bounds are smaller than the four upper bounds bounds, \ie the eight inequalities
\beqs
y_2\ge0,\qquad y_2+y_3-y_4\ge 0, \qquad 2y_1 - y_3 -y_4\ge 0,\qquad y_1-y_4\ge 0, \nono\\ 
y_4\ge0,\qquad y_3\ge 0,\qquad 2y_1-y_2-y_3\ge 0,\qquad 2y_1-y_2-y_4
\ge 0.\label{eq:143-ineq2}
\eeqs

Now we are ready to work out the fragile indices. 
First we look at the inequality-type.
We notice that the first, fourth, fifth, sixth, and seventh inequalities in \cref{eq:143-ineq2} are identical to the inequalities in \cref{eq:143-ineq1} obtained from $B\ge 0$ and hence do not bring any new index. 
But the second, third, and last inequalities are not included in \cref{eq:143-ineq1}.
Therefore we can get three inequality-type fragile criteria as the second, third and last inequalities in \cref{eq:143-ineq1} (for which $k_2$ solutions do not exist.)
\beqs
y_4 - y_2 - y_3 >& 0, \\
y_3 + y_4 -2y_1 >& 0, \\
y_2 + y_4 -2y_1 >& 0. \\
\eeqs

Now we look at the $\mbb{Z}_n$-type indices. 
First let us derive the condition for $k_2$ to have fractional solutions but no integer solution.
\cref{eq:143-k2} can be thought as the intersection of the following five equations
\begin{equation}
0 \le k_2 \le \min\pare{\frac12 y_2, \frac12 y_2 + \frac12 y_3 - \frac12 y_4, y_1 - \frac12 y_3 - \frac12 y_4, y_1 - y_4}, \label{eq:143-k2-s0}
\end{equation}
\begin{equation}
\frac12 y_2 - \frac12 y_4 \le k_2 \le \frac12 y_2, \label{eq:143-k2-s1}
\end{equation}
\begin{equation}
\frac12 y_2 - \frac12 y_4 \le k_2 \le \frac12 y_2 + \frac12 y_3 - \frac12 y_4, \label{eq:143-k2-s2}
\end{equation}
\begin{equation}
\frac12 y_2 - \frac12 y_4 \le k_2 \le y_1 - \frac12 y_3 - \frac12 y_4, \label{eq:143-k2-s3}
\end{equation}
\begin{equation}
\frac12 y_2 - \frac12 y_4 \le k_2 \le y_1 - y_4. \label{eq:143-k2-s4}
\end{equation}
If \cref{eq:143-k2-s0} has solutions, the solutions must contain the integer $0$, which would \textit{not} be a fractional solution of $k_2$ but an integer solution.
Similarly, if \cref{eq:143-k2-s4} has solutions, the solutions must contain the integer $y_1 - y_4$.
Thus neither \cref{eq:143-k2-s0} nor \cref{eq:143-k2-s4} can bring new indices, since the bounds of these equations are integers, and we are looking for non-integer solutions but no integer solutions.
Therefore we only need to consider the cases where \cref{eq:143-k2-s1} or \cref{eq:143-k2-s2} or \cref{eq:143-k2-s3} has only fractional solutions but no integer solution.
For \cref{eq:143-k2-s1} to have no integer solution, we set the lower bound as half integer and set the interval to be smaller than $\frac12$, \ie $y_2-y_4 = 1\mod2$ and $0\le\frac12y_2 - (\frac12 y_2 - \frac12 y_4) < \frac12$.
Considering $y$'s are integers, we can rewrite this condition as (A).
\begin{equation}
y_2-y_4 = 1\mod2,\qquad y_4=0. \label{eq:143-Z2-1}
\end{equation}
Similarly, for \cref{eq:143-k2-s2,eq:143-k2-s3} to have no integer solution, we get (B)
\begin{equation}
y_2-y_4 = 1\mod2,\qquad y_3=0, \label{eq:143-Z2-2}
\end{equation} 
and (C) 
\begin{equation}
y_2-y_4 = 1\mod2,\qquad 2y_1-y_2-y_3=0, \label{eq:143-Z2-3}
\end{equation}
respectively.
Then we consider the case where $k_2$ can be integer but $k_1$ cannot.
The solution \cref{eq:143-k1} can be thought as the intersection of the following three solutions
\begin{equation}
-\frac12 y_2 - \frac12 y_3 + \frac12 y_4 \le k_1 \le -k_2, \label{eq:143-k1-s1}
\end{equation}
\begin{equation}
-\frac12 y_2 - \frac12 y_3 + \frac12 y_4 \le k_1 \le -k_2 - \frac12 y_3 + \frac12 y_4, \label{eq:143-k1-s2}
\end{equation}
\begin{equation}
-y_1 + y_4 \le k_1 \le
\min\pare{-k_2, -k_2 - \frac12 y_3 + \frac12 y_4}. \label{eq:143-k1-s3}
\end{equation}
Since we are looking for non-integer solutions, \cref{eq:143-k1-s1,eq:143-k1-s3} cannot bring new indices because if they have solutions, they must have integer solutions.
(If there is an integer solution, then there is always at least one way of writing the $B$ as a sum of EBRs with nonnegative coefficients.)
For example, when \cref{eq:143-k1-s1} has solutions,  $-k_2$ must be a solution; when \cref{eq:143-k1-s3} has solutions, $-y_1+y_4$ must be a solution.
Thus we only need to consider the case where \cref{eq:143-k1-s2} has no integer solution.
We set the lower bound of \cref{eq:143-k1-s2} as half integer, \ie $-y_2-y_3+y_4 = 1\mod 2$, and set the interval to be smaller than $\frac12$, \ie $ -k_2-\frac12 y_3+\frac12 y_4 -\pare{-\frac12 y_2-\frac12 y_3 +\frac12 y_4} = -k_2 + \frac12 y_2 <\frac12$ for arbitrary integer $k_2$ allowed by \cref{eq:143-k2}.
The interval smaller than $\frac12$ condition can be equivalently written as $  \frac12 y_2 -\frac12 < \min(k_2)$.

Due to \cref{eq:143-k2}, $\min(k_2)$ can be either $0$ or $\lceil \frac{y_2-y_4}{2}\rceil$.
There are three cases: (D) if $y_2-y_4\le 0$, then $\min(k_2) = 0$, (E) if $y_2-y_4\ge 0$ and $y_2-y_4=0\mod2$, then $\min(k_2) = \frac12 (y_2-y_4)$, and (F) if $y_2-y_4\ge 0$ and $y_2-y_4=1\mod2$, then $\min(k_2) = \frac12 (y_2-y_4) + \frac12$.
(D) with the condition $\frac12 y_2 -\frac12 < \min(k_2)$ implies $y_2<1$ and $y_4\ge 0$ ($y_4\ge0$ is already contained in \cref{eq:143-ineq1}). Since $y_2\ge0$ (\cref{eq:143-ineq1}) we have $y_2=0$ and the fragile criterion
\begin{equation}
-y_2-y_3+y_4 = 1\mod 2,\qquad y_2=0. \label{eq:143-Z2-4}
\end{equation}
(E) with the condition $\frac12 y_2 -\frac12 < \min(k_2)$ implies $y_4<1$ and $y_2\ge y_4$. Since $y_4\ge 0$ (\cref{eq:143-ineq1}) have $y_4=0$ and $y_2\ge 0$ ($y_2\ge0$ is already contained in \cref{eq:143-ineq1}). Thus the fragile criterion is
\begin{equation}
-y_2-y_3+y_4 = 1\mod 2,\qquad y_2-y_4=0\mod2, \qquad y_4=0. \label{eq:143-Z2-5}
\end{equation}
(F) with the condition $\frac12 y_2 -\frac12 < \min(k_2)$ implies $y_4<2$ and $y_2\ge y_4$. Since $y_4\ge 0$ (\cref{eq:143-ineq1}), we have $y_4=0,1$.
For $y_4=0$, we have $y_2\ge 0$, and the fragile criterion 
\begin{equation}
-y_2-y_3+y_4 = 1\mod 2,\qquad y_2-y_4=1\mod2, \qquad y_4=0. \label{eq:143-Z2-6}
\end{equation}
For $y_4=1$, we have $y_2\ge 1$, and the fragile criterion
\begin{equation}
-y_2-y_3+y_4 = 1\mod 2,\qquad y_2-y_4=1\mod2, \qquad y_4=1. \label{eq:143-Z2-7}
\end{equation}
Therefore, \cref{eq:143-Z2-1,eq:143-Z2-2,eq:143-Z2-3,eq:143-Z2-4,eq:143-Z2-5,eq:143-Z2-6,eq:143-Z2-7} are all the $\mbb{Z}_2$-type criteria in SG 143.

%Here we give and example where each $k_i$ has an integer solution but the $k_i$'s cannot be all integers at the same time due to the constraints between them. 
%We take the EFP point $y=(3,2,2,1)$, which satisfies \cref{eq:143-ineq1,eq:143-Z2-7}.
%Due to \cref{eq:143-k1,eq:143-k2}, we obtain
%\begin{equation}
%\frac12\le k_2\le 1,\qquad -\frac32 \le k_1 \le -k_2-\frac12\le -1
%\end{equation}
%If we take $k_2$ to be integer, \ie $k_2=1$, then $k_1$ must be $-\frac32$; if we take $k_1$ to be integer, \ie $k_1=-1$, then $k_2$ must be $\frac12$.

\section{Fragile phases as affine monoids}\label{sec:polyhedron}
% In this subsection, we introduce the polyhedron description of the fragile phases, which provides a geometric picture of the structures of fragile phase classification. It allows us to find all the fragile indices (\cref{sec:N-type,sec:Zn-type}) and to enumerate of root states (\cref{sec:rootY}) in a systematic framework.

\begin{figure}
\begin{centering}
\includegraphics[width=0.9\linewidth]{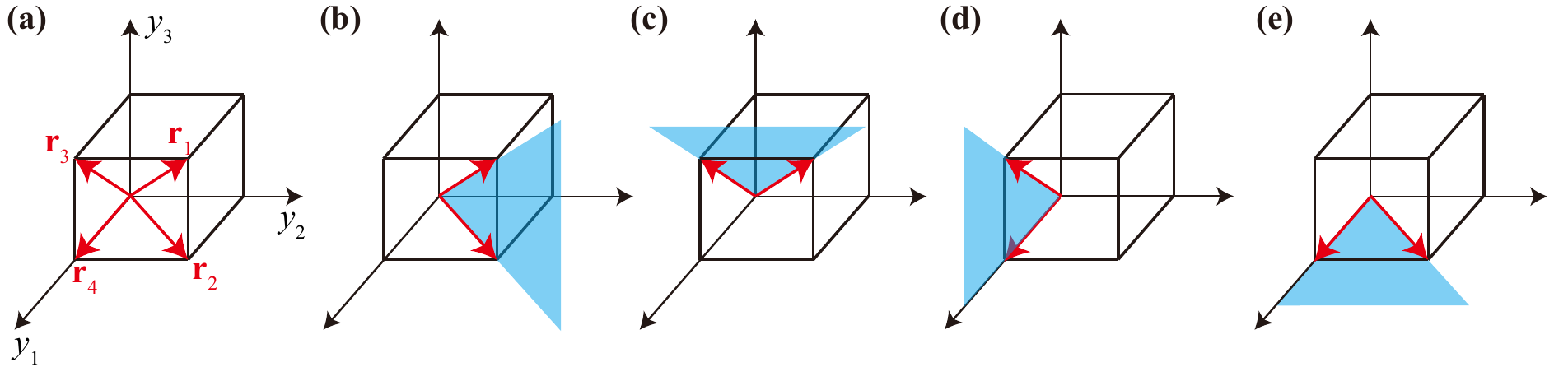}
\par\end{centering}
\protect\caption{The rays and boundary planes in of the polyhedral cone $Y$ in SG 150. (a) $\mbf{r}_1=(1,1,1)^T$,  $\mbf{r}_2=(1,1,0)^T$,  $\mbf{r}_3=(1,0,1)^T$,  $\mbf{r}_4=(1,0,0)^T$. (b) the $y_1-y_2=0$ plane. (c) the $y_1-y_3=0$ plane. (d) the $y_2=0$ plane. (e) the $y_3=0$ plane.
\label{fig:150Y}}
\end{figure}

\subsection{Examples of $Y$ and $X$}\label{sec:example-Y-X}

Here we take $Y$ of SG 150 as an example to show the two representations of the polyhedral cone.
Due to \cref{eq:Y-150}, the H-representation of $Y$ can be written as
\beq
Y = \{ y\in \mbb{R}^3\ |\ y_1\ge y_2,\ y_1\ge y_3,\ y_2\ge 0,\ y_3\ge0 \}. \label{eq:Y-150-2}
\eeq
As shown in \cref{fig:150Y}, the 2-dimensional faces of the polyhedral cone are the subsets of $Y$ where a single inequality is saturated, and the 1-dimensional faces, or the rays, of the polyhedral cone are where two of the inequalities are saturated.
To be specific, the six pairs of the four inequalities in \cref{eq:Y-150-2} set (i) $y_1=y_2=y_3$ and $y_1\ge 0$, (ii) $y_1=y_2=0$ and $0\ge y_3\ge 0$ (or $y_3=0$), (iii) $y_1=y_2$ and $y_3=0$ and $y_2\ge 0$ (iv) $y_1=y_3$ and $y_2=0$ and $y_1\ge 0$, (v) $y_1=y_3=0$ and $0\ge y_2\ge 0$, (vi) $y_2=y_3=0$ and $y_1\ge 0$; we find that (i), (iii), (iv) and (vi) are rays and (ii) and (v) are points.
Therefore, the ray matrix in the V-representation of $Y$ is given by
\beq
Ray = \begin{pmatrix}
1 & 1 & 1 & 1 \\    
1 & 1 & 0 & 0 \\
1 & 0 & 1 & 0
\end{pmatrix}. \label{eq:150-Y-Ray}
\eeq

\begin{figure}
\begin{centering}
\includegraphics[width=0.9\linewidth]{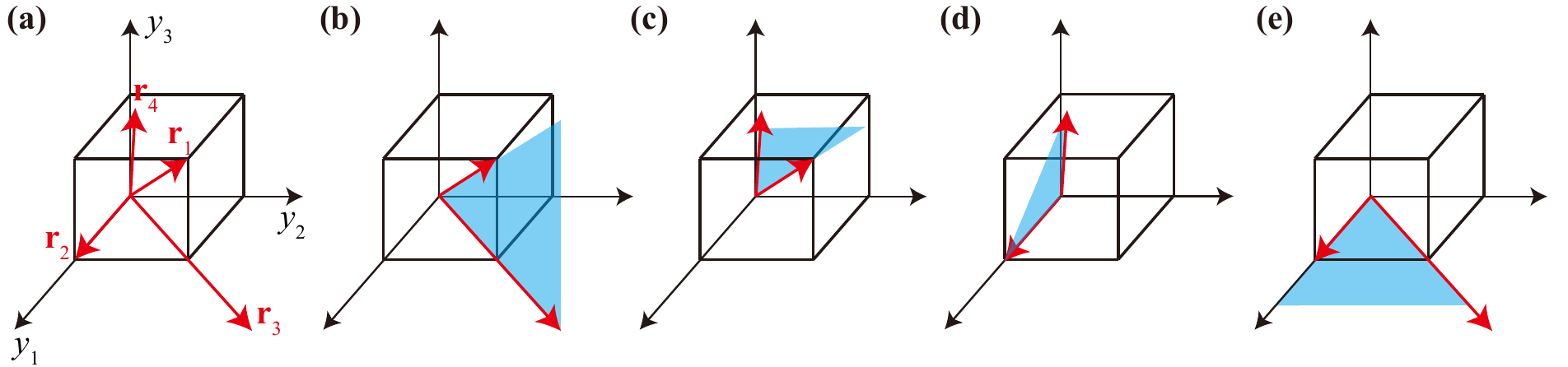}
\par\end{centering}
\protect\caption{The rays and boundary planes in of the polyhedral cone $X$ in SG 150. (a) $\mbf{r}_1=(1,1,1)^T$,  $\mbf{r}_2=(1,0,0)^T$,  $\mbf{r}_3=(2,2,0)^T$,  $\mbf{r}_4=(2,1,2)^T$. (b) the $y_1-y_2=0$ plane. (c) the $y_1-y_3=0$ plane. (d) the $2y_2-y_3=0$ plane. (e) the $y_3=0$ plane.
\label{fig:150X}}
\end{figure}

Here we take $X$ in SG 150 as another example to show the two representations of the polyhedral cone.
The rays of $X$ are given by the first $r$ rows of each column of the $R$ matrix in \cref{eq:EBR-150}, as shown in \cref{fig:150X}a.
Due to \cref{thm:cone}, there will be an H-representation of $X$. 
Let us work out the H-representation. 
As $R_{1:r,:}$ is a $3\times 4$ matrix (\cref{eq:EBR-150}), $X$ has four rays, where each pair sets a plane: (i) the first two rays set the plane $ y_2-y_3=0$, (ii) the first and the third set $ y_1-y_2=0$ (\cref{fig:150X}b), (iii) the first and the last set $ y_1-y_3=0$ (\cref{fig:150X}c), (iv) the second and the third set $y_3=0$ (\cref{fig:150X}d), (v) the second and the last set $2y_2-y_3= 0$ (\cref{fig:150X}e), and (vi) the last two set $ 2y_1-2y_2 +y_3=0$. 
It is direct to verify that (ii)-(v) are boundaries of $X$, whereas (i) and (vi) are not.
For example, all points (except origin) on the third ray and the fourth ray satisfy $y_2-y_3>0$ and $y_2-y_3<0$, respectively; because the points on the rays are on different sides of the $y_2-y_3=0$,  $y_2-y_3=0$ is not a boundary. 
On the other hand, all the rays satisfy $y_1-y_2\ge 0$, thus $y_1-y_2=0$ is a boundary.
Therefore, we obtain
\beq
X =\{y\in \mbb{R}^3\ |\  y_1\ge y_2,\ y_1\ge y_3,\ 2y_2-y_3\ge0,\ y_3\ge 0 \}. \label{eq:X150-0}
\eeq

\subsection{Hilbert bases of $\ovl{Y}$ and EFP roots}\label{sec:rootY}

An affine monoid $M$ is called \textit{positive} if $\forall a,b \in M -\{0\} \Rightarrow a+b\neq 0$.
\cref{thm:Gordan} in \cref{sec:math} tells us that the intersection of a pointed polyhedral cone and the integer lattice is a positive monoid.
Therefore, $\ovl{Y}$ is indeed a positive affine monoid.
Since $\ovl{X}$ is a subset of $\ovl{Y}$, $\ovl{X}$ is also a positive affine monoid.

Due to \cref{thm:Hilbert}, any positive affine monoid has a unique minimal set of generators, called the Hilbert bases.
All the elements in the monoid can be written as sum of the Hilbert bases with positive coefficients.
It should be noticed that any of the Hilbert basis cannot be written as a sum of other nonzero elements in the positive affine monoid with positive coefficients.
As shown in the following examples, in some cases the vectors of the Hilbert bases are linearly dependent on each other, but writing anyone of them as a linear combination of others will involve negative coefficients.
%In \cref{sec:polyhedron-method} we showed that, for a given SG, any band with vanishing stable/strong indicator, fragile or not, is represented by a point in the positive affine monoid $\ovl{Y}$.
%And, as shown in \cref{sec:Hb}, any element in $\ovl{Y}$ can be written as a sum of Hilbert bases of $\ovl{Y}$ with positive coefficients.
Here we divide the Hilbert bases into two parts: the fragile phase bases and the trivial bases.
The trivial bases actually correspond to EBRs because they are trivial (BR) and cannot be written as a sum of other elements with positive coefficients (elementary basis).
We call the fragile phase bases as \textit{EFP roots}.
From the aspect of symmetry data, the fragile roots are the ``representative" phases of EFPs, as any EFP can be obtained by either stacking the roots or stacking the roots with EBRs (trivial bands).

\textbf{Example.}
We take $Y$ in SG 150 as an example to derive the Hilbert bases.
The $Y$ polyhedron is given in \cref{eq:Y-150-2}.
To derive the Hilbert bases, we further divide the points in $\ovl{Y}$ into two cases: (i) $0 \le y_3 \le y_2 \le y_1$, (ii) $0 \le y_2 \le y_3 \le y_1$.
For case-(i), we can rewrite the $y$-vector as
\beq
y=\begin{pmatrix} y_1 \\ y_2\\ y_3 \end{pmatrix} = 
y_3\begin{pmatrix} 1\\1\\1 \end{pmatrix} +
(y_2-y_3)\begin{pmatrix} 1\\1\\0 \end{pmatrix} +
(y_1-y_2) \begin{pmatrix} 1\\0\\0 \end{pmatrix}. \label{eq:150-Ybasis-tmp1}
\eeq
For case-(ii), we can rewrite the $y$-vector as
\beq
y=\begin{pmatrix} y_1 \\ y_2\\ y_3 \end{pmatrix} = 
y_2\begin{pmatrix} 1\\1\\1 \end{pmatrix} +
(y_3-y_2)\begin{pmatrix} 1\\0\\1 \end{pmatrix} +
(y_1-y_3) \begin{pmatrix} 1\\0\\0 \end{pmatrix}. \label{eq:150-Ybasis-tmp2}
\eeq
Therefore there are four Hilbert bases
\begin{equation}
b_1 =\begin{pmatrix} 1 \\ 1 \\ 1 \end{pmatrix},\qquad
b_2 =\begin{pmatrix} 1 \\ 0 \\ 0 \end{pmatrix},\qquad
b_3 =\begin{pmatrix} 1 \\ 1 \\ 0 \end{pmatrix},\qquad
b_4 =\begin{pmatrix} 1 \\ 0 \\ 1 \end{pmatrix}. \label{eq:150-Y-Hb}
\end{equation}
$b_1$, $b_2$, $b_3$, $b_4$ are also the rays of the polyhedral cone $Y$ (\cref{eq:150-Y-Ray}).
The four bases are linearly dependent, but they are not redundant for the monoid as none of them can be written as a sum of the other three with positive coefficients.
To be specific, $b_1 = b_3 + b_4 - b_2$, $b_2 = b_3+b_4-b_1$, $b_3 = b_1+b_2-b_4$, $b_4=b_1+b_2-b_3$.
Applying the fragile criteria of SG 150, \ie \cref{eq:150-inequality,eq:150-Z2-1,eq:150-Z2-2,eq:150-Z2-3} to be four bases, we find that $b_1$ and $b_2$ are trivial, $b_3$ satisfies the $\mbb{Z}_2$-type criterion \cref{eq:150-Z2-1}, and $b_4$ satisfies the inequality-type criterion \cref{eq:150-inequality}.
In fact, $b_1$ and $b_3$ are the first and second columns of the right transformation matrix $R$ (the first three rows) in the Smith Decomposition of the EBR matrix (\cref{eq:EBR-150}), respectively, and thus present two EBRs of SG 150.
Therefore SG 150 has only two EFP roots: $b_3$ and $b_4$.

There are two commonly used algorithms to calculate the Hilbert bases of positive affine monoid, \ie the Normaliz algorithm \cite{Bruns2010} and the Hemmecke algorithm \cite{Hemmecke2002}, which are available in the \href{https://www.normaliz.uni-osnabrueck.de/}{\it Normaliz} package and the \href{https://4ti2.github.io}{\it 4ti2} package, respectively.
In this work, we mainly use the 4ti2 package to solve the Hilbert bases.
Applying this algorithm for each $\ovl{Y}$, we are able to calculate all the fragile roots in all SGs, as tabulated in Table S3 of \cite{SM}.
In Table 1 in the main text, we summarized the numbers of fragile roots in all the SGs.

\subsection{Hilbert bases of $\mathbb{Z}^r\cap X$}\label{sec:rootX}

As introduced in \cref{sec:polyhedron-method}, $\ovl{X} = \{y\in \mbb{Z}^r \ |\ y_i = (Rp)_i,\ p\in \mbb{N}^{N_{EBR}} \}$ represent all the trivial points in $Y$.
Thus the fragile phases are represented by points in $\ovl{Y}-\ovl{X}$.
For convenience we introduced the auxillary polyhedral cone ${X} = \{y\in \mbb{R}^r \ |\ y_i = (Rp)_i,\ p\in \mbb{R}^{N_{EBR}} \}$ and divide the points in $\ovl{Y}-\ovl{X}$ into $\ovl{Y} - \mbb{Z}^r\cap X$ and $\mbb{Z}^r\cap X - \ovl{X}$.
Here we discuss a special issue of the Hilbert bases of $\mbb{Z}^r\cap X$, which will be used in deriving the fragile indices in \cref{sec:Zn-type}.
A basis $b_l \in \mrm{Hil}(\mbb{Z}^r \cap X)$ is either trivial ($\in \ovl{X}$) or nontrivial ($\not \in \ovl{X}$), depending whether it can be written as a sum of columns of $R_{1:r,:}$, \ie $\exists q_l \in \mbb{N}^{N_{EBR}}\;s.t.\; b_l = R_{1:r,:}q_l$.
Now we define the order of $b_l$ as the smallest positive integer $\kappa_l$ that makes $\kappa_l b_l \in \ovl{X}$.
We first consider the solutions of the equation $b_l = R_{1:r,:}q_l$, where $q_l$ is vector with $N_{EBR}$ components.
The general solution of  $b_l = R_{1:r,:}q_l$ is given as $q_l = \sum_{i}^r (b_l)_i R^{-1}_i + \sum_{j=1}^{N_{EBR}-r}  k_j R^{-1}_{j+r} $, where $R_i^{-1}$ is th $i$-th column of $R^{-1}$, $r$ is the rank of the EBR matrix, and $k$'s are free parameters.
For convenience, we introduce the auxillary polyhedron
\begin{equation}
K_l = \brace{ k\in \mbb{Q}^{N_{EBR}}\; \bigg|\; \sum_{i}^r (b_l)_i R^{-1}_i + \sum_{j=1}^{N_{EBR}-r}  k_j R^{-1}_{j+r} \ge 0 }. \label{eq:Kl}
\end{equation}
Notice that $R$ is a unimodular matrix, thus $ q_l \in \mbb{N}^{r} \Leftrightarrow k \in \mbb{Z}^{N_{EBR}-r}$.
If $K_l$ contains integer points, we can take an integer point in it, $k$, such that the corresponding $q_l\in \mbb{N}^{r} $ and hence $b_l = R_{1:r,:}q_l$ is a combination with nonnegative coefficients of columns in $R_{1:r,:}$.
If $K_l$ contains only fractional points but no integer point, the corresponding $q_l$'s are nonnegative but fractional, and hence $b_l$ can be only be written as a combination with nonnegative fractional coefficients of columns in $R_{1:r,:}$.
For this second case we can introduce a (minimal) positive integer $\kappa_l$ such that $\kappa_l K_l$, \ie
\begin{equation}
\kappa_l K_l = \brace{ \kappa_l k | k\in K_l },
\end{equation}
contains at least one integer point.
Such a $\kappa_l$ always exists: suppose $k$ is a fractional vector in $K_l$, then we can take $\kappa_l$ as the least common multiplier of the denominators of the components of $k$ such that $\kappa_l k$ is an integer vector.
We always choose $\kappa_l$ as the minimal integer that makes $\kappa_l K_l$ contains at least one integer point.
$\kappa_l$ can be thought as the ``order'' of a nontrivial Hilbert basis because $\kappa_l b_l $ can be written as a combination with nonnegative integer coefficients of columns in $R_{1:r,:}$ and hence belongs to $\ovl{X}$.

\textbf{Example.}
Now we derive the Hilbert bases of $\mbb{Z}^3 \cap X$ in SG 150.
$X$ is shown in \cref{fig:150X}, and its H-representation is derived in \cref{eq:X150-0}.
To derive the Hilbert bases, we further divide the points in $\mbb{Z}^3\cap X$ into two cases: (i) $y_2-y_3\ge 0$, (ii) $y_2-y_3\le 0$.
For case-(i), we can rewrite the $y$-vector as
\begin{equation}
y=\begin{pmatrix} y_1 \\ y_2\\ y_3 \end{pmatrix} = 
y_3\begin{pmatrix} 1\\1\\1 \end{pmatrix} +
(y_2-y_3)\begin{pmatrix} 1\\1\\0 \end{pmatrix} +
(y_1-y_2) \begin{pmatrix} 1\\0\\0 \end{pmatrix}.
\end{equation}
The three bases in case-(i) are same with the bases in case-(i) of $\ovl{Y}$ (\cref{eq:150-Ybasis-tmp1}).
For case-(ii), where $y_2-y_3\le0$, we can rewrite the $y$-vector as
\begin{equation}
y=\begin{pmatrix} y_1 \\ y_2\\ y_3 \end{pmatrix} = 
(y_1-y_3)\begin{pmatrix} 1\\1\\1 \end{pmatrix} +
(2y_2-y_3)\begin{pmatrix} 1\\0\\0 \end{pmatrix} +
(y_3-y_2) \begin{pmatrix} 2\\1\\2 \end{pmatrix}.
\end{equation}
The three bases in case-(ii), where $y_2-y_3\le0$, are different with the bases in case-(ii) of $\ovl{Y}$ (\cref{eq:150-Ybasis-tmp2}) because here we cannot decompose $y$ into $(1,0,1)^T$ since $(1,0,1)^T \not \in \ovl{X}$.
Therefore there are four Hilbert bases
\begin{equation}
b_1 =\begin{pmatrix} 1\\1\\1\end{pmatrix},\qquad
b_2 =\begin{pmatrix} 1\\0\\0\end{pmatrix},\qquad
b_3 =\begin{pmatrix} 1\\1\\0\end{pmatrix},\qquad
b_4 =\begin{pmatrix} 2\\1\\2\end{pmatrix}. \label{eq:150-X-Hb}
\end{equation}
$b_1$, $b_2$, $b_4$ are the (first three rows of) columns of $R$ (\cref{eq:EBR-150}) and hence are trivial.
$b_3$, on the other hand, is half of the (first three rows of) third column in $R$.
Thus we obtain $\kappa_1=\kappa_2=\kappa_4=1$, and $\kappa_3=2$.
To check, we calculate $\kappa_3$ using the algorithm described in the last paragraph.
The inverse of the $R$ matrix (\cref{eq:EBR-150}) is 
\begin{equation}
R^{-1}=\pare{\begin{array}{rrrr}
0 & 2 &-1 & 4\\
1 & 0 &-1 & 2\\
0 & 0 & 0 &-1\\
0 &-1 & 1 &-2
\end{array} },
\end{equation}
and hence due to \cref{eq:Kl} we obtain
\begin{equation}
K_3 = \brace{k\in \mbb{Q}\Big| 2 + 4k\ge 0,\; 1+2k\ge 0,\; -k\ge 0,\; -1-2k\ge 0}=\{ -\frac12 \}.
\end{equation}
Therefore, $\kappa_3=2$ is the minimal integer that makes $\kappa_3K_3$ to have an integer point.

\section{Fragile indices} \label{sec:index}

\subsection{Removing un-allowed inequality-type indices}\label{sec:removing-inequality}
In \cref{sec:N-type} we have introduced the general method to derive inequality-type criteria in form of $ay<0$.
Here we describe how to judge whether $ay<0$ is allowed.
For a given row $a$ in $A$, we define $Y^\pr =\{ y\in \mbb{R}^r \ |\ L\Lambda_{:,1:r}y\ge 0\ \text{and}\ ay< 0 \}$.
Notice that $Y^\pr$ is an open set (due to the condition $ay<0$). 
(A set is open if it does not contain any of its boundary points.)
Clearly, $Y^\pr= \emptyset$ implies that $ay<0$ is not allowed in $Y$.
However, we do not use $Y^\pr$ in practical calculation because it is complicated to store and process an open set in our group calculations; instead, we make use of the closed extention of $Y^\pr$, \ie $Y^\prpr =\brace{ y\in \mbb{R}^r \ |\ \begin{pmatrix} L\Lambda_{:,1:r} \\ -a \end{pmatrix} y\ge 0}$.
Since $Y^\prpr$ is a superset of $Y^\pr$, obviously, $Y^\prpr=\emptyset \Rightarrow Y^\pr=\emptyset$. Thus $Y^\prpr=\emptyset$ implies $ay<0$ is forbidden by the $B\ge 0$ condition. 
Now we show how to detect the case $Y^\pr=\emptyset$ but $Y^\prpr\neq \emptyset$.
We notice that $Y^\pr = \emptyset$ implies that $ Y^\prpr = Y^\prpr - Y^\pr = \{ y\in \mbb{R}^r \ |\ L\Lambda_{:,1:r}y\ge 0\ \text{and}\ ay= 0 \}\neq \emptyset$.
The presence of equation $ay=0$ will reduce the dimension of the polyhedron. 
Thus this case can be diagnosed by $\mrm{dim}(Y^\prpr)<r$.

\textbf{Example.}
In the paragraphs above we have described a general algorithm to derive the inequality-type fragile criteria.
As an example, here we re-derive the inequality-type fragile criteria of SG 150 using the polyhedron method.
The polyhedral cone $Y$ is given by \cref{eq:Y-150-2}, and the polyhedral cone $X$ is given by \cref{eq:X150-0}.
We rewrite \cref{eq:X150-0} in terms of the $A$ matrix as
\begin{equation}
X = \{ y\in \mbb{R}^3\ |\ Ay\ge 0\},\qquad
A=\pare{\begin{array}{rrr}
1 &-1 & 0\\
1 & 0 &-1\\
0 & 2 &-1\\
0 & 0 & 1
\end{array}},\label{eq:X150}
\end{equation}
The four rows in $A$ correspond to four possible inequality-type fragile indices, \ie, $y_2-y_1$, $y_3-y_1$, $y_3-2y_2$, and $ -y_3$, the positive values of which imply fragile phases.
However, the first, second, and last inequalities are not allowed in $Y$.
For the first index $y_2-y_1$, the auxillary polyhedral cone $Y^\prpr=\{y_1\ge y_2,\ y_1\ge y_3,\ y_2\ge0,\ y_3\ge0,\ y_2-y_1\ge 0\} = \{y_1=y_2,\ y_1\ge y_3,\ y_2\ge0,\ y_3\ge0\}$ has a dimension 2 ($<3$), implying $y_2-y_1>0$ is not allowed in $Y$.
For the second index $y_3-y_1$, the auxillary polyhedral cone $Y^\prpr=\{y_1\ge y_2,\ y_1 = y_3,\ y_2\ge0,\ y_3\ge0\}$ has a dimension 2 ($<3$), implying $y_3-y_1>0$ is not allowed in $Y$.
For the last index $-y_3$, the auxillary polyhedral cone $Y^\prpr=\{y_1\ge y_2,\ y_1 \ge y_3,\ y_2\ge0,\ y_3=0\}$ has a dimension 2 ($<3$), implying $-y_3>0$ is not allowed in $Y$, so the corresponding fragile index is not necessary.
Therefore, the only  inequality-type index is $y_3-2y_2$, consistent with the result (\cref{eq:150-inequality}) in \cref{sec:P321}.

\subsection{$\mbb{Z}_2$-type fragile indices} \label{sec:Zn-type}
In this subsection we consider the type-II fragile phases, \ie symmetry data vectors represented by points in $\mbb{Z}^r\cap X-\ovl{X}$, and derive the corresponding $\mbb{Z}_n$-type fragile criteria.
It turns out that all the  $\mbb{Z}_n$-type criteria are of $\mbb{Z}_2$-type.

\subsubsection{$\mbb{Z}^r\cap X-\ovl{X}$ is close to the boundary of $X$} \label{sec:ZX-Xbar}
A key property allowing us to derive the general $\mbb{Z}_n$-indices is that the points in $\mbb{Z}^r\cap X-\ovl{X}$ are all close to the boundaries of $X$, as will be explained more clearly below.
Here we present a heuristic description of this conclusion and leave the proof for the following paragraphs.
As we will prove, there always exists a finite integer vector $\Delta y \in \ovl{X}$ such that for $\forall y \in \mbb{Z}^r\cap X-\ovl{X}$, $y + \Delta y \in \ovl{X}$.
In other words, the shifted monoid $\Delta y + \mbb{Z}^r \cap X = \{y+\Delta y | y\in \mbb{Z}^r \cap X\}$ is a subset of $\ovl{X}$ and hence 
\begin{align}
& \Delta y + \mbb{Z}^r \cap X \subset \ovl{X} \nono \\
\Rightarrow \quad & \mbb{Z}^r\cap X-\ovl{X}\quad \subset\quad \mbb{Z}^r\cap X - (\Delta y + \mbb{Z}^r \cap X) = \mbb{Z}^r \cap (X - (\Delta y + X)), \label{eq:ZX-Xbar_in_X-shiftX}
\end{align}
\ie $X - (\Delta y + X)$ is superset of $\mbb{Z}^r\cap X-\ovl{X}$.
We show that $X - (\Delta y + X)$ is close to the boundary of $X$.
We assume that $X$ has the H-representation $X=\{y\in \mbb{R}^r\ | Ay\ge 0\}$, then if $x\in \Delta y + X$, there is $x-\Delta y \in X$ and hence $A(x-\Delta y)\ge 0$. 
Thus we obtain the H-representation of $\Delta y + X$
\begin{equation}
\Delta y + X = \{y\in \mbb{R}^r\ |\ A(y-\Delta y)\ge 0\}.
\end{equation}
and hence obtain $X-(\Delta y +X)$ as 
\begin{equation}
X-(\Delta y +X)= \{y\in \mbb{R}^r\ |\ Ay\ge 0,\text{and}\ \exists i,\  s.t.\ (Ay)_i < (A \Delta y)_i\}. \label{eq:X-shiftX}
\end{equation}
Since the $i$-th boundary of $X$ is given by $(Ay)_i=0$, for a given point $y$, $(A y)_i$ can be thought as the distance from $y$ to the $i$-th boundary of $X$. 
\cref{eq:X-shiftX} means that for $\forall y \in X-(\Delta y +X)$, there always exists some boundary of $X$ such that the distance from $y$ to this boundary is smaller than the distance from $\Delta y$ (the existence of which we will prove) to this boundary. 
That means the points $X-(\Delta y +X)$ is close to the boundary of $X$.

Before going to prove the existence of $\Delta y$, here we first study the property of points in $\mbb{Z}^r \cap X$.
$\mbb{Z}^r \cap X$ is generated from the so-called Hilbert bases, denoted as $ \mrm{Hil}(\mbb{Z}^r \cap X)$ (\cref{thm:Hilbert} in \cref{sec:math}).
To be specific, we can rewrite $\mbb{Z}^r \cap X$ as
\beq
\mbb{Z}^r \cap X = \{ y = b_1 p_1 +b_2 p_2 + \cdots + b_{N_H} p_{N_{H}}\ |\ b_{1},b_2\cdots b_{N_H} \in \mrm{Hil}(\mbb{Z}^r \cap X),\ p_1,p_2\cdots p_{N_H} \in \mbb{N}\}, \label{eq:Z-cap-X}
\eeq 
where $N_H$ is the number of Hilbert bases.
As shown in \cref{sec:rootX}, for each basis $b_l$, there is a positive integer $\kappa_l$ - the order of $b_l$ - such that $\kappa_l b_l \in \ovl{X}$ (a trivial point).
With the concept of order $\kappa_l$ of Hilbert basis, we can decompose a general point in $\mbb{Z}^r\cap X$ in \cref{eq:Z-cap-X} into two parts
\begin{equation}
y = \sum_{l} (p_l\ \mrm{mod}\ \kappa_l) b_l + \sum_l \lfloor p_l/\kappa_l \rfloor \kappa_l b_l, \label{eq:y-2parts}
\end{equation}
where $\lfloor a \rfloor$ is the largest integer equal or smaller to $a$.
The second part in this decomposition belongs to $\ovl{X}$ by construction, since $\kappa_l b_l \in \ovl{X}$ and $\lfloor p_l/\kappa_l \rfloor \in \mbb{N}$.
Therefore, to shift $y$ to $\ovl{X}$, we only need to shift the first part to $\ovl{X}$.

%It should be noticed, as the decomposition of $y$ on the Hilbert bases is not unique, the decomposition into two parts in \cref{eq:y-2parts} is also not unique.

Now we prove the existence of $\Delta y$.
As shown in \cref{eq:y-2parts}, the nontrivial part of any point in $ \mbb{Z}^r\cap X - \ovl{X}$ is of the form $\sum_{l} (p_l\ \mrm{mod}\ \kappa_l) b_l$, thus to shift it to $\ovl{X}$, we only $\Delta y$ to satisfy
\begin{equation}
\forall p \in \mbb{N}^{N_{H}},\qquad \Delta y + \sum_{l} (p_l\mod \kappa_l) b_l \in \ovl{X}. \label{eq:y0+plbl_condition}
\end{equation}
As $b_l$ represent either a trivial state (EBR) or an EFP, both of which can be written as an integer combination of EBRs, we can write $b_l$ as $b_l = R_{1:r,:}q_l$ for some $q_l \in \mbb{Z}^{N_{H}}$.
If $b_l \in \ovl{X}$, $q_l$ can be a nonnegative vector; whereas if $b_l \not\in \ovl{X}$, at least one component of $q_l$ is negative.
The choice of $q_l$ is not unique. 
In general $q_l$ can be written as $ \sum_{i=1}^r R^{-1}_i (b_l)_i + \sum_{j=1}^{N_{EBR}-r} R^{-1}_j k_j $, where $R^{-1}_i$ is the $i$-th column of the $R^{-1}$ matrix, and $k$'s are free parameters.
For now, for each $b_l$ we just pick a specific $q_l$.
We decompose $q_l$ into two parts: the nonnegative part $q_l^{+}$ and the negative part $q_l^{-}$, \ie 
\begin{equation}
(q^{+}_l)_i = \begin{cases}
(q_l)_i,\qquad & \text{if } (q_l)_i\ge 0\\
0,\qquad & \text{if } (q_l)_i<0
\end{cases},\qquad
(q^{-}_l)_i = \begin{cases}
0,\qquad & \text{if } (q_l)_i\ge 0 \\
(q_l)_i,\qquad & \text{if } (q_l)_i< 0
\end{cases}. \label{eq:qpr+-}
\end{equation}
Then we have
\begin{equation}
\Delta y + \sum_{l} (p_l\mod \kappa_l) b_l = \Delta y + \sum_l(p_l\mod \kappa_l) R_{1:r,:} q^{+}_l + \sum_l(p_l\mod \kappa_l) R_{1:r,:} q^{-}_l. \label{eq:y0+plbl}
\end{equation}
Notice that $p_l$ is a number, $q_l$, $q^+_l$, and $q^{-}_l$ are vectors.
The second term in the right hand side of \cref{eq:y0+plbl} is already in $\ovl{X}$ as it is a nonnegative integer combination of columns of $R_{1:r,:}$.
Hence $\Delta y$ only need to shift the third term in \cref{eq:y0+plbl} to $\ovl{X}$.
We can choose $\Delta y$ as
\begin{equation}
\Delta y = - \sum_l (\kappa_l-1) R_{1:r,:} q^{-}_l \label{eq:y0}
\end{equation}
such that
\begin{equation}
\Delta y + \sum_l (p_l\mod \kappa_l) R_{1:r,:}q_l^- = \sum_l ((p_l\mod \kappa_l) -\kappa_l+1) R_{1:r,:} q^-_l
\end{equation}
is always a nonnegative integer combination of columns of $R_{1:r,:}$, because $(p_l\mod \kappa_l) -\kappa_l+1\le 0$ and hence $((p_l\mod \kappa_l) -\kappa_l+1) q^{-}_l \ge 0$.
Therefore $\Delta y$ defined \cref{eq:y0} satisfies the condition in \cref{eq:y0+plbl_condition}.

\begin{figure}
\begin{centering}
\includegraphics[width=0.2\linewidth]{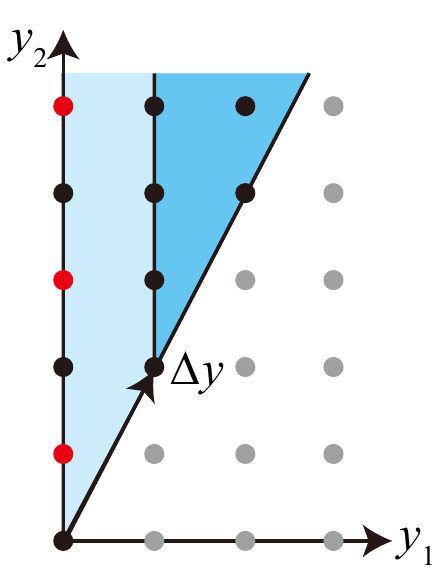}
\par\end{centering}
\protect\caption{The shift vector $\Delta y$ in SG 199. Points in $\ovl{X}$ is represented by the black dots, the polyhedral cone $X$ is shaded by light blue, the shifted polyhedral cone $\Delta y + X$ is shaded by blue, points in $\mbb{Z}^2\cap X - \ovl{X}$ are represented by red dots. All the points in $\mbb{Z}^2\cap X - \ovl{X}$ are in $X - (\Delta y+X)$ and are close to the boundary of $X$.
\label{fig:199shift}}
\end{figure}

\textbf{Example.} 
Here we take SG 199 as an example to show how to determine $\Delta y$.
As discussed in the main text (Eq. (8)) and shown in \cref{fig:199shift}, $\ovl{X}$ is given as
\begin{equation}
\ovl{X} = \{ p_1 (0,2)^T + p_2 (1,2)^T + p_3 (1,3)^T\ |\ p_{1,2,3}\in \mbb{N}\} = \{ R_{1:r,:} p\ |\ p\in \mbb{N}^3\}, \label{eq:Xbar199}
\end{equation}
where $R_{1:r,:}$ is given by (Eq. (4) in the main text)
\begin{equation}
R_{1:r,:} = \begin{pmatrix}
0 & 1 & 1\\
2 & 2 & 3
\end{pmatrix}. \label{eq:R1r-199}
\end{equation}
On the other hand, the polyhedron cone $X$, which is identical to $Y$ (Eq. (6) in the main text), is given as
\begin{equation}
X = \{ y\in \mbb{R}^2\ |\ y_2\ge 2y_1\ge 0\}.
\end{equation}
The inequality $y_1\ge 0$ can be rewritten as $(1,0) y\ge 0$, and the inequality $y_2\ge 2y_1$ can be rewritten as $(-2, 1)y\ge 0$. 
Thus $X$ can be rewritten as
\begin{equation}
X = \{ y\in\mbb{R}^2\ |\ Ay\ge 0\}, \qquad A=\begin{pmatrix} 1 & 0 \\ -2 & 1 \end{pmatrix}. \label{eq:A-199}
\end{equation}
For any point $y$ in $\mbb{Z}^2\cap X$, we can decompose it as $y=y_1(1,2)^T + (y_2-2y_1)(0,1)^T$. 
Thus the Hilbert bases of $\mbb{Z}^2\cap X$ are $b_1=(0,1)^T$ and $b_2=(1,2)^T$ and the monoid $\mbb{Z}^r \cap X$ is
\begin{equation}
\mbb{Z}\cap X = \{ p_1(0,1)^T + p_2 (1,2)^T\ |\ p_{1,2}\in \mbb{N}\}, \label{eq:ZX199}
\end{equation}
On the other hand, $\ovl{X}$ can be obtained by adding the (first two rows of) columns of $R$ in \cref{eq:R1r-199}, \ie $\ovl{X}=\{p_1(0,2)^T + p_2 (1,2)^T + p_3 (1,3)^T\ |\ p_{1,2,3}\in \mbb{N}\}$ (\cref{fig:199shift}).
As shown in \cref{fig:199shift} and proved in the main text (around Eq. (9)), the set $\mbb{Z}^r\cap X-\ovl{X}$ is given as
\begin{equation}
\mbb{Z}^r\cap X-\ovl{X} = \{(0,2p+1)^T\ |\ p\in\mbb{N}\} = \{(0,1)^T,\ (0,3)^T,\ (0,5)^T\cdots \} . \label{eq:ZX-Xbar_199}
\end{equation}
One can immediately observe that the vector $\Delta y =(1,2)^T$ shift all the points in $\mbb{Z}^r\cap X-\ovl{X}$ to $\ovl{X}$, \eg $(0,1)\to (1,3)^T$, $(0,3)\to (1,5)^T$, $(0,5)\to (1,7)^T$, \etc

Now let us pretend that we do not know $\Delta y$ and use the algorithm described in last paragraph to determine $\Delta y$.
Twice of $b_1$ belongs to $\ovl{X}$ (\cref{eq:Xbar199}) and hence $\kappa_1 = 2$; $b_2$ is already in $\ovl{X}$ and hence $\kappa_2=1$.
To obtain the $q_l^+$ and $q_l^-$ (\cref{eq:qpr+-}), which will be used to determine $\Delta y$, we write $b_1$ $b_2$ in terms of columns of $R_{1:r,:}$ with integer coefficients as
\begin{equation}
b_1 = (1,3)^T - (1,2)^T =  R_{1:r,1} (0,-1,1)^T,\qquad
b_2 = (1,2)^T  =  R_{1:r,1} (0,1,0)^T,
\end{equation}
\ie $q_1 = (0,-1,1)^T$ and $q_2=(0,1,0)^T$.
Due to \cref{eq:qpr+-}, $q_1^{-} = (0,-1,0)^T$, $q_2^- = 0$.
Then according to \cref{eq:y0+plbl}, we obtain
\begin{equation}
\Delta y = - (\kappa_1-1) R_{1:r,:} q_1^{-}= (1,2)^T
\end{equation}
which is identical with direct observation.
In the end let us verify \cref{eq:X-shiftX}. 
%Since the points in $\mbb{Z}^r\cap X-\ovl{X}$ are all on the first boundary $y_1=0$ and the distance from $\Delta y$ to the first boundary is 1 ($(A\Delta y)_1=1$), the condition $\exists i,\  s.t.\ (Ay)_i < (A \Delta y)_i$ in \cref{eq:X-shiftX} is fulfilled by taking $i=1$.
Since the distances from $\Delta y$ to the first and second boundaries are $(A\Delta y)_1 = 1$ and $(A\Delta y)_2=0$, respectively.
Thus \cref{eq:X-shiftX} can be written as $X-(\Delta y +X) = \{y\in\mbb{R}^4\ |\ Ay\ge0,\ (Ay)_1< 1\} = \{y\in\mbb{R}^4\ |\ 0\le y_1<1,\ -2y_1+y_2\ge 0\}$, which is consistent with \cref{fig:199shift}.

\textbf{Example.}
We take SG 150 as a nontrivial example to show how to determine $\Delta y$.
The $R_{1:r,:}$ matrix can be directly read from \cref{eq:EBR-150}.
Thus we can write $\ovl{X}$ as
\begin{equation}
\ovl{X} = \{ R_{1:r,:} p\ |\ p\in \mbb{N}^4\},\qquad
R_{1:r,:} = \begin{pmatrix}
1 & 1 & 2 & 2\\
1 & 0 & 2 & 1\\
1 & 0 & 0 & 2
\end{pmatrix}. \label{eq:Xbar150} 
\end{equation}
On the other hand, from the example analyses in \cref{sec:polyhedron-method,sec:N-type,sec:rootX}, we obtain the H-representation of $X$ as \cref{eq:X150-0,eq:X150}, and the four Hilbert bases of $\mbb{Z}^3\cap X$ as $b_1=(1,1,1)^T$, $b_2=(1,0,0)$, $b_3=(1,1,0)^T$, and $b_4=(2,1,2)^T$ (\cref{eq:150-X-Hb}), respectively.
$b_1$, $b_2$, $b_4$ are the first, second, and fourth column of $R_{1:r,:}$ shown in \cref{eq:Xbar150}, and hence belong to $\ovl{X}$.
So we have the order $\kappa_1=\kappa_2=\kappa_4=1$.
$b_3$ is half of the second column of $R_{1:r,:}$  and thus the order $\kappa_3=2$.
To obtain the $q_3^+$ and $q_3^-$ (\cref{eq:qpr+-}), which will be used to determine $\Delta y$, we write $b_3$ as an integer combination of the columns of $R_{1:r,:}$ as
\begin{equation}
b_3 = 2\begin{pmatrix} 1\\1\\1\end{pmatrix} + \begin{pmatrix} 1\\0\\0\end{pmatrix} -\begin{pmatrix} 2\\1\\2\end{pmatrix} =  R_{1:r,:} q_3,
\end{equation}
with
\begin{equation}
q_3 = (2,1,0,-1)^T.
\end{equation}
Due to \cref{eq:qpr+-}, we have $q_3^-=(0,0,0-1)^T$.
According to \cref{eq:y0} we have
\begin{equation}
\Delta y = -(\kappa_3-1) R_{1:r,:} q_3^- = (2,1,2)^T. \label{eq:Dy-150}
\end{equation}
Now we calculate the distances from $\Delta y$ to the boundaries of $X$.
Using the $A$ matrix in \cref{eq:X150}, we obtain (i) distance from $\Delta y$ to the boundary $y_1-y_2=0$ is $(A y)_1 = 2-1=1$, (ii) distance from $\Delta y$ to the boundary $y_1-y_3=0$ is $(A y)_2 = 2-2=0$, (iii) distance from $\Delta y$ to the boundary $2y_2-y_3=0$ is $(A y)_3 = 2-2=0$, and (iv) distance from $\Delta y$ to the boundary $y_3=0$ is $(A y)_4 =2$.
This means the points in $\mbb{Z}^r\cap X-\ovl{X}$ satisfy either $y_1-y_2=0$ or $ y_3=0,1$, which is consistent with \cref{eq:150-Z2-1,eq:150-Z2-2,eq:150-Z2-3}.

%In the above, we picked a specific $q_l \in \mbb{Z}^{N_{EBR}}$ satisfying $b_l = R_{1:r,:}q_l$ to determine $\Delta y$.
%However, in practice, we can make use of the nonuniqueness of $q_l$ to minimize the distances from $\Delta y$ to the boundaries of $X$.

\subsubsection{Determining the $\mbb{Z}_2$-type indices} \label{sec:determine-Z2}
\begin{figure}
\begin{centering}
\includegraphics[width=0.7\linewidth]{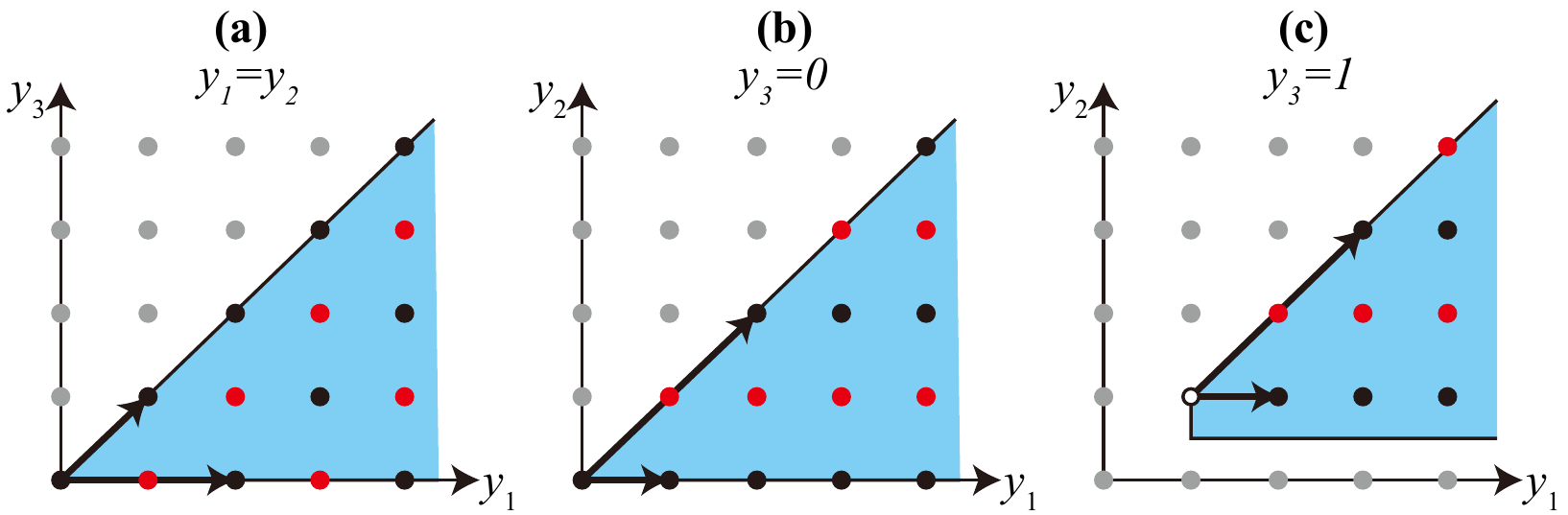}
\par\end{centering}
\protect\caption{ $W^{(i,d)}$'s in SG 150. (a) $W^{(1,0)}$ in the $y_1=y_2$ plane. (b) $W^{(4,0)}$ in the $y_3=0$ plane. In (a)-(b) the polyhedral cone $W^{(i,0)}$ is represented by the shaded area, the generators of $\ovl{W}^{(i,0)}$ are represented by the bold black arrows, the points in $\ovl{W}^{(i,0)}$ are represented by black dots, and the points in $\mbb{Z}^3\cap W^{(i,0)} - \ovl{W}^{(i,0)}$ are represented by red dots. 
(c) $W^{(4,1)}$ in the $y_3=1$ plane. The polyhedron $W^{(4,1)}$ is represented by the shaded area, the points (only one) in the set $\ovl{V}^{(4,1)}$ are represented by the hollow circle, the points in $\ovl{W}^{(4,1)}$ are represented by black dots, and the points in $\mbb{Z}^3\cap W^{(4,1)} - \ovl{W}^{(4,1)}$ are represented by red dots. 
\label{fig:150W}}
\end{figure}

We emphasize that in general $X-(\Delta y+X)$ defined in \cref{eq:X-shiftX} is \textit{not} a polyhedron, because the \cref{eq:X-shiftX} does not match the definition of polyhedron in \cref{thm:polyhedron}.
For example if we take $X$ as $X=\{y\in \mbb{R}^2\ |\ y_1\ge0,\ y_2\ge 0\}$ and $\Delta y=(1,1)$. 
Then $X- (\Delta y + X) = \{ y\in \mbb{R}^2\ |\ 0\le y_1\le1,\ y_2\ge0\} + \{ y\in \mbb{R}^2\ |\ 0\le y_2\le1,\ y_1\ge0\}$ is obviously not a polyhedron.
Nevertheless, the integer points in $X-(\Delta y+X)$ belong to some lower-dimensional polyhedra, \ie
\beq
\mbb{Z}^r \cap (X-(\Delta y+X)) = \bigoplus_i \bigoplus_{d=0}^{{(A\Delta y)_i-1}} \mbb{Z}^r \cap W^{(i,d)}, \label{eq:ZX-shiftX-decompose}
\eeq
with 
\beq 
W^{(i,d)} = \{x \in \mbb{R}^r | (Ax)_i=d\ \text{and}\ Ax\ge 0\} \label{eq:W(id)}
\eeq
a $(r-1)$-dimensional polyhedron.
Since $\Delta y$ shifts all the points in $\mbb{Z}^r\cap X$ into $\ovl{X}$, \cref{eq:ZX-shiftX-decompose} sums over all the lower-dimensional polyhedra close to the boundaries with distances up to the distances from $\Delta y$ to the boundaries.
By definition (\cref{thm:cone} in \cref{sec:math}), the H-representation of a polyhedral cone consists of a set of inequalities and a set of homogeneous equations, \ie $P = \{ x\in \mbb{R}^r\ |\ Ax\ge 0,\ Cx=0\}$, with $A$ some $n\times r$ matrix and $C$ some $m\times r$ matrix for some $n$ and $m$.
Thus $W^{(i,0)}$ is a polyhedral cone in the $d=0$ subspace, but in general, $W^{(i,d)}$ ($d>0$) is neither polyhedral cone nor a shifted polyhedral cone, \ie $v + P = \{ A (x-v)\ge 0, C(x-v)=0\}$.
$W^{(i,d)}$ is a shifted polyhedral cone only if we can find some $v \in \mbb{R}^r$ such that $(Av)_j=\delta_{ij} d$ and hence $W^{(i,d)}$ can be written as $\{(A(x-v))_i=0\ \text{and}\ A(x-v)\ge 0 \}$. However, such $v$ does not exist in general case where $d>0$. For example, if $A_{j,:}$ ($j=1\cdots,i-1,i+1,\cdots r+1$) are all linearly independent, then $v=0$ due to $(Av)_j=0$ ($j\neq i$), which is in contradiction with the condition $(Av)_i=d$.
In the example discussed in the end of this section, as shown in \cref{fig:150W}c, the  $W^{(4,1)}$ is not a shifted polyhedral cone.

The trivial integer points in $W^{(i,d)}$ are given by
\begin{equation}
\ovl{W}^{(i,d)} = \ovl{X} \cap W^{(i,d)} = \{ R_{1:r,:} p\ |\ p\in\mbb{N}^{N_{EBR}}\ \text{and}\ (AR_{1:r,:} p)_i =d\ \text{and}\ AR_{1:r,:}p\ge0\} 
\end{equation}
As each column in $R_{1:r,:}$ represents a point in $X$ and hence certainly satisfies the inequalities of $X$, \ie $\forall j,\;A R_{1:r,j}\ge 0$, and hence $AR_{1:r,:}p\ge0$ is redundant, we can rewrite $\ovl{W}^{(i,d)}$ as
\begin{equation}
\ovl{W}^{(i,d)} = \{ R_{1:r,:} p\ |\ p\in\mbb{N}^{N_{EBR}}\ \text{and}\ (AR_{1:r,:}p)_i =d\}. \label{eq:Wbar(id)}
\end{equation}
In the following we will derive the criterion for a point $x \in \mbb{Z}^r \cap W^{(i,d)}$ to not belong to $\ovl{W}^{(i,d)} $ (such that $x$ represents a fragile state).
We first consider the case $d=0$.
Due to \cref{eq:Wbar(id)}, the monoid $\ovl{W}^{(i,0)}$ is generated from the columns of $R_{1:r,:}$ that satisfy $ (Ax)_i= 0$.
We denote the columns of $R_{1:r,:}$ satisfying $(Ac)_i= 0$ columns as $\{C_1^{(i)}, C_2^{(i)},\cdots\}$. 
In principle, there are two kinds of points in $\mbb{Z}^r\cap W^{(i,0)} - \ovl{W}^{(i,0)}$, which include the points representing EFPs: (i) points that cannot be written as any integer combinations of $\{C_1^{(i)}, C_2^{(i)},\cdots\}$ but can only written as fractional combinations of $\{C_1^{(i)}, C_2^{(i)},\cdots\}$, and (ii) points can be written as some integer combinations of $\{C_1^{(i)}, C_2^{(i)},\cdots\}$ but cannot be written as nonnegative integer combinations of $\{C_1^{(i)}, C_2^{(i)},\cdots\}$.
However, as will be explained in \cref{sec:observation}, case-(ii) does not exist in practice.
Thus we only need to consider case-(i).
We denote the matrix consisting of the columns $\{C_1^{(i)}, C_2^{(i)},\cdots\}$. as $C^{(i)}$, \ie $C^{(i)}=(C_1^{(i)},C_2^{(i)},\cdots)$.
Then a point $x\in\mbb{Z}^r\cap W^{(i,0)}$ belongs to case-(i) only if $x = C^{(i)}p$ has \textit{no} integer solution, where $p$ is regarded as the variable.
To see whether such integer solutions exist, here we apply the Smith Decomposition technique again.
We write $C^{(i)}$ as $C^{(i)} = L^{(i)} \Lambda^{(i)} R^{(i)}$, where $L^{(i)}$, $R^{(i)}$ are unimodular integer matrices, and $\Lambda^{(i)}$ is a diagonal integer matrix.
Here we assume the rank of $C$ is $r^{(i)}$, and thus the first $r^{(i)}$ diagonal elements of $\Lambda^{(i)}$ are nonzero.
(The Smith Decomposition matrices of $C^{(i)}$ are indexed by $i$. One should not confuse them with the Smith Decomposition matrices of the EBR matrix, \ie $L\Lambda R$.)
Then the equation $x = C^{(i)}p$ has integer solutions only if 
\begin{equation}
(L^{(i)-1} x)_j =0 \mod \Lambda^{(i)}_{jj},\qquad \text{for}\; j\le r^{(i)}  \label{eq:Zn-criterion-tmp}
\end{equation}
because when \cref{eq:Zn-criterion-tmp} is true we can write the integer solution as 
\begin{equation}
p_j= \sum_{k=1}^{r^{(i)}} R^{(i)-1}_{jk} \frac{1}{\Lambda_{kk}^{(i)}} (L^{(i)-1} x)_k.
\end{equation}
Therefore, we conclude that the fragile criterion to diagnose points in $ \mbb{Z}^r \cap W^{(i,0)} - \ovl{W}^{(i,0)}$ is
\begin{equation}
(A x)_i=0,\qquad \text{and}\quad \delta^{(i)}(x) \neq 0 \qquad (\text{for } \Lambda^{(i)}_{jj}>0). \label{eq:Zn-criterion}
\end{equation}
where $\delta^{(i)}(x)$ is a vector consisting of the $\mbb{Z}_n$-type fragile indices
\begin{equation}
\delta^{(i)}_j(x) = (L^{(i)-1} x)_j \mod \Lambda^{(i)}_{jj} \label{eq:Zn-index}
\end{equation}
As the components where the corresponding $\Lambda_{jj}^{(i)}=1$ always vanish ($0\mod1=0$, $1\mod1=0$), in the following we only keep the components where the corresponding $\Lambda_{jj}^{(i)}>1$.
In practice, $\Lambda_{jj}^{(i)}=2$ is the only case where $\Lambda_{jj}^{(i)}>1$.
Thus all the $\mbb{Z}_n$-type indices are $\mbb{Z}_2$-type indices.

\textbf{Example.}
Here we take SG 199 as an example to show the algorithm described above.
First as discussed in the example in \cref{sec:ZX-Xbar} and shown in \cref{fig:199shift}, $\Delta y$ is $(1,2)^T$. 
Its distance to the first boundary $y_1=0$ is $(A\Delta y)_1 = 1$, and its distance to the second boundary $y_2-2y_1=0$ is $(A\Delta y)_2 = 0$, where $A$ is given in \cref{eq:A-199}.
Due to \cref{eq:ZX-shiftX-decompose} we only need to consider the sub-polyhedron $W^{(1,0)} = \{ y\in\mbb{R}^2\ |\ (Ay)_1=0,\ (Ay)_2\ge0 \}$, which, due to $A$ in \cref{eq:W(id),eq:A-199}, is given as
\begin{equation}
W^{(1,0)} = \{ y\in \mbb{R}^{2}\;|\; y_1=0,\;y_2\ge 0\}.
\end{equation}
It contains the integer points
\begin{equation}
\mbb{Z}^2\cap W^{(1,0)} = \{(0,0),\ (0,1),\ (0,2),\cdots \}.
\end{equation}
Among the three columns of $R_{1:r,:}$ (\cref{eq:R1r-199}), only the first column satisfies $ (A c)_1 = 0$.
Thus, due to \cref{eq:Wbar(id)}, $\ovl{W}^{(1,0)}$, which represent trivial points in $W^{(1,0)}$, is given as
\begin{equation}
\ovl{W}^{(1,0)} = \{ p(0,2)^T\; |\; p\in \mbb{N}\},
\end{equation}
and the $C^{(1)}$ matrix is given as $C^{(1)} = (0,2)^T$.
Following the algorithm described in last paragraph, we calculate the Smith Decomposition of $C^{(1)}$
\begin{equation}
C^{(1)} = L^{(1)} \Lambda^{(1)} R^{(1)} = 
\begin{pmatrix} 0 & 1 \\ 1 & 0  \end{pmatrix}
\begin{pmatrix} 2 \\ 0  \end{pmatrix}
(1).
\end{equation}
Substituting $L^{(1)}$, $\Lambda^{(1)}$, and \cref{eq:A-199} into \cref{eq:Zn-criterion,eq:Zn-index}, we obtain the fragile criterion as
\begin{equation}
y_1=0,\qquad \text{and}\quad \delta^{(1)}(y) = y_2 \neq 0 \mod 2, 
\end{equation}
which is identical with Eq. (9) in the main text.

\textbf{Example.}
We take SG 150 as another example to show the criteria to diagnose points in $\mbb{Z}^r \cap W^{(i,0)} - \ovl{W}^{(i,0)}$.
According to \cref{eq:ZX-shiftX-decompose}, we only need to analyse the $W^{(i,d)}$'s with $(A\Delta y)_i-1\ge d$.
As discussed in the example in \cref{sec:ZX-Xbar}, the shift vector is $\Delta y=(2,1,2)^T$ (\cref{eq:Dy-150}), and its distances to the four boundaries defined by $A$ in \cref{eq:X150} (\cref{fig:150X}b-e) are $(A\Delta y)_1=1$, $(A\Delta y)_2=0$, $(A\Delta y)_3=0$, $(A\Delta y)_4=2$, respectively.
Thus, we only need to consider the subpolyhedron $W^{(1,0)}$, $W^{(4,0)}$, $W^{(4,1)}$.
Here we only calculate the criteria in $W^{(1,0)}$ and $W^{(4,0)}$.
Due to the $A$ matrix for SG 150 in \cref{eq:X150}, $(Ay)_1=0$ gives the equation $y_1=y_2$, and $(Ay)_4=0$ gives the condition $y_3=0$.
Then, following the definition of $W^{(i,d)}$ in \cref{eq:W(id)}, we obtain
\begin{align}
W^{(1,0)} &= \{ y\in \mbb{R}^3\;|\; y_1=y_2,\; y_1\ge y_3,\; y_3\ge 0\}, \label{eq:W10-150} \\ 
W^{(4,0)} &= \{ y\in \mbb{R}^3\;|\; y_3=0,\; y_1\ge y_2,\; y_2\ge 0 \}. \label{eq:W40-150}
\end{align}
Now let us determine the trivial point monoids $\ovl{W}^{(1,0)}$ and $\ovl{W}^{(4,0)}$ due to \cref{eq:Wbar(id)}.
For $\ovl{W}^{(1,0)}$, among the four columns of $R_{1:r,:}$ (\cref{eq:Xbar150}), only the first $(1,1,1)^T$ and third $(2,2,0)^T$ satisfy $(Ac)_1=0$.
For $\ovl{W}^{(4,0)}$, among the four columns of $R_{1:r,:}$, only the second $(1,0,0)^T$ and third $(2,2,0)^T$ satisfy $(Ac)_4=0$.
Thus 
\begin{align}
\ovl{W}^{(1,0)} &= \{ p_1(1,1,1)^T + p_2 (2,2,0)^T \;|\; p_1, p_2 \in \mbb{N} \}, \\
\ovl{W}^{(4,0)} &= \{ p_1(1,0,0)^T + p_2 (2,2,0)^T \;|\; p_1, p_2 \in \mbb{N} \}. \label{eq:Wbar40-150}
\end{align}
From \cref{fig:150W}a, where $W^{(1,0)}$ and $\ovl{W}^{(1,0)}$ are plotted, one can  conclude the criterion in $W^{(1,0)}$ is $y_1=y_2$ and $y_1-y_3=1\mod2$, which is identical with \cref{eq:150-Z2-3}.
Similarly, from \cref{fig:150W}b, where $W^{(4,0)}$ and $\ovl{W}^{(4,0)}$ are plotted, we can conclude the criterion in $W^{(4,0)}$ is $y_3=0$ and $y_2=1\mod2$, which is identical to \cref{eq:150-Z2-1}.
In the following we show how to get these criteria by following the algorithm from \cref{eq:Wbar(id)} to \cref{eq:Zn-criterion}.
For $W^{(1,0)}$, the $C^{(1)}$ matrix and its Smith Decomposition are 
\begin{equation}
C^{(1)} =\begin{pmatrix}
1 & 2\\ 
1 & 2\\
1 & 0
\end{pmatrix} = 
\begin{pmatrix}
1 & 1 & 0\\
1 & 1 & 1\\
1 & 0 & 0
\end{pmatrix}
\begin{pmatrix}
1 & 0\\ 0 & 2
\end{pmatrix}
\begin{pmatrix}
1 & 0 \\ 0 & 1
\end{pmatrix}.
\end{equation}
The inversion of $L^{(1)}$ is 
\begin{equation}
L^{(1)-1}=\pare{\begin{array}{rrr}
 0 & 0 & 1\\
 1 & 0 &-1\\
-1 & 1 & 0
\end{array}}.
\end{equation}
Substituting $L^{(1)-1}$ and $\Lambda^{(1)}$ into \cref{eq:Zn-criterion,eq:Zn-index} we obtain
\begin{equation}
y_1-y_2=0,\qquad\text{and}\quad \delta^{(1)}(y)= y_1-y_3\neq0\mod2. \label{eq:y1-y2=0-criterion-150}
\end{equation}
For $W^{(4,0)}$, the $C^{(4)}$ matrix and its Smith Decomposition are 
\begin{equation}
C^{(4)} =\begin{pmatrix}
1 & 2\\ 
1 & 2\\
1 & 0
\end{pmatrix} = 
\begin{pmatrix}
1 & 0 & 0\\
0 & 1 & 0\\
0 & 0 & 1
\end{pmatrix}
\begin{pmatrix}
1 & 0\\ 0 & 2
\end{pmatrix}
\begin{pmatrix}
1 & 2 \\ 0 & 1
\end{pmatrix},
\end{equation}
The inversion of $L^{(4)}$ is 
\begin{equation}
L^{(4)-1}=\pare{\begin{array}{rrr}
1 & 0 & 0\\
0 & 1 & 0\\
0 & 0 & 1
\end{array}}.
\end{equation}
Substituting $L^{(4)-1}$ and $\Lambda^{(4)}$ into \cref{eq:Zn-criterion,eq:Zn-index} we obtain
\begin{equation}
y_3=0,\qquad\text{and}\quad  \delta^{(4)}(y)=y_2\neq0\mod2. \label{eq:y3=0-criterion-150}
\end{equation}

Now we consider the remaining part: the points in $\mbb{Z}^r \cap W^{(i,d)} - \ovl{W}^{(i,d)}$ for $d>0$.
In general, a point $x \in \mbb{Z}^r\cap {W}^{(i,d)}$ decomposes into two parts $x^\pr + x^\prpr$: the first part is generated from $\{C^{(i)}_1,C^{(i)}_2,\cdots\}$ such that $(Ax^\pr)_i=0$, and the second part is generated by the other columns of $R_{1:r,:}$ and $(Ax^\prpr)_i=d$.
We denote the columns of $R_{1:r,:}$ that satisfy $(Ax)_i>0$ as $\{D^{(i)}_1, D^{(i)}_2,\cdots\}$, and the matrix consisting of these columns as $D^{(i)} = (D^{(i)}_1, D^{(i)}_2,\cdots)$.
(This decomposition is in general not unique due to the possible linear dependencies between columns of $R_{1:r,:}$.
For example, if there is $C^{(i)}_1 = D^{(i)}_1 - D^{(i)}_2$, then $x=D^{(i)}_1$ has two at least different decompositions: $x^\pr=0$, $x^\prpr=D^{(i)}_1$ or $x^\pr=C^{(i)}_1$, $x^\prpr=D^{(i)}_2$. 
)
Then we can rewrite $\ovl{W}^{(i,d)}$ as
\begin{equation}
\ovl{W}^{(i,d)} = \brace{x=x^\pr + x^\prpr\ | x^\pr\in \ovl{W}^{(i,0)},\; x^\prpr \in \ovl{V}^{(i,d)}},
\end{equation}
where
\begin{equation}
\ovl{V}^{(i,d)} = \brace{ v \ |\ v = p_1 D^{(i)}_1 + p_2 D^{(i)}_2+\cdots,\; p_{1,2\cdots}\in \mbb{N} \; \text{and}\;  (Av)_i =d }. \label{eq:V(id)}
\end{equation}
Since all the columns in $D^{(i)} $ satisfy $ (A D^{(i)}_j) >0$ and the combination coefficients, $p_j$'s, are nonnegative integers and $(Av)_i=d$ is finite, $\ovl{V}^{(i,d)}$ is always finite.
Particularly, $\ovl{V}^{(i,0)}=\{0\}$.
By this construction, $\ovl{W}^{(i,d)}$ can be thought as a sum of shifted $\ovl{W}^{(i,0)}$'s shifted by vectors in $\ovl{V}^{(i,d)}$, \ie
\beq
\ovl{W}^{(i,d)} = \bigoplus_{v\in \ovl{V}^{(i,d)}} v + W^{(i,0)}, \label{eq:W(id)-decompose}
\eeq
where $v + W^{(i,0)} = \{v + y\ |\ y\in \ovl{W}^{(i,0)}\}$.
The sum in \cref{eq:W(id)-decompose} is finite because $\ovl{V}^{(i,d)}$ is a finite set.
A point $x$ belongs to $v+\ovl{W}^{(i,0)}$ only if $x-v$ belongs to $\ovl{W}^{(i,0)}$.
For $x-v$ to belong to $\ovl{W}^{(i,0)}$, first it should belong to the polyhedral cone $W^{(i,0)}$, \ie $A(x-v)\ge0$, and second it should have vanishing $Z_n$-type indices such that it is a trivial point in $\mbb{Z}^r\cap W^{(i,0)}$.
Thus for a point $x\in W^{(i,d)}$ we have
\begin{equation}
x \in v+\ovl{W}^{(i,0)}\quad\Leftrightarrow\quad A(x-v) \ge 0,\;\text{and}\; \delta^{(i)}(x-v) = 0,
\end{equation}
where $\delta^{(i)}(x)$ is defined in \cref{eq:Zn-index}, and $\delta^{(i)}(x-v)=0$ means $\delta^{(i)}_j(x-v)=0$ for \textit{all} $j$.
As the $\mbb{Z}_n$-type fragile indices are additive (\cref{eq:Zn-index}), this condition can be equivalently written as
\begin{equation}
x \in v+\ovl{W}^{(i,0)}\quad\Leftrightarrow\quad  A(x-v) \ge 0,\;\text{and}\; \delta^{(i)}(x) = \delta^{(i)}(v).  \label{eq:x-in-W(i0)-cond}
\end{equation}
For a point $x \in \mbb{Z}^r\cap W^{(i,d)}$ to be outside $\ovl{W}^{(i,d)}$, which is a sum of some shifted $\ovl{W}^{(i,0)}$ (\cref{eq:W(id)-decompose}), $x$ needs to be outside of all of the shifted $\ovl{W}^{(i,0)}$'s.
In other words, for a point $x$ outside $\ovl{W}^{(i,d)}$, the condition in \cref{eq:x-in-W(i0)-cond} is violated for any $W^{(i,0)}$.
Mathematically, the condition for $x\not\in \ovl{W}^{(i,d)}$ is
\begin{equation}
x\not\in \ovl{W}^{(i,d)}\qquad \Leftrightarrow\qquad
\forall v \in \ovl{V}^{(i,d)}\quad \text{either}\quad \exists j\ s.t.\ (A(x-v))_j<0\quad \text{or}\quad \delta^{(i)}(x) \neq \delta^{(i)}(v). \label{eq:Zn-criterion-W(id)}
\end{equation}
For simplicity, here we consider a sufficient condition (which will be shown also necessary later) for $x\in W^{(i,d)}$ to not belong to $x\not\in \ovl{W}^{(i,d)}$
\begin{equation}
x\not\in \ovl{W}^{(i,d)}\quad \Leftarrow \quad
\forall v \in \ovl{V}^{(i,d)}\quad \delta^{(i)}(x) \neq \delta^{(i)}(v).  \label{eq:Zn-criterion-W(id)-s0}
\end{equation}
This condition is obtained by abandoning the $\exists j\ s.t.\ (A(x-v))_j<0$ condition in \cref{eq:Zn-criterion-W(id)}.
A point with the indices ($\delta^{(i)}$) which cannot be realized by any $v\in\ovl{V}^{(i,d)}$ fullfills \cref{eq:Zn-criterion-W(id)-s0}.
Thus we rewrite it as
\begin{equation}
\delta^{(i)}(x) \not\in \brace{ \delta^{(i)}(v) \;\big|\; v\in \ovl{V}^{(i,d)}} \label{eq:Zn-criterion-W(id)-s}
\end{equation} 
In principle, the criterion in \cref{eq:Zn-criterion-W(id)-s} will miss some cases, where $\delta^{(i)}(x)$ equals to $\delta^{(i)}(v)$ for some $v$ in $\ovl{V}^{(i,d)}$, but $x$ does not satisfy $A(x-v)\ge 0$.
However, as will be discussed in \cref{sec:observation}, such cases never appear in practical calculation with TRS and SOC.
Therefore, we will treat \cref{eq:Zn-criterion-W(id)-s} as the fragile criterion in $W^{(i,d)}$ for $d>0$.

\textbf{Example.}
We take SG 150 as an example to show how the algorithm described above works.
As discussed from \cref{eq:W10-150} to \cref{eq:y3=0-criterion-150}, points in $\mbb{Z}^r\cap X-\ovl{X}$ are included in (at least) one of the three subpolyhedra $W^{(1,0)}$, $W^{(4,0)}$, and $W^{(4,1)}$.
The fragile criteria in $W^{(1,0)}$ and $W^{(4,0)}$ are shown in \cref{eq:y1-y2=0-criterion-150,eq:y3=0-criterion-150}, respectively.
Now we are going to work out the fragile criterion in $W^{(4,1)}$.
First, due to the $A$ matrix in \cref{eq:X150} and the $W^{(i,d)}$ definition in \cref{eq:W(id)}, we obtain
\begin{equation}
W^{(4,1)} = \brace{ y\in \mbb{R}^3\;\Big|\; y_3=1,\; y_1\ge y_2,\; y_1\ge1,\; y_2\ge \frac12 }.
\end{equation}
Second, we need to determine the set $\ovl{V}^{(4,1)}$ and $\ovl{W}^{(4,1)}$.
Among the four columns in $R_{1:r,:}$ (\cref{eq:Xbar150}), only the first $(1,1,1)^T$ and the fourth $(2,1,2)^T$ satisfy $(Ax)_4>0$.
To be specific, the first gives $(Ax)_4=1$ and the fourth gives $(Ax)_4=2$.
Then due to \cref{eq:V(id)}, we obtain
\begin{equation}
V^{(4,1)} = \{ (1,1,1)^T \}, \label{eq:V41-150}
\end{equation}
and due to \cref{eq:W(id)-decompose,eq:Wbar40-150}, we obtain
\begin{equation}
\ovl{W}^{(4,1)} = \{ (1,1,1)^T + p_1(1,0,0)^T + p_2 (2,2,0)^T \;|\; p_1, p_2 \in \mbb{N} \}. 
\end{equation}
In \cref{fig:150W}c we plot the $W^{(4,1)}$ and $\ovl{W}^{(4,1)}$.
From \cref{fig:150W}c we can see that the points with odd $y_2$ can always be reached by adding $(1,0,0)^T$ and $(2,2,0)^T$ to $(1,1,1)^T$, while the points even $y_2$ cannot.
Thus we conclude that the criterion to diagnose the points in $\mbb{Z}^3\cap W^{(4,1)} - \ovl{W}^{(4,1)}$ is $y_3=1$ and $y_2=0\mod2$, which is identical with \cref{eq:150-Z2-2}.
Now we apply the algorithm described in the last paragraph to re-derive \cref{eq:150-Z2-2}.
As we already have $V^{(4,1)}$, to get \cref{eq:Zn-criterion-W(id)-s} we only need to calculate the $\mbb{Z}_2$ indices of the points in $V^{(4,1)}$.
Due to \cref{eq:V41-150} and $\delta^{(4)}(y)$ shown in \cref{eq:y3=0-criterion-150}, we obtain $\delta^{(4)}(v)= (v_2\mod2) = 1$.
Then \cref{eq:Zn-criterion-W(id)-s} gives the criterion $ \delta^{(4)}(y) = (y_2\mod2) \not \in \{1\}$, which can be equivalently written as
\begin{equation}
y_3=1,\quad y_2=0\mod2.
\end{equation}

\subsection{Two observations about the results}\label{sec:observation}
In this subsection we discuss two observations about the results obtained from applying our algorithm for every SGs.
These observations have been used to support some conclusions in \cref{sec:Zn-type}.
As discussed in \cref{sec:determine-Z2}, the type-II nontrivial points, \ie $\mbb{Z}^r\cap X - \ovl{X}$, are close to the boundary of $X$ and hence belong to some lower dimensional subpolyhedron of $X$ (\cref{eq:ZX-shiftX-decompose}.)
In each of the subpolyhedron $W^{(i,d)}$ (\cref{eq:W(id)}), the trivial points are denoted as $\ovl{W}^{(i,d)}$ (\cref{eq:Wbar(id)}).
First we focus on the $d=0$ case.
$\ovl{W}^{(i,0)}$ is generated from the columns of $R_{1:r,:}$ that are exactly on the $i$-th boundary of $X$ ($d=(A c)_i=0$).
We denote these columns as $\{C^{(i)}_1,C^{(i)}_2,\cdots\}$.
In general, there are two kinds of nontrivial points in $\mbb{Z}^r\cap W^{(i,0)}$: (i) the points cannot be written as any integer combination of $\{C^{(i)}_1,C^{(i)}_2,\cdots\}$, (ii) the points can be written an integer combination $\{C^{(i)}_1,C^{(i)}_2,\cdots\}$, but at least one of the coefficients is necessarily negative.
However, we found by exhaustive computation that case-(ii) does not exists in practical calculation with TRS and SOC.
Now we prove this statement based on an observation about the $\{C^{(i)}_1,C^{(i)}_2,\cdots\}$.
Let us assume there is a point $x$ belonging to case-(ii).
On one hand, as said above, $x$ can be written as an integer combination $\{C^{(i)}_1,C^{(i)}_2,\cdots\}$, but at least one of the coefficients is negative.
On the other hand, as $x$ belongs to $W^{(i,0)}$, $x$ can be written as a linear combination of the columns of $\{C^{(i)}_1,C^{(i)}_2,\cdots\}$ where the coefficients are positive and rational.
$x$ can in principle have two different decompositions because $\{C^{(i)}_1,C^{(i)}_2,\cdots\}$ are not linearly independent. 
Now let us see whether the linear dependencies can change positive and rational coefficients into integers coefficients where at least one is negative. 
We enumerate all the linear dependencies of $\{C^{(i)}_1,C^{(i)}_2,\cdots\}$ in all SGs, and we find there are only two kinds of dependencies: (A) $c_1+c_2=c_3+c_4$ and (B) $\frac12 c_1 + \frac12 c_2=c_3$, where $c_{1,2,3,4}$ represent different vectors in $\{C^{(i)}_1,C^{(i)}_2,\cdots\}$.
And, we find that for each $W^{(i,0)}$, different linear dependence equations involve completely different sets of vectors, \ie no $C^{(i)}_j$ is contained in two or more linear dependence equations.
For example, in SG 188 ($P\bar{6}c2$), for a particular $\ovl{W}^{(i,0)}$, there are two linear dependencies: $\frac12 C_1^{(i)} +\frac12 C_2^{(i)} = C_3^{(i)}$ and $\frac12 C_4^{(i)} +\frac12 C_5^{(i)} = C_6^{(i)} $.
%For a given decomposition of $x$ on $\td{R}_{1:r,:}$, the first linear dependence can change the coefficients on $c_{1,2,3}$ but leaves the coefficients on $c_{4,5,6}$ invariant; whereas the second linear dependence can change the coefficients on $c_{4,5,6}$ but leaves the coefficients on $c_{1,2,3}$ invariant.
Thus we only need to deal with the linear dependencies separately.
Obviously, $c_1+c_2=c_3+c_4$ can only change an integer coefficient to another integer coefficient.
Then we consider the linear dependence $\frac12 c_1 + \frac12 c_2=c_3$, which in principle could change rational coefficients to integer coefficients.
Now we prove this is not the case. 
Let $x$ be a fragile phase spaned by three columns having dependence $\frac12 c_1 + \frac12 c_2=c_3$.
We notice that only the coefficients of $c_1$ and $c_2$ are fractions ($\frac12$), and thus we consider three cases 
\beq x=\frac12 c_1 + p_1c_1 + p_2 c_2 + p_3 c_3,\eeq 
\beq x=\frac12 c_2+p_1c_1 + p_2 c_2 + p_3 c_3,\eeq 
and
\begin{equation}
x=\frac12 c_1 + \frac12 c_2+p_1c_1 + p_2 c_2 + p_3 c_3
\end{equation}
where $p_{1,2,3}\in \mbb{N}$.
Due to the $\frac12 c_1 + \frac12 c_2=c_3$, the first case can be equivalently written as 
\beqs
x= p_1c_1 + (p_2-\frac12) c_2 + (p_3-1) c_3 = (p_1-\frac12)c_1 + (p_2-1) c_2 + (p_3-2) c_3 = \cdots,
\eeqs
all of which are not integer combinations.
Similarly the second case cannot be written as an integer combination.
The third case is a trivial point as it can be written as 
\begin{equation}
x =p_1c_1 + p_2c_2 +  (p_3+1)c_3.
\end{equation}
Therefore, we conclude that the linear dependencies cannot change positive and rational coefficients into integers coefficients where at least one is negative. 
In other words, the points in $\mbb{Z}^r\cap W^{(i,0)} - \ovl{W}^{(i,0)}$ can never be written as an integer combination $\{C^{(i)}_1,C^{(i)}_2,\cdots\}$ with at least one negative coefficient.

Now we consider the $d>0$ case.
In last section we derive a sufficient condition (\cref{eq:Zn-criterion-W(id)-s}) for a point in $\mbb{Z}^r\cap W^{(i,d)}$ to be nontrivial ($\not \in \ovl{W}^{(i,d)}$).
Here we show that this condition is necessary.
As discussed in \cref{sec:Zn-type}, $\Delta y$ sets the upper bound of $d$. 
We find that in most SGs the maximal $d$ determined by $\Delta y$ is 0, and only for nine exceptions, \ie SGs 150 ($P321$), 157 ($P31m$), 185 ($P6_3cm$), 143 ($P3$), 149 ($P312$), 156 ($P3m1$), 158 ($P3c1$), 165 ($P\bar{3}c1$), 188 ($P\bar{6}c2$), the maximal $d$ is 1.
No higher value is found.
Hence we only need to check the $d=1$ sub-polyhedra in the nine SGs.
Due to the proof in \cref{sec:equivalent}, SGs 157 and 185 are equivalent with SG 150, and SGs 149, 156, 158 are equivalent with SG 143.
Here ``equivalent'' means that there is a one-to-one mapping between the fragile criteria in equivalent SGs \cite{PaperOnTheInductionMethor}.
Thus in fact we only need to check the four inequivalent SGs 150, 143, 165, and 188.
In \cref{sec:P321,sec:P3} we have derived all the fragile criteria in SG 150 and 143 by hand, which are all included in the polyhedron method based criteria, as shown in Table S2 of \cite{SM}.
Therefore the only cases left to be checked are SGs 165 and 188.
Because of the high-rank - ranks of SG 165 and SG 188 are 6 and 7, respectively - we did not derive all the criteria by hand.
Instead, we apply numerical checks: we enumerate all the fragile phases up to a number of bands and then check whether they can be diagnosed by \cref{eq:Zn-criterion-W(id)-s}.
We use a very large of number of bands - six times the largest number of bands of band structures represented by the Hilbert bases of $\ovl{Y}$ - and find no fragile phase is missed by \cref{eq:Zn-criterion-W(id)-s}.
Here the symmetry data vector generators are the $B$ vectors corresponding to the Hilbert bases of $\ovl{Y}$.

\subsection{Equivalent SGs}\label{sec:equivalent}
In this subsection we denote the $\ovl{Y}$ ($\ovl{X}$) monoid for a given SG $G$ as $\ovl{Y}_G$ ($\ovl{X}_G$). 
The definition for two SGs to be equivalent is given as
\begin{mydef}
For two given SGs $G$ and $H$, if there exists an isomorphism between $\ovl{Y}_G$ and $\ovl{Y}_H$, \ie a linear one-to-one mapping $f: \ovl{Y}_H \to \ovl{Y}_G$ (\cref{def:homo}), such that $f$ is also an isomorphism between $\ovl{X}_H$ and $\ovl{X}_G$, then we say $G$ and $H$ are equivalent SGs.
\end{mydef}
If $G$ and $H$ are equivalent, then there is a one-to-one mapping between the fragile phases in $\ovl{Y}_G-\ovl{X}_G$ and $\ovl{Y}_H-\ovl{X}_H$.
Now we derive the equivalence condition. 
First we rewrite $\ovl{Y}_G$ and $\ovl{Y}_H$ as $\mbb{Z}^{r} \cap Y_G$ and $\mbb{Z}^{r} \cap Y_H$, respectively, where $Y_G = \{ Ray\cdot p | p \in \mbb{R}_+^n \} \subset \mbb{R}^r$ and $Y_H = \{ Ray^\pr\cdot p | p \in \mbb{R}_+^{n} \} \subset \mbb{R}^{r}$ are two polyhedral cones. 
Here we assume both $Ray$ and $Ray^\pr$ are $r\times n$ matrices, and $\mrm{rank}(Ray)=\mrm{rank}(Ray^\pr)=r$.
(If $Ray$ and $Ray^\pr$ have different shapes or ranks, $G$ and $H$ cannot be equivalent.)
If $\ovl{Y}_G$ and $\ovl{Y}_H$ are isomorphic, we can represent the isomorphism map $f$ by an $r\times r$ unimodular matrix $F$, the inverse of which is also an integer matrix, such that each column of $F\cdot Ray^\pr$ gives a different column of $Ray$, and every column of $Ray$ is given by some column of $F\cdot Ray^\pr$.
In other words, the columns of $F\cdot Ray^\pr$ are given by an rearrangement of the columns of $Ray$.
Mathematically, there exists an $n\times n$ permutation matrix $S$ such that ${Ray}\cdot S = F\cdot Ray^\pr$.
Given a point $y^\pr= {Ray}^\pr \cdot p^\pr \in \ovl{Y}_H$, $F$ maps it to $y=F y^\pr = {Ray}\cdot (S p^\pr) \in \ovl{Y}_G$; given a point $y={Ray}\cdot p\in \ovl{Y}_G$, $F^{-1}$ maps it to $ y^\pr = S^{-1} y = {Ray}^\pr \cdot (S^{-1})p \in \ovl{Y}_H$.
If there does not exist such $F$ and $S$, $\ovl{Y}_G$ and $\ovl{Y}_H$ cannot be not isomorphic.
Let us assume we have found the matrices $F$ and $S$.
Then we need to check whether $F$ maps $\ovl{X}_H$ to $\ovl{X}_G$.
The condition for $\ovl{X}_G$ to $\ovl{X}_H$ to be isomorphic under $F$ is that the Hilbert bases of $\ovl{X}_G$ and $\ovl{X}_H$ transform to each other under $F$.
%As $F$ is a linear one-to-one correspondence between $\ovl{Y}_G-\ovl{X}_G$ and $\ovl{Y}_H-\ovl{X}_H$, $F$ also gives a one-to-one correspondence between the EFP roots (\cref{sec:rootY}) between $\ovl{Y}_G$ and $\ovl{Y}_H$.

In practice, given two SGs with $Ray$ and $Ray^\pr$ two $r\times n$ matrices, we enumerate all the $n\times n$ permutation matrices, and for each permutation matrix $S$ we try to solve the matrix equation $Ray\cdot S = F \cdot Ray^\pr $.
To solve this matrix equation, we write the Smith Decomposition of ${Ray}^\pr$ as ${Ray}^\pr = L \pare{\Lambda\; 0_{r\times (n-r)} } R$, where $\Lambda$ is an $r$ by $r$ diagonal integer matrix. (One should not confuse this with the Smith Decomposition of the EBR matrix.)
All the $r$ diagonal elements in $\Lambda$ are nonzero because the rank of $Ray$ is $r$, which is true because the polyhedral cone $Y_H$ spanned by $Ray^\pr$ has the dimension $r$. 
We can define the right inverse of $Ray^\pr$ as $\ovl{Ray^\pr} = R^{-1} \begin{pmatrix} \Lambda^{-1} \\ 0_{(n-r)\times r} \end{pmatrix} L^{-1}$ such that $Ray^\pr\cdot \ovl{Ray^\pr} = \mathbbm{1}_{r\times r}$.
Then a necessary condition of $Ray\cdot S = F\cdot Ray^\pr$ is that 
\begin{equation}
Ray\cdot S = F \cdot Ray^\pr \qquad \Rightarrow \qquad F = {Ray}\cdot S \cdot \ovl{Ray^\pr} = {Ray}\cdot S\cdot R^{-1}\begin{pmatrix} \Lambda^{-1} \\ 0_{(n-r)\times r} \end{pmatrix} L^{-1}. \label{eq:F-Ray}
\end{equation}
When $Ray\cdot S \cdot \ovl{Ray^\pr} \cdot {Ray^\pr} =Ray\cdot S $, the right side of \cref{eq:F-Ray} becomes sufficient
\begin{equation}
Ray\cdot S \cdot \ovl{Ray^\pr} \cdot {Ray^\pr} =Ray\cdot S\quad \text{and}\quad F = {Ray}\cdot S \cdot \ovl{Ray^\pr} \quad\Rightarrow\quad Ray\cdot S = F\cdot Ray^\pr. \label{eq:F-Ray2}
\end{equation} 
However, this is not true in general since we usually have $\ovl{Ray^\pr} \cdot {Ray^\pr} \neq \mathbbm{1}_{n\times n}$.
Therefore, the equation $Ray\cdot S = F\cdot Ray^\pr$ has either no solution (when  $Ray\cdot S \cdot \ovl{Ray^\pr} \cdot {Ray^\pr} \neq Ray\cdot S $) or a unique solution (when  $Ray\cdot S \cdot \ovl{Ray^\pr} \cdot {Ray^\pr} =Ray\cdot S $).
On the other hand, even if $F$ in \cref{eq:F-Ray} is a solution of $Ray\cdot S = F\cdot Ray^\pr$, we still need to check whether $F$ is an isomorphism between $\ovl{X}_H$ and $\ovl{X}_G$.
We change to different permutation matrix $S$ until $Ray\cdot S \cdot \ovl{Ray^\pr} \cdot {Ray^\pr} =Ray\cdot S $ and $F$ in \cref{eq:F-Ray} become an isomorphism between $\mrm{Hil}(\ovl{X}_H)$ and $\mrm{Hil}(\ovl{X}_G)$.
As there are $n!$ distinct permutation matrices, this brute force algorithm takes a factorially long time as the $n$ increases.
We have to stop at some finite step.
Therefore, for a given pair of SGs, with finite steps, we cannot guarantee to successfully find the possible equivalent relation.

\begin{figure}
\begin{centering}
\includegraphics[width=0.4\linewidth]{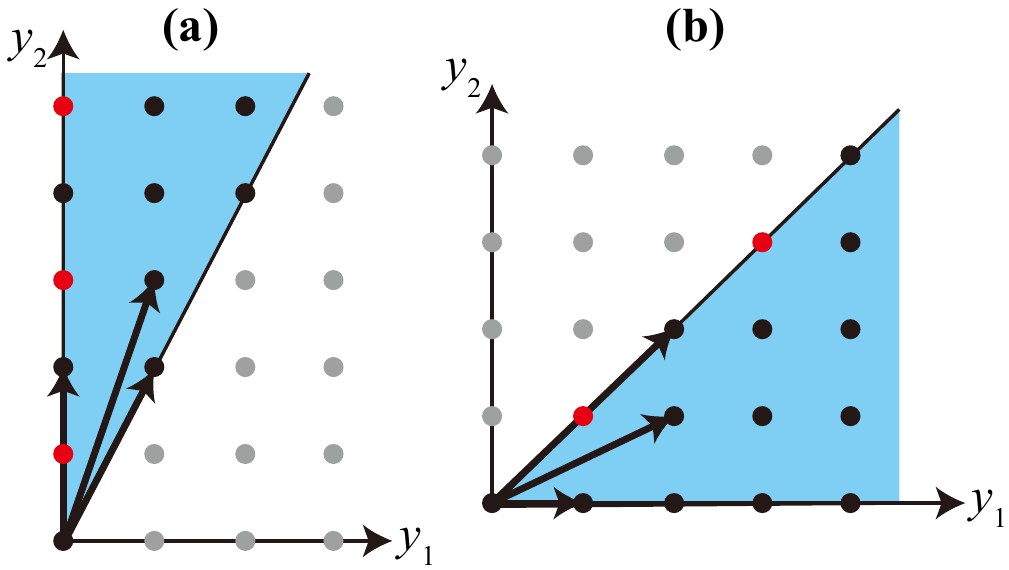}
\par\end{centering}
\protect\caption{ 
Example of equivalent SGs. (a) SG 199. (b) SG 208.
The polyhedral cone $Y$'s are represented by the shaded area, the generators of $\ovl{X}$ are represented by the bold black arrows, the points in $\mbb{Z}^2\cap Y$ are represented by black dots, and the points in $\mbb{Z}^2\cap Y - \ovl{X}$ are represented by red dots. 
The transformation $F=\begin{pmatrix}1 & -1\\ 2 & -1\end{pmatrix}$ transforms $Y$ and $\ovl{X}$ in SG 208 to the $Y$ and $\ovl{X}$ in SG 199.
\label{fig:199208}}
\end{figure}

\textbf{Example.}
We take the equivalent SGs 199 and 208 as examples to show the algorithm.
As shown in \cref{fig:199208}, the $Ray$ matrices of $Y_{199}$ and $Y_{208}$ are 
\begin{equation}
Ray = \begin{pmatrix} 0 & 1\\ 1 & 2 \end{pmatrix},\qquad
Ray^\pr = \begin{pmatrix} 1 & 1\\ 0 & 1 \end{pmatrix},
\end{equation}
respectively.
On the other hand, the generators of $\ovl{X}_{199}$ and $\ovl{X}_{208}$ are 
\begin{equation}
b_1 = (0,2)^T,\qquad b_2=(1,2)^T,\qquad b_3=(1,3)^T,
\end{equation}
and
\begin{equation}
b_1^\pr = (2,2)^T,\qquad b_2^\pr=(1,0)^T,\qquad b_3^\pr=(2,1)^T,
\end{equation}
respectively.
We first choose the permutation matrix as $\begin{pmatrix} 1 & 0\\0 & 1 \end{pmatrix}$, and, due to \cref{eq:F-Ray}, obtain the trial solution
\begin{equation}
F = \begin{pmatrix}
0 & 1\\ 1 & 1
\end{pmatrix}.
\end{equation}
This solution maps $Ray^\pr$ to $Ray\cdot S$. 
But this $F$ is not an isomorphism between $\ovl{X}_{199}$ and $\ovl{X}_{208}$, for example, $F b_1^\pr = (2,4)^T$ is not a generator of $\ovl{X}_{199}$.
Secondly, we choose the permutation matrix as $\begin{pmatrix} 0 & 1\\1 & 0 \end{pmatrix}$ and obtain the trial solution
\begin{equation}
F = \pare{\begin{array}{rr}
1 & -1\\ 2 & -1
\end{array}}.
\end{equation}
This solution maps $Ray^\pr$ to $Ray\cdot S$, and is an isomorphism between $\ovl{X}_{199}$ and $\ovl{X}_{208}$.
To be specific, there are $F b_1^\pr = b_1$, $F b_2^\pr=b_2$, $F b_3^\pr = b_3$.
Therefore, SG 199 and 208 are equivalent.

In the following are the equivalences found by the brute force algorithm (each line is a class of equivalent SGs)
\begin{enumerate}[label=(\arabic*)]
\item 1, 3, 4, 5, 6, 7, 8, 9, 16, 17, 18, 19, 20, 21, 22, 23, 24, 25, 26, 27, 28, 29, 30, 31, 32, 33, 34, 35, 36, 37, 38, 39, 40, 41, 42, 43, 44, 45, 46, 76, 77, 78, 80, 91, 92, 93, 94, 95, 96, 98, 101, 102, , 105, 106, 109, 110, 144, 145, 151, 152, 153, 154, 169, 170, 171, 172, 178, 179, 180, 181.
\item 79, 97, 104, 107, 146, 155, 160, 161, 195, 196, 197, 198, 212, 213.
\item 90, 100, 108.
\item 199, 208, 214, 210.
\item 48, 50, 59, 68.
\item 52, 54, 56, 57, 60, 62, 73, 112, 113, 116, 117, 118, 120
\item 61, 75, 89, 99, 103, 114, 122.
\item 133, 142
\item 150, 157, 185.
\item 159, 173, 182, 186.
\item 209, 211.
\item 63, 72.
\item 135, 138.
\item 143, 149, 156, 158.
\item 168, 177, 183, 184.
\item 218, 219
\item 11, 13, 49, 51, 67.
\item 14, 53, 55, 58, 81, 82, 111, 115, 119.
\item 15, 66
\item 86, 134
\item 85, 125, 129.
\item 12, 65
\item 2, 10, 47
\item 162, 164.
\end{enumerate}

Notice that the SGs equivalent to SG 1 are all the rank-1 SGs.
Thus all the rank-1 SGs do not have EFPs.
This will be explained in more detail in Ref. \cite{PaperOnTheInductionMethor}.
% Suppose the symmetry data vector $B$ is parameterized as $B_i = c_i y$ with $y$ the only parameter and $c_i$'s nonnegative coefficients. 
% Then the condition $B\ge 0$ implies $y\ge0$.
% Thus the $Y$ polyhedral cone of any rank-1 SG is $Y = \{y\in \mbb{R} | y\ge 0\}$.
% All the symmetry data vectors are represented by integers in $Y$, \ie points in $\ovl{Y}=\mbb{Z}\cap Y=\{0,1,2\cdots\}$.
% On the other hand, since $\ovl{X}$, the set of points corresponding to trivial phases, is a subset of $\ovl{Y}$, it has to have the form $\ovl{X} = \{ ap | p\in \mbb{N} \}$, for some postive integer $a$.
% If $a>1$, then $ y=1 $ represents some robust topological state as $y=1$ cannot be written as linear combination of $a$ with integer coefficients.
% Therefore this enforces that $a=1$, and leading to $\ovl{Y}=\ovl{X}=\mbb{N}$.

\section{Twisted bilayer graphene}\label{app:TBG}
In this section, we apply our scheme to the twisted bilayer graphene (TBG).
The single-valley Hamiltonian of TBG has the magnetic SG $P6^\pr2^\pr2$ \cite{Song2018c}.
The irreps of $P6^\pr2^\pr2$ are given in \cref{tab:irreps-TBG}.
We define the symmetry-data-vector as
\begin{equation}
B = (m(\Gamma_1), m(\Gamma_2), m(\Gamma_3), m(\mrm{K}_1), m(\mrm{K_2K_3}), m(\mrm{M}_1), m(\mrm{M}_2) )^T.
\end{equation}
The EBRs of $P6^\pr2^\pr2$ can be found in Table 1 in the supplementary material of \cite{Song2018c}.
From the EBRs we construct the EBR matrix as
\begin{equation}
EBR = 
\left(\begin{array}{rrrrrr}
1 & 0 & 0 & 2 & 0 & 0 \\
0 & 1 & 0 & 0 & 2 & 0 \\
0 & 0 & 1 & 0 & 0 & 2 \\
1 & 1 & 0 & 0 & 0 & 2 \\
0 & 0 & 1 & 1 & 1 & 1 \\
1 & 0 & 1 & 2 & 0 & 2 \\
0 & 1 & 1 & 0 & 2 & 2
\end{array}\right).
\end{equation}
Following the method introduced in \cref{sec:inequality}, we can parameterize the symmetry-data-vector as
\begin{equation}
B = (y_1-y_4,\ y_2-y_4,\ y_4,\ y_1+y_2-2y_3,\ y_3,\ y_1,\ y_2)^T,
\end{equation}
where $y_{1,2,3,4}$ are 
\begin{equation}
y_1 = m(\Gamma_1) + m(\Gamma_3),\qquad
y_2 = m(\Gamma_2) + m(\Gamma_3),\qquad
y_3 = m(\mrm{K_2K_3}),\qquad
y_4 = m(\Gamma_3).
\end{equation}
Following the machinery of the polyhedron method introduced in \cref{sec:index} we obtain two criteria 
\begin{equation}
2y_3-y_4<0,
\end{equation}
\begin{equation}
y_1+y_2-2y_3 = 0,\qquad y_2-y_4=1\mod2.
\end{equation}
Following the algorithm in \cref{sec:rootY}, we obtain two EFP roots
\begin{equation}
b_1 = (1,1,0,1)^T,\qquad b_2 = (1,1,1,0)^T.
\end{equation}

\section{Fu's topological crystalline insulator state and a generalized symmetry eigenvalue criterion}

\subsection{Symmetry eigenvalues of Fu's state}\label{app:Fu-irrep}
\begin{figure}[t]
\centering
\includegraphics[width=0.8\linewidth]{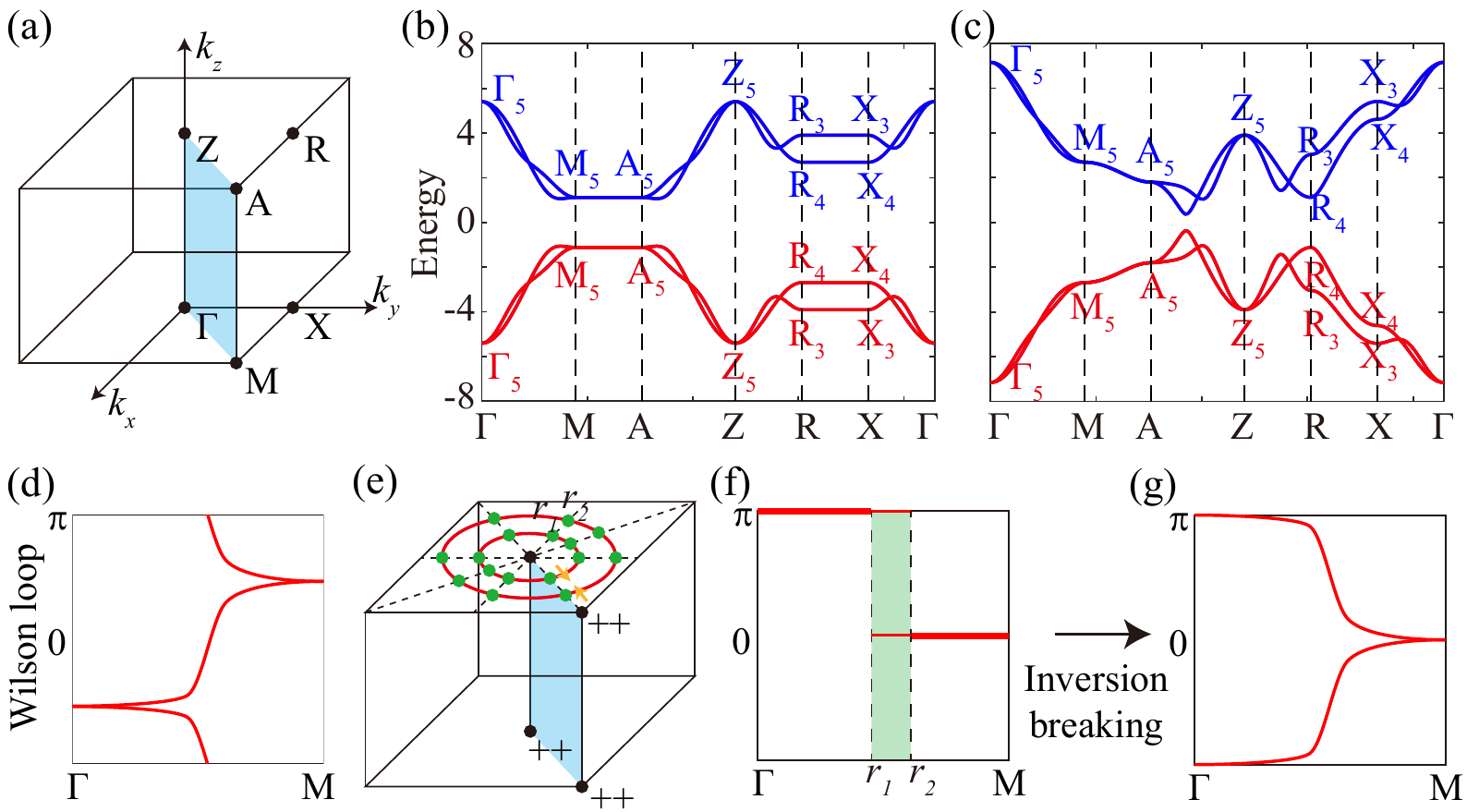}    
\caption{Fu's topological crystalline insulator \cite{Fu2011} and the generalized symmetry eigenvalue criterion. 
(a) Brillouin zone of the SG $P4$. 
(b) Band structure and the irreps in the trivial phase of Fu's model ($t^\pr=0$). 
(c) Band structure and the irreps in the topological phase of Fu's model ($t^\pr=2$). 
The irreps in (b,c) are same as the EBR induced from $p_{x,y}$ orbitals at the $1a$ position.
(d) An illustration of the nontrivial Wilson loop spectrum. 
The Wilson loop operator $W(k_x,k_y)$ is calculated along the $k_z$ direction. The spectrum is plotted along the line $(k_x,k_y) = \rm \Gamma \to M$. 
The crossing at $\Gamma$ and $\rm M$ are protected by $C_4$ and $T$.
(e) For SG $P4/mmm$, which is a supergroup of $P4$, a $\mbb{Z}_2$ invariant can be defined based on the inversion eigenvalues (\cref{eq:Fu-Z2}). 
This $\mbb{Z}_2$ invariant implies a nodal ring semimetal. 
The red circles represent the two nodal rings.
The four dashed lines represent the four $C_2$ rotation axes.
(f) The discontinuous Wilson loop spectrum of the nodal ring semimetal. 
(g) If the symmetry is slightly broken such that $P4/mmm$ reduces to $P422$ and no gap closing happens at $\rm \Gamma,M,Z,A$, the Wilson loop will have a winding protected by $C_4$ and $T$.}
\label{fig:Fu}
\end{figure}

Here we explain why Fu's model cannot be diagnosed through usual symmetry eigenvalue analysis.
Fu's model is
{\small
\begin{align}
H =& \tau_z \sigma_0(\cos k_x + \cos k_y + \cos k_x \cos k_y) +\tau_z\sigma_z(\cos k_x - \cos k_y) + \tau_z \sigma_x \sin k_x \sin k_y +  \tau_x\sigma_0 \pare{\frac52 + \cos k_x + \cos k_y} \nono\\
+& t^\pr ( \tau_x\sigma_0 \cos k_z + \tau_y \sigma_0 \sin k_z).\label{eq:Fu-model}
\end{align}
}%
This model has TRS $T = K$, and $C_4$-rotation symmetry $C_4 = i\sigma_y$, and an mirror symmetry $M_{1\bar10} = \sigma_x$.
The corresponding space group is $P4mm$.
The model is a trivial insulator for $t^\pr=0$.
As $t^\pr$ is increased, a phase transition happens at $t^\pr=3/2$, and then the state becomes topological.
As shown in \cref{fig:Fu}b,c, the trivial phase and the topological phase have the same irreps.
These irreps are same as the EBR induced from $p_{x,y}$ orbitals at the $1a$ position.
(See the \href{http://www.cryst.ehu.es/cgi-bin/cryst/programs/representations.pl?tipogrupo=dbg}{Irreducible representations of the Double Point Groups} and \href{http://www.cryst.ehu.es/cgi-bin/cryst/programs/bandrep.pl}{Band representations of the Double Space Groups} on BCS \cite{Elcoro2017} for the definitions of the irreps and EBRs.)
Thus this state cannot be diagnosed through symmetry eigenvalues.

\subsection{Topological invariant protected by $C_4$ and $T$}
We define the topological invariant based on the Wilson loop. 
The Wilson loop matrix $W(k_x,k_y)$ is given as $W(k_x,k_y) = \lim_{N\to \infty}\prod_{i=0}^{N-1} U^\dagger_{k_x,k_y,\frac{2\pi}{N}i} U_{k_x,k_y,\frac{2\pi}{N}(i+1)}$.
Here $U_{\kk}$ is the matrix $(u_{1\kk}, u_{2\kk}, \cdots )$ with $u_{n\kk}$ the periodic part the $n$th occupied Bloch wavefunction at the wavevector $\kk$.
Now we show that the spectrum of $W(\kk_0)$ for $\kk_0=(0,0),\; (\pi,\pi)$ must be doubly degenerate.
We denote the $C_4$ representation matrix at $(\kk_0,0)$ as $D_{\kk_0}$. 
Since $C_4^2=-1$ (because the model consists of $p_{x,y}$ orbitals), $D_{\kk_0}$ has only two eigenvalues, $i$ and $-i$, which transform to each other under the action of TRS.
As $W(\kk_0)$ commutes with $D_{\kk_0}$, $W(\kk_0)$ is block diagonal in the bases of eigenvectors of $D_{\kk_0}$.
We denote the blocks in the $i$ and $-i$ sectors as $W_i(\kk_0)$ and $W_{-i}(\kk_0)$, respectively.
Since $W_i(\kk_0)$ and $W_{-i}(\kk_0)$ are related by TRS, they must have the identical eigenvalues.
Therefore, each eigenvalue of $W(\kk_0)$ is doubly degenerate.
Now we consider the spectra of $W(\kk_\perp)$ for a continuous path from $\kk_\perp=\Gamma$ to $ \kk_\perp= {\rm M}$.
There are two possible types of connectivity: for the trivial phase, a doublet at $\kk_\perp=\Gamma$ splits into two branches in the intermediate process and then the two branches connect to the same doublet at $\kk_\perp=\rm M$; for the topological phase, a doublet at $\kk=\Gamma$ splits into two branches in the intermediate process and the two branches connect to two adjacent doublets at $\kk=\rm M$, as shown \cref{fig:Fu}d.

\subsection{Generalized symmetry eigenvalue criterion} \label{app:Fu-criterion}
In order to obtain the generalized symmetry eigenvalue criterion for the fragile topology protected by $C_4$ and TRS, we consider two additional symmetries, inversion ($P$) and $C_{2x}$ rotation. 
(\cref{eq:Fu-model} does not have these symmetries.)
With the additional symmetries, the SG is enhanced to $P4/mmm$.
%We will make use of the inversion eigenvalues to determine the spectra of $W(0,0)$ and $W(\pi,\pi)$.
We can think the subsystem in the line $(0,0,k_z)$ as a 1D system with TRS, $C_4$, and $P$ symmetries. 
Since $C_4^2=-1$, the 1D system decomposes into a $C_4=i$ sector and a $C_4=-i$ sector.
The Berry's phase $\theta_1$ in the $C_4=i$ sector can be calculated from the inversion eigenvalues as $e^{i\pi \theta_1} = \prod_{n} \xi^{(i)}_{n,\Gamma} \xi^{(i)}_{n,\rm Z}$, where $\xi^{(i)}_{n \kk}$ is the inversion eigenvalue of the $n$th occupied state in the $C_4=i$ sector at $\kk$.
Similarly, the Berry's phase $\theta_2$ in the $C_4=i$ sector in the line $(\pi,\pi,k_z)$ can be calculated as $e^{i\pi \theta_{2}} = \prod_{n} \xi^{(i)}_{n,\rm M} \xi^{(i)}_{n,\rm A}$.
Due to the TRS, the Berry's phases in the $C_4=-i$ sectors are same as $\theta_{1,2}$.
Then we define the $\mbb{Z}_2$ invariant $\delta$ as the difference of $\theta_1$ and $\theta_2$
\begin{equation}
e^{i\pi \delta} = \prod_{n} \xi^{(i)}_{n,\rm \Gamma} \xi^{(i)}_{n,\rm Z}  \xi^{(i)}_{n,\rm M} \xi^{(i)}_{n,\rm A}.\label{eq:Fu-Z2}
\end{equation}
%As will be explained in the following, $\delta=1$ indeed implies a nodal ring semimetal.
%However, if we slightly break the inversion symmetry, the nodal rings will be gapped and the $\delta=1$ phase will become the fragile phase.
For $\delta=1$, either $\Gamma$ and $\rm M$ or $\rm Z$ and $\rm A$ will have opposite products of inversion eigenvalues in each $C_4$ sector. 
Since $M_z = C_2 P$ and $C_2=C_4^2=-1$, opposite inversion eigenvalues imply opposite $M_z$ eigenvalues.
Therefore, the $\delta=1$ phase has nodal rings protected by $M_z$.
To be specific, we consider the parities shown in \cref{fig:Fu}e, where the two nodal rings are denoted as $r_1$ and $r_2$.
%The two parities at each momentum comes from the $C_4=i$ and $C_4=-i$ sectors.
The $M_z$ eigenvalues $(m_{1}^{k_z=0} m_{2}^{k_z=0}, m_1^{k_z=\pi} m_2^{k_z=\pi})$ inside $r_1$, between $r_1$ and $r_2$, and outside $r_2$ are $(--,++)$, $(--,+-)$, and $(--,--)$, respectively.
According to the correspondence between inversion eigenvalues and Berry's phases \cite{alexandradinata_2014}, the Wilson loop matrices in the three regions have the spectra $(\pi,\pi)$, $(0,\pi)$, and $(0,0)$, respectively, as shown in \cref{fig:Fu}f.

Now we consider to break the inversion symmetry such that the SG reduces to $P422$.
Since the mirror symmetry is absent, the nodal rings in \cref{fig:Fu}e will be gapped. 
However, 16 exceptional gapless points in the four $C_2$ ($C_{2x}$, $C_{2y}$, $C_{2xy}$, $C_{2x\bar y}$) rotation axes will remain.
These gapless crossing points are locally protected by the $C_{2}T$ symmetries and are pinned in the four $C_2$ axes. 
There are two ways to gap out these gapless points.
The first way is to annihilate two crossings in the same $C_2$ axes pairwise, which  will not close the gap at the high symmetry points.
This way is indicated by the yellow arrows in \cref{fig:Fu}e.
%We will show this process in next paragraph in detail.
The second way is to annihilate the eight crossings from the same ring at the Z point or the A point.
The second way will close the gap at the high symmetry points.
Now we prove that the first way gives the topologically nontrivial phase.
As we annihilate the two gapless points, the discontinuous region in the Wilson loop (green region in \cref{fig:Fu}f) will be removed and the Wilson loop spectrum will become continuous.
Due to the $C_{2x\bar y}T$ symmetry, the Wilson loop must be ``particle-hole'' symmetric \cite{Song2018c}.
Therefore, the Wilson loop must have the connectivity shown in \cref{fig:Fu}g, which has a nontrivial winding protected by $C_4$ and $T$.

In the end, by a k$\cdot$p model, we show that the two crossings in the same $C_2$ axes do annihilate each other.
We consider a band inversion of two doublets at the Z point. 
Each of the two doublets has the $C_4$ eigenvalues $\pm i$, and the two doublets have opposite inversions.
Thus the symmetries can be represented as $C_4=i\tau_0\sigma_z$, $P=\tau_z\sigma_0$, $C_{2x}=\tau_0\sigma_x$, $T=K$.
Then the Hamiltonian for the mirror protected nodal ring semimetal is
\begin{equation}
H = (M-q_x^2-q_y^2)\tau_z\sigma_0 + q_z\tau_y \sigma_z,
\end{equation}
where $\mbf{q} = \kk-(0,0,\pi)$.
In this Hamiltonian, the two nodal rings are degenerate.
One can add perturbation terms to split them.
But in order to show that the gapless points can be gapped symmetrically, this Hamiltonian is good enough.
The term $m \tau_x \sigma_0$, which breaks $P$ but preserves $C_4$ and $C_{2x}$, will fully gap the nodal rings.

\section{Related mathematical theorems}\label{sec:math}
In this section, we summarize the mathematical theorems used in the paper.
The theorems are given without proof.
Interested readers might look at Ref. \cite{wiki:snf} for \cref{thm:snf}, Ref. \cite{Fukuda2013Lecture} for \cref{thm:polyhedron,thm:cone}, Refs. \cite{Renner2006,Bruns2009polytopes,Bruns2010} for \cref{def:affine-monoid,thm:Hilbert,thm:Gordan}.

\begin{mythm}\label{thm:snf}
(Smith Decomposition.) If $A$ is an $n\times m$ integer matrix, then there is an $n\times n$ unimodular matrix $L$ and an $m\times m$ unimodular matrix $R$ such that $A = L \Lambda R$, where $\Lambda_{ij} = \delta_{ij} \lambda_i$ is an $n\times m$ integer matrix.
$\lambda_i$ is positive integer for $1\le i\le \mrm{rank}(A)$ and zero for $i>\mrm{rank}(A)$.
$\Lambda$ is referred to as the Smith Normal Form of $A$.
\end{mythm}

\begin{mythm} \label{thm:polyhedron} (Minkowski-Weyl theorem for polyhedra.)
For $P \subseteq \mbb{R}^d$, the following two statements are equivalent:
\begin{enumerate}
\item (H-representation) $P$ is a polyhedron, \ie there exist $A \in \mbb{R}^{m\times d}$, $C \in \mbb{R}^{m^\pr\times d}$, $b \in \mbb{R}^{m}$, and $f \in \mbb{R}^{m^\pr}$ for some $m,m^\pr$, such that $P = \brace{x\in \mbb{R}^d | Ax\ge b,\; Cx = f}$.
\item (V-representation) $P$ is finitely generated, \ie there exist $V \in \mbb{R}^{d \times n}$, $Ray \in \mbb{R}^{d\times n^\pr}$, and $Line \in \mbb{R}^{d\times n^\prpr}$, for some $n,n^\pr,n^\prpr$, such that $
P = \{ V u + Ray\cdot p + Line\cdot q\  \big|\ u \in \mbb{R}^n_+,\ u_1+\cdots+u_n=1,\ p \in \mbb{R}_{+}^{n^\pr},\ q\in \mbb{R}^{n^\prpr} \}$.
\end{enumerate}
\end{mythm}
The dimension of the polyhedron $P$, which is given as $d-\mrm{rank}(C)$, is denoted as $\dim(P)$.
The algorithm to get V-repsentation from H-repsentation or vise versa is available in many mathematical packages.
In this work, we use the \href{http://www.sagemath.org/}{\it SageMath} package \cite{Sage}. 
A special kind of polyhedron is polyhedral cone, where $b=0$, $f=0$, and $V=0$.
For polyhedral cone, \cref{thm:polyhedron} becomes
\begin{mythm} \label{thm:cone} (Minkowski-Weyl theorem for polyhedral cones.)
For $P \subseteq \mbb{R}^d$, the following two statements are equivalent:
\begin{enumerate}
\item (H-representation) $P$ is a polyhedral cone, \ie there exist $A \in \mbb{R}^{m\times d}$ and $C \in \mbb{R}^{m^\pr\times d}$ for some $m,m^\pr$, such that $P = \brace{x\in \mbb{R}^d | Ax\ge 0,\; Cx = 0}$.
\item (V-representation) $P$ is a finitely generated cone, \ie there exist $Ray \in \mbb{R}^{d\times n}$, and $Line \in \mbb{R}^{d\times n^\pr}$, for some $n,n^\pr$, such that 
$P = \{ Ray\cdot p + Line\cdot q\  \big|\ p \in \mbb{R}_{+}^{n},\ q\in \mbb{R}^{n^\pr} \}$.
\end{enumerate}
\end{mythm}
A polyhedral cone is called \textit{pointed} if it dose not contain lines, \ie $Line=0$.
$Line=0$ if $\begin{pmatrix} C\\A\end{pmatrix}$ is a full-rank matrix.
In the case $C=0$, $Line=0$ if the $A$ is a full-rank matrix.
\begin{mydef} \label{def:affine-monoid}
An affine monoid, denoted as $M$, is a finitely generated sub-monoid of a lattice $\mbb{Z}^d$, \ie there exist $r_1, r_2, \cdots r_n \in \mbb{Z}^d$ such that $M=\{r_1 p_1 + r_2 p_2 + \cdots r_n p_n | p_1\cdots p_n \in \mbb{N}\}$. $M$ is called positive if $a,-a\in M \Rightarrow a=0$.
\end{mydef}
\begin{mythm} (Van der Corput theorem.) \label{thm:Hilbert}
Let $M$ be a positive affine monoid. The elements in $M$ that cannot be written as a sum of other elements with positive coefficients are referred to as irreducible elements. Then (i) every element of $M$ is a sum of irreducible elements with positive coefficients, (ii) $M$ has only finitely many irreducible elements, (iii) the irreducible elements form the unique minimal system of generators $\mrm{Hil}(M)=\{b_1,b_2,\cdots\}$ of M, the Hilbert bases.
\end{mythm}
Algorithms to find the Hilbert bases include the Normaliz algorithm \cite{Bruns2010} and the Hemmecke algorithm \cite{Hemmecke2002}, which are available in the \href{https://www.normaliz.uni-osnabrueck.de/}{\it Normaliz} package and the \href{https://4ti2.github.io}{\it 4ti2} package, respectively.
\begin{mythm} (Gordan's Lemma.) \label{thm:Gordan}
Let $P\subseteq \mbb{R}^d$ be a polyhedral cone. Then $P\cap \mbb{Z}^d$ is an affine monoid. And when $P$ is pointed, $P\cap \mbb{Z}^d$ is a positive affine monoid.
\end{mythm}
\begin{mydef} (Monoid homomorphisms.) \label{def:homo}
A homomorphism between two affine monoids $M$ and $N$ is a function $f: M\to N$ such that (i) $f(x+y) = f(x) + f(y) $ for all $x,y$ in $M$, and (ii) $f(0)=0$. A bijective monoid homomorphism is called a monoid isomorphism.
\end{mydef}

% \begin{table}
% \caption{Fragile bands in materials.
% In the first three columns the chemical formulae, the space group numbers, and the
% ICSD numbers of the materials are tabulated.
% The fourth column gives the number of fragile branches in the band structure of the corresponding material.
% In the fifth to tenth columns the information of the fragile branch closest to the Fermi level are tabulated.
% ``Bands'' gives the band indices of the fragile branch. 
% Here we refer the highest occupied band as the $0$th band, and the lowest empty band as the 1st band, \etc
% ``Irreps'' gives the irreps formed by the fragile bands at high symmetry momenta.
% $\Delta_l$ ($\Delta_u$) is the indirect gap between the fragile bands and the lower (upper) bands.
% $\Delta_l^\pr$ ($\Delta_u^\pr$) is the direct gap between the fragile bands and the lower (upper) bands.
% The complete table is available in \cite{SM}.
% \label{tab:material}}
% \end{table}

% \begin{table}
% \caption{Parameterization of the symmetry data. The complete table is available in \cite{SM}. \label{tab:para}}\\
% \end{table}

% % \twocolumngrid
% % \LTcapwidth=0.45\textwidth
% % \clearpage
% % \include{table_para}

% % \clearpage
% % \include{table_criteria}

% \begin{table}
% \caption{Fragile roots in all space groups with significant spin-orbit coupling and time-reversal symmetry. The complete table is available in \href{Supplemantary materials}{http://www.cryst.ehu.es/html/doc/FragileRoots.pdf} \label{tab:root}}
% \end{table}

\end{document}